\documentclass[aps,prl,twocolumn,twoside,a4paper,10pt,amsmath,amssymb,superscriptaddress]{revtex4}
\usepackage{dcolumn}
\usepackage{bm}
\usepackage{CJK}
\usepackage{color}
\usepackage{epstopdf}

\usepackage{graphicx}
\usepackage{dcolumn}
\usepackage{bm}
\usepackage{upgreek}
\usepackage{amsfonts,amssymb}
\usepackage{ulem}
\usepackage{tikz}%

\usepackage[mathlines]{lineno}
\usepackage{float}

\usepackage[colorlinks, linkcolor=blue, anchorcolor=blue, citecolor= blue]{hyperref}

\newcommand{\e}{\mathrm{e}}
\newcommand{\mi}{\mathrm{i}}

\newcommand\myeq{\mkern0.5mu{=}\mkern0.5mu}
\newcommand\myeqv{\mkern0.5mu{\equiv}\mkern0.5mu}
\newcommand\mygg{\mkern0.5mu{\gg}\mkern0.5mu}
\newcommand\myap{\mkern0.5mu{\approx}\mkern0.5mu}

\let\oldaddcontentsline\addcontentsline
\renewcommand{\addcontentsline}[3]{}

\begin{document}
	
	\title{Temporal Spinwave Fabry-P\'{e}rot Interferometry via Coherent Population Trapping}
\author{Ruihuan Fang}
\affiliation{Guangdong Provincial Key Laboratory of Quantum Metrology and Sensing $\&$ School of Physics and Astronomy, Sun Yat-Sen University (Zhuhai Campus), Zhuhai 519082, China}
\affiliation{State Key Laboratory of Optoelectronic Materials and Technologies, Sun Yat-Sen University (Guangzhou Campus), Guangzhou 510275, China}

\author{Chengyin Han}
\affiliation{Guangdong Provincial Key Laboratory of Quantum Metrology and Sensing $\&$ School of Physics and Astronomy, Sun Yat-Sen University (Zhuhai Campus), Zhuhai 519082, China}

\author{Xunda Jiang}
\author{Yuxiang Qiu}
\author{Yuanyuan Guo}
\author{Minhua Zhao}
\affiliation{Guangdong Provincial Key Laboratory of Quantum Metrology and Sensing $\&$ School of Physics and Astronomy, Sun Yat-Sen University (Zhuhai Campus), Zhuhai 519082, China}
\affiliation{State Key Laboratory of Optoelectronic Materials and Technologies, Sun Yat-Sen University (Guangzhou Campus), Guangzhou 510275, China}

\author{Jiahao Huang}
\email{hjiahao@mail2.sysu.edu.cn}
\affiliation{Guangdong Provincial Key Laboratory of Quantum Metrology and Sensing $\&$ School of Physics and Astronomy, Sun Yat-Sen University (Zhuhai Campus), Zhuhai 519082, China}

\author{Bo Lu}
\email{lubo3@mail.sysu.edu.cn}
\affiliation{Guangdong Provincial Key Laboratory of Quantum Metrology and Sensing $\&$ School of Physics and Astronomy, Sun Yat-Sen University (Zhuhai Campus), Zhuhai 519082, China}

\author{Chaohong Lee}
\email{lichaoh2@mail.sysu.edu.cn}
\affiliation{Guangdong Provincial Key Laboratory of Quantum Metrology and Sensing $\&$ School of Physics and Astronomy, Sun Yat-Sen University (Zhuhai Campus), Zhuhai 519082, China}
\affiliation{State Key Laboratory of Optoelectronic Materials and Technologies, Sun Yat-Sen University (Guangzhou Campus), Guangzhou 510275, China}

\date{\today}

\begin{abstract}
	Ramsey spectroscopy via coherent population trapping (CPT) is essential in precision measurements.
	The conventional CPT-Ramsey fringes contain numbers of almost identical oscillations and so that it is difficult to identify the central fringe.
	Here, we experimentally demonstrate a temporal spinwave Fabry-P\'{e}rot interferometry via double-$\Lambda$ CPT of laser-cooled $^{87}$Rb atoms.
	Due to the constructive interference of temporal spinwaves, the transmission spectrum appears as a comb of equidistant peaks in frequency domain and thus the central Ramsey fringe can be easily identified.
	From the optical Bloch equations for our five-level double-$\Lambda$ system, the transmission spectrum is analytically explained by the Fabry-P\'{e}rot interferometry of temporal spinwaves.
	%
	%
	Due to small amplitude difference between the two Land\'{e} factors, each peak splits into two when the external magnetic field is not too weak.
	This peak splitting can be employed to measure an unknown magnetic field without involving magneto-sensitive transitions.
\end{abstract}

\maketitle

Coherent population trapping (CPT)~\cite{Gray:78}, a result of destructive quantum interference between different transition paths, is of great importance in quantum science and technology.
CPT spectroscopy has been extensively employed in quantum engineering and quantum metrology, such as, all-optical manipulation~\cite{PhysRevLett.113.263602, PhysRevA.97.033838, PhysRevLett.115.093602, PhysRevLett.97.247401, PhysRevLett.116.043603, Ni231}, atomic cooling~\cite{ PhysRevLett.61.826}, atomic clocks~\cite{Vanier2005, Merimaa:03, PhysRevApplied.7.014018, PhysRevApplied.8.054001}, and atomic magnetometers~\cite{PhysRevLett.69.1360, Nagel_1998, doi:10.1063/1.1839274, Tripathi2019}.
To narrow the CPT resonance linewidth, one may implement Ramsey interferometry in which two CPT pulses are separated by an integration time of the dark state for a time duration \textit{T}~\cite{PhysRevLett.94.193002,Merimaa:03, PhysRevA.67.065801}.
In a CPT-Ramsey interferometry, the fringe-width $\Delta\upsilon\myeq 1/(2T)$ is independent of the CPT laser intensity and so that one may narrow the linewidth via increasing the time duration \textit{T}~\cite{Merimaa:03, PhysRevA.67.065801}.
However, it becomes difficult to identify the central CPT-Ramsey fringe from adjacent ones, since the adjacent-fringe amplitudes are almost equal to the central-fringe amplitude~\cite{doi:10.1063/1.5001179, Warren2018}.
Thus it becomes very important to suppress the non-central fringes.

In order to suppress the non-central fringes, a widely used and highly efficient way is inserting a  CPT pulse sequence between the two CPT-Ramsey pulses.
By employing the techniques of multi-pulse phase-stepping~\cite{4126869, Yun_2012} or repeated query~\cite{Warren2018}, the non-central fringes have been successfully suppressed.
Similarly, high-contrast transparency comb~\cite{PhysRevA.98.033802} has been achieved via electromagnetically-induced-transparency multi-pulse interference~\cite{Nicolas2018}.
The existed experiments of multi-pulse CPT interference are almost performed under the $\sigma$-$\sigma$ configuration, in which the two-photon transition occurs between states of the same magnetic quantum number.
However, under the $\sigma$-$\sigma$ configuration, atoms will gradually accumulate in a ``trap" state that does not contribute ground-state coherence~\cite{Taichenachev2005a}.
To eliminate undesired atomic accumulations with no contributions to ground-state coherence, one may employ the lin$||$lin configuration~\cite{PhysRevA.79.063837, PhysRevA.81.013833, Mikhailov:10, PhysRevA.88.042120}.
Under the lin$||$lin configuration, a five-level double-$\Lambda$ system is constructed by simultaneously coupling two sets of ground states to a common excited state.
Up to now, the multi-pulse CPT interference has never been demonstrated in experiments under the lin$||$lin configuration.

Moreover, by employing multi-beam interference, the optical Fabry-P\'{e}rot (FP) interferometer has been widely used as a bandpass filter that transmits light of certain frequencies~\cite{2016OExpr..2416366I, Poirson:97}.
In analogy to multi-beam interference in spatial domain, multi-pulse interference in temporal domain has been proposed for two-level systems~\cite{Akkermans2012} and three-level $\Lambda$ systems~\cite{RN5, Nicolas2018}.
The multi-pulse interferences, such as Carr-Purcell decoupling~\cite{PhysRev.94.630} and periodic dynamical decoupling~\cite{Viola_1998}, have enabled versatile applications in quantum sensing~\cite{RevModPhys.89.035002} from narrower spectral response, sideband suppression, to environmental noise filtering.
To the best of our knowledge, it is the first time that we demonstrate the temporal spinwave FP interferometry via multi-pulse CPT-Ramsey interference in a double-$\Lambda$ system.

In this Letter, based upon the double-$\Lambda$ CPT in an ensemble of laser-cooled $^{87}$Rb atoms under the lin$||$lin configuration, we experimentally demonstrate the temporal spinwave FP interferometry.
The interferometry is carried out with the multi-pulse CPT-Ramsey interference.
Due to the temporal spinwave interference, the transmission spectrum appears as a comb with multiple equidistant interference peaks and the central CPT-Ramsey fringe can be easily identified.
The distance between adjacent peaks is exactly the repeated frequency of the applied CPT pulses, analogous to the free spectral range (FSR) of an optical cavity.
Accordingly, side fringes between two interference peaks are suppressed by destructive interference.
Based upon the optical Bloch equations for the five-level double-$\Lambda$ system, we develop an analytical theory for the temporal spinwave FP interferometry and well explain the transmission spectrum in experiments.

\begin{figure}[ht]
	\centering
	\includegraphics[width=1\linewidth]{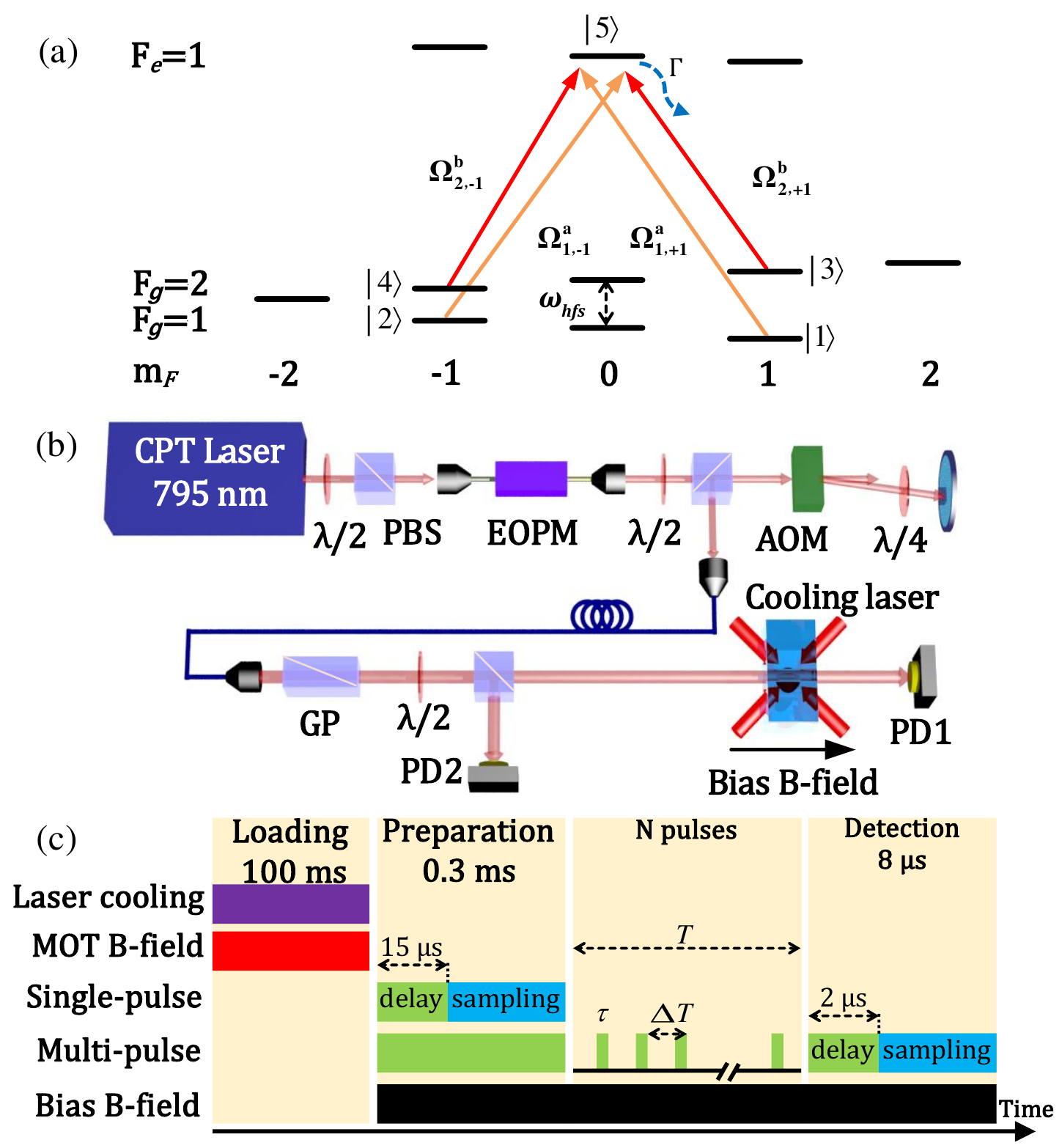}
	\caption{(Color online) Experimental schematic. (a) Energy levels for double-$\Lambda$ CPT of $^{87}$Rb under the lin$||$lin configuration. The symbols are detailedly defined in the main texts. (b) The schematic diagram of the experimental apparatus. PBS: polarization beam splitter, EOPM: electro-optic phase modulator, PD: photodetector, AOM: acousto-optic modulator, GP: Glan prism. (c) Experimental timing sequence. A periodic CPT pulse sequence with pulse length $\uptau$ and pulse period $\Delta T$ is inserted between two CPT-Ramsey pulses respectively called as preparation and detection.}
	\label{fig1}
\end{figure}

The experimental schematic is shown in Fig.~\ref{fig1}.
Under the lin$||$lin configuration, two CPT fields are linearly polarized to the same direction orthogonal to the applied magnetic field.
We choose the double-$\Lambda$ system constructed by the D1 line of $^{87}$Rb, see Fig.~\ref{fig1}(a).
A bichromatic field with frequencies of $\omega_a$ and $\omega_b$ simultaneously couples two sets of ground states $\left\{|1\rangle = |F_{g}\myeq 1, m_{F}\myeq +1\rangle, |2\rangle = |F_{g}\myeq 1, m_{F}\myeq -1\rangle\right\}$ and $\left\{|3\rangle = |F_{g}\myeq 2, m_{F}\myeq +1\rangle, |4\rangle = |F_{g}\myeq 2, m_{F}\myeq -1\rangle\right\}$ to the common excited state $|5\rangle = \left|F_{e}\myeq 1, m_{F}\myeq 0\right\rangle$.
%
%
The eigenfrequencies for five involved levels are respectively $\omega_{1,\pm 1}$, $\omega_{2,\pm 1}$ and $\omega_e$.
%
The two-photon detuning is $\delta=(\omega_a-\omega_b)-\omega_{hfs}$.
$\Gamma$ is the excited-state decay rate.
The Rabi frequencies for transitions from four ground states to the common excited state are respectively denoted by $\Omega_{1,\pm1}^a$ and $\Omega_{2,\pm1}^b$.

We perform the temporal spinwave FP interferometry with laser-cooled atoms released from a magneto-optical trap (MOT).
The schematic diagram of our experimental apparatus is shown in Fig.~\ref{fig1}(b).
Within an ultra-high vacuum cell with the pressure of $10^{-8}$ Pa, the $^{87}$Rb atoms are cooled and trapped via a three-dimensional MOT which is created by laser beams and a quadruple magnetic field produced by a pair of magnetic coils.
Two external cavity diode lasers (ECDL) are used as the cooling and repumping lasers that are locked to the D2 cycling transition with a saturated absorption spectrum (SAS).
In order to eliminate the stray magnetic field, three pairs of Helmholtz coils are used to cancel ambient magnetic fields.
In addition, a pair of Helmholtz coil is used to apply a bias magnetic field aligned with the propagation direction of CPT laser beam to split the Zeeman sublevels.

The CPT laser source is provided by an ECDL locked to the $|F_{g}\myeq 2\rangle \leftrightarrow |F_{e}\myeq 1\rangle$ transition of $^{87}$Rb D1 line at 795 nm.
The CPT beam is generated by modulating a single laser with a fiber-coupled electro-optic phase modulator (EOPM).
The positive first-order sideband forms the $\Lambda$ systems with the carrier.
The 6.835-GHz modulated frequency matches the two hyperfine ground state.
We set the powers of the first-order sidebands equal to the carrier signal by monitoring their intensities with a FP cavity.
Following the EOPM, an acousto-optic modulator (AOM) is used to generate the CPT pulse sequence.
The modulated laser beams are coupled into a polarization maintaining fiber and collimated to an 8-mm-diameter beam after the fiber.
A Glan prism is used to purify the polarization.
Then the CPT beam is equally separated into two beams by a half-wave plate and a polarization beam splitter (PBS).
One beam is detected by the photodetector [PD2 in Fig.~\ref{fig1}(b)] as a normalization signal $S_{N}$ to reduce the effect of intensity noise on the CPT signals.
The other beam is sent to interrogate the cold atoms and collected on the CPT photodetector as $S_{T}$ [PD1 in Fig.~\ref{fig1}(b)].
The transition signal~(TS) are given by $S_{TS} \myeq S_{T}$/$S_{N}$.

\begin{figure}[ht]
	\centering
	\includegraphics[width= 1\linewidth]{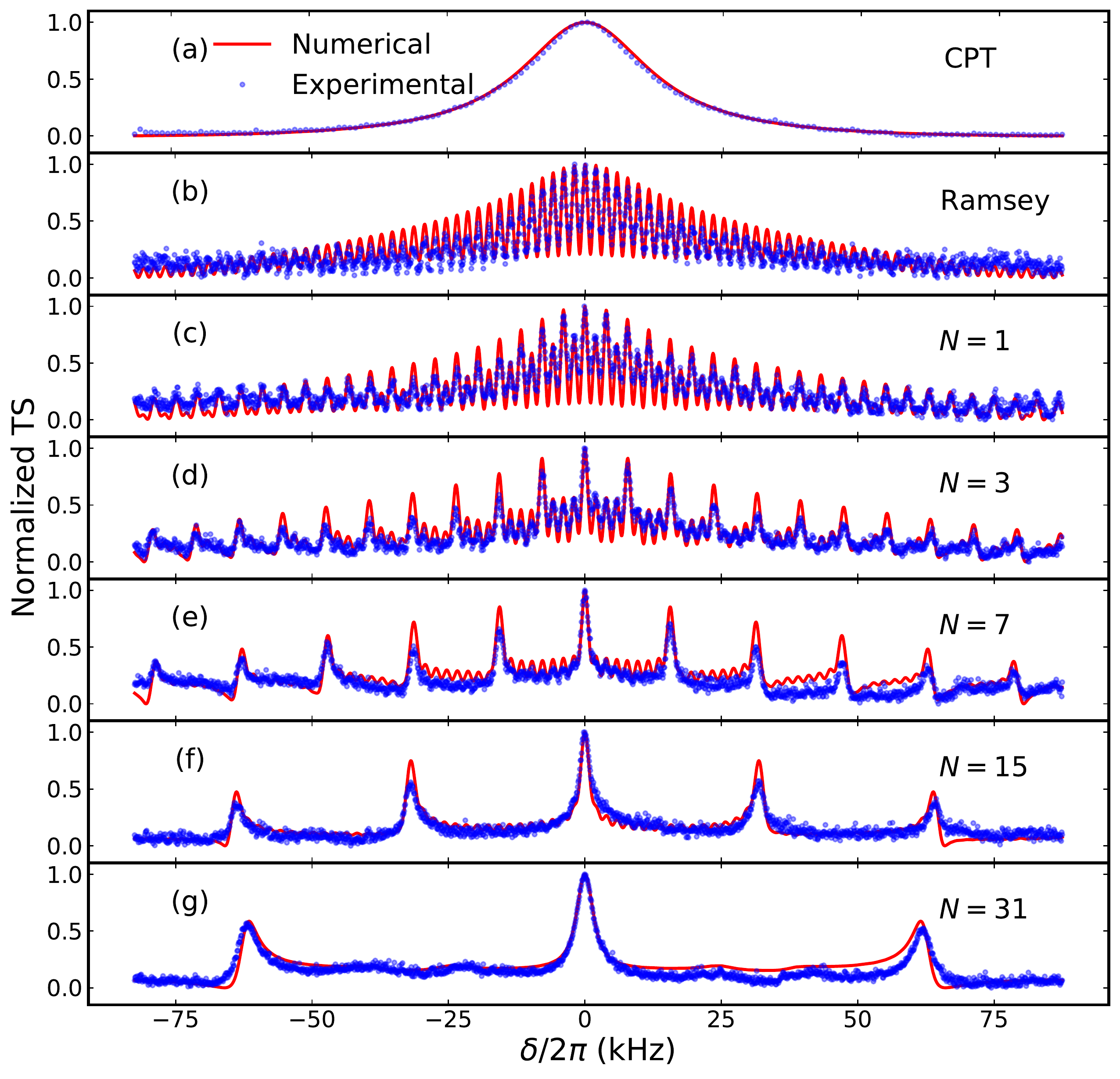}
	\caption{(Color online) Experimental transmission signals (TS) for different schemes. (a) A Lorentz fitting of single-pulse CPT spectrum shows a full width at half maximum (FWHM) is 27 kHz. (b) Two-pulse CPT-Ramsey spectrum is obtained with a integration time of 0.5 ms. (c)-(g) Multi-pulse CPT-Ramsey spectra via temporal spinwave FP interferometry of $N$ equidistant pulses with a length $\uptau \myeq $2 $\mu$s into the integration time of 0.5 ms.}
	\label{fig2}
\end{figure}

To implement the CPT-Ramsey interferometry, about $10^{7}$ $^{87}$Rb atoms are cooled and trapped within a 100-ms cooling period.
Then the atoms are interrogated under free fall after turning off the MOT magnetic field and the cooling laser beams.
In ensure the MOT magnetic field decays to zero, the CPT beams and a bias magnetic field are simultaneously applied after 1 ms waiting time.
The first CPT-Ramsey pulse with a duration of 0.3 ms is used to pump the atoms into the dark state and here called as preparation.
If no following pulses are applied, through averaging the collected voltage signals of this pulse after a delay of $15~\mu\text{s}$, the single-pulse CPT spectrum is obtained by scanning the modulation frequency of EOPM, see Fig.~\ref{fig2}(a).
%
%
By fitting the spectrum with a Lorentz shape, its full width at half maximum (FWHM) is given as 27 kHz.
By comparing the spectrum with the numerical results of optical Bloch equations, the four Rabi frequencies are estimated as $\pm\Omega_{1,\pm 1}^a\myeq \frac{1}{\sqrt{3}}\Omega_{2,\pm 1}^b\myeq 1.25e6~\mathrm{s^{-1}}$, which well agrees with the measured value through CPT light intensity of $I\myeq 1.48 \mathrm{W/m^2}$.
The CPT-Ramsey spectra are given by sampling the signal voltages during the detection pulse (the second CPT-Ramsey pulse) after a delay of $2~\mu\text{s}$.
Without additional CPT pulses between the two CPT-Ramsey pulses, the central fringe in the conventional CPT-Ramsey spectrum is difficult to be distinguished from neighboring fringes, see Fig.~\ref{fig2}(b).

The temporal spinwave FP interferometry is performed by inserting a periodic CPT pulse sequence between the two CPT-Ramsey pulses.
The minimum length of a single pulse is limited to 1 $\mu$s, given by the precision of the controlled digital I/O devices.
In Fig.~\ref{fig2}(c)-(g), we show the spectra for different pulse number $N$.
Given the pulse length $\uptau \myeq $2 $\mu$s and the integration time $T=0.5~\mathrm{ms}$, because of the constructive interference, high-contrast transmission peaks gradually appear when the pulse number $N$ increases.
The distance between two neighboring peaks is exactly given by the repeated frequency.
Due to the destructive interference, the background becomes more and more flat as the pulse number $N$ increases, which makes the central fringe more distinguishable at a small expense of the linewidth.
%

We theoretically analyze the aspects of temporal spinwave FP interference using the Bloch equations and obtain an analytical expression for the spectra.
When the bias magnetic field is applied, near the magneto-insensitive two-photon resonance, only $\left|F_g\myeq1,m_F\myeq-1\right\rangle \leftrightarrow \left|F_g\myeq2,m_F\myeq+1\right\rangle$, $\left|F_g\myeq2,m_F\myeq-1\right\rangle \leftrightarrow \left|F_g\myeq1,m_F\myeq+1\right\rangle$, and $\left|F_g\myeq1,m_F\myeq0\right\rangle \leftrightarrow \left|F_g\myeq2,m_F\myeq0\right\rangle$ are likely to occur two-photon $\Lambda$ resonances.
Due to the destructive interference of two-photon transitions in the case of lin$||$lin configuration, the $\left|F_g\myeq1,m_F\myeq0\right\rangle \leftrightarrow \left|F_g\myeq2,m_F\myeq0\right\rangle$ resonance is absent.
Thus, one can efficiently describe the system via a five-level model, although a complete description is an eleven-level model (see the Supplementary Material~\cite{SM}).
By ignoring the ground states exchange, the time-evolution is governed by a Liouville equation~\cite{Shahriar2014,SM}
\begin{equation}
	\frac{\partial \rho}{\partial t} \myeq  -\frac{\mathrm{i}}{\hbar}(\hat{H}\rho-\rho\hat{H}^\dagger) + \dot{\rho}_{trans-decay}+\dot{\rho}_{src},
\end{equation}
with the density matrix $\rho\myeq\sum_{j=1}^{5}\sum_{i=1}^{5}\rho_{ij}| i \rangle \langle j|$, the decoherence between ground states $\dot{\rho}_{trans-decay} = \sum_{j=2}^{4}\sum_{i=1}^{j-1} (-\gamma_{ij}\rho_{ij}| i \rangle \langle j| + h.c.)$ with the decoherence rates $\gamma_{ij}$, the population decay $\dot{\rho}_{src} = \sum_{i=1}^{4}\frac{\Gamma}{4}\rho_{55}|i\rangle \langle i|$, and the Hamiltonian $\hat{H} = \hbar\left[\right.(\delta+g_1\frac{\mu_B}{\hbar} B_z)|1\rangle \langle1|+(\delta-g_1\frac{\mu_B}{\hbar} B_z)|2\rangle \langle2| +g_2\frac{\mu_B}{\hbar} B_z |3\rangle\langle3|-g_2\frac{\mu_B}{\hbar} B_z|4\rangle\langle4|- \frac{\mathrm{i}\Gamma}{2}|5\rangle\langle5| + \frac{\Omega_{1}}{2}\left(|1\rangle\langle5|+|2\rangle\langle5|+h.c.\right) +\frac{\Omega_{2}}{2}\left(|3\rangle\langle5|+|4\rangle\langle5|+h.c.\right)\left.\right]$.
Here, $B_z$ is the bias magnetic field along the light propagation direction, $\mu_B$ is the Bohr magneton, $\{\Omega_{1}=\Omega_{1,+1}^a=-\Omega_{1,-1}^a,\Omega_{2}= \Omega_{2,\pm 1}^b\}$ are two Rabi frequencies, and $\{g_1, g_2\}$ are  respectively the Land\'{e} $ g $ factors for ground states $F\myeq \{1, 2\}$.
For $^{87}\mathrm{Rb}$, $g_1$ and $g_2$ have a tiny different value but opposite signs, therefore there are two magneto-insensitive transitions: $|1\rangle \leftrightarrow |4\rangle$ and $|2\rangle \leftrightarrow |3\rangle$.
Experimentally, each density matrix element should be summed over the atoms contributing signals, i.e. $\rho_{ij}\myeq \langle \hat{\rho}_{ij} \rangle$~\cite{PhysRevLett.98.123601}.
%

\begin{figure}[!ht]
	\centering
	\includegraphics[width=1.0\linewidth]{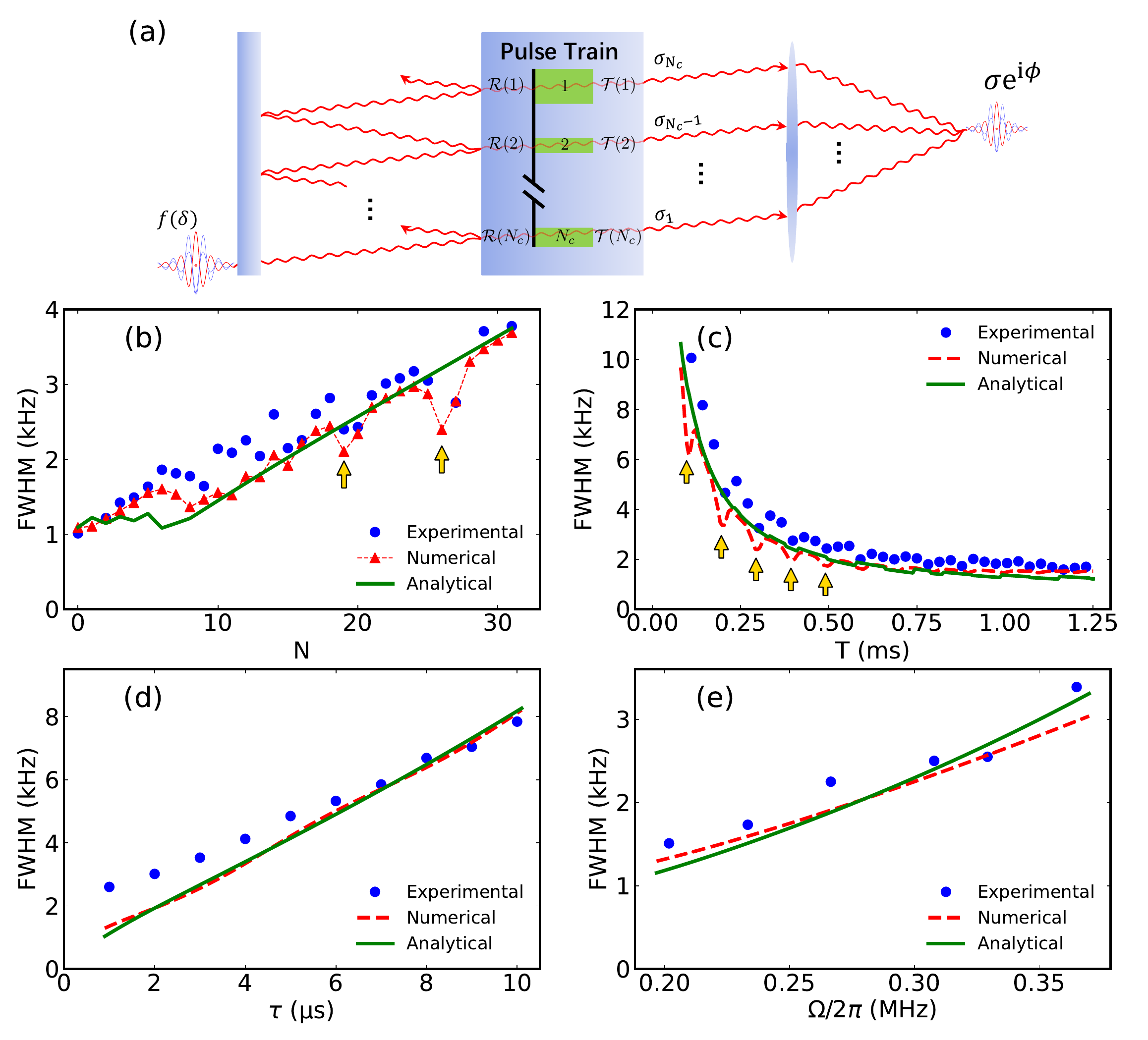}
	\caption{(Color online) Schematic of FP interferometry and linewidth aspects of the temporal spinwave FP interferometry.
		(a) The analog of our temporal spinwave FP interferometry to the light transmission in a FP cavity.
		(b) The linewidth versus the pulse number $N$ with $\Omega\myeq 1.77e6~\mathrm{s^{-1}}$, $T\myeq 0.5$ ms and $\uptau \myeq $2 $\mu$s. 	
		(c) The linewidth versus the integration time $T$ with $\Omega\myeq 1.77e6~\mathrm{s^{-1}}$, $N\myeq$ 15 and $\uptau \myeq $2 $\mu$s.	
		(d) The linewidth versus the pulse length $\uptau$ with $\Omega \myeq  1.77e6~\mathrm{s^{-1}}$, $ N\myeq $15 and $\Delta T\myeq 0.031$ ms.
		(e) The linewidth versus the average Rabi frequency $\Omega$ with $\uptau \myeq 2$ $ \mu $s, $ N\myeq $15 and $\Delta T\myeq 0.031$ ms.
		The bias magnetic field is chosen as $B_z\myeq 0.116$ G and the yellow arrows label the exotic dips due to magneto-sensitive transitions.}
	\label{experiment_pulses}
\end{figure}

The spectra are experimentally obtained from the transmission of CPT light.
The transmission signal is proportional to $\left(1-\rho_{55}\right)$, that is, the absorption is proportional to the excited-state population $\rho_{55}$.
For simplicity, we do not consider degenerate Zeeman sublevels and set all four Rabi frequencies as the average Rabi frequency~\cite{Chen_PRA_2000}.
To compare with the experimental observation, the average Rabi frequency can be given as $\Omega \myeq \sqrt{\left(\Omega_{1}^2 +\Omega_{2}^2\right)/2}$~\cite{Hemmer_1989}.
Due to the large decoherence rates $\{\gamma_{12},\gamma_{13},\gamma_{24},\gamma_{34}\}$ of magneto-sensitive transitions~\cite{Baumgart2016}, the corresponding density matrix elements $\{\rho_{12},\rho_{13},\rho_{24},\rho_{34}\}$ can be ignored near the magneto-insensitive two-photon resonance.
Thus, using adiabatic elimination and resonant approximation~\cite{Chuchelov2019}
\begin{equation}
	\label{rho55}
	\begin{aligned}
		\rho_{55}=& \frac{\Omega^2}{\Gamma^2}+\frac{2\Omega^2}{\Gamma^2}\mathrm{Re}\left(\rho_{14}+\rho_{23}\right).
	\end{aligned}
\end{equation}
Under the lin$||$lin configuration, for a weakly magnetic field, the two CPT resonances are nearly identical~\cite{Taichenachev2005a,PhysRevA.88.042120} as $\left(g_1+g_2\right)\mu_B B_z \to 0$.
Applying $N$ multiple CPT pulses, we analytically obtain~\cite{SM}
\begin{equation}
	\rho_{55}\myeq  \frac{\Omega^2}{\Gamma^2}+\frac{4\Omega^2}{\Gamma^2}\mathrm{Re}[\sigma(\delta)],
\end{equation}
with
\begin{equation*}
	\sigma(\delta) \myeq f(\delta)\sum_{l=1}^{N_c}\left[\prod_{k=l+1}^{N_c}\mathcal{R}(k)\right]\mathcal{T}(l)\mathrm{e}^{-\mathrm{i} \delta\sum_{k=l+1}^{N_c}\Delta T(k)},
	\label{sigma}
\end{equation*}
where $\uptau(l)$ and $\Delta T(l)$ denote the pulse length and the pulse interval for the $l$-th pulse, respectively.
The $l$-th pulse maps onto the $(N_c-l+1)$-th reflection event in a FP interferometer [see Fig.~3(a)], in which the corresponding local reflection and transmission coefficients are respectively given as $\mathcal{R}(l)\myeqv \mathrm{e}^{-{\Omega^2\over \Gamma}\uptau(l)}$ and $\mathcal{T}(l)\equiv 1-\mathrm{e}^{-\left(\frac{\Omega^2}{\Gamma}+\mathrm{i}\delta\right)\uptau(l)}$.
Here, $f(\delta) \myeq - \frac{\Omega^2}{4\Gamma\left(\mathrm{i}\delta + \frac{\Omega^2}{\Gamma}\right)}$,
and $N_c\myeq N+2$ is the total number of CPT pulses including the preparation and detection pulses.
In our experiment, the pulse length and the pulse interval are chosen as $\uptau(l)\myeq \uptau$ and $\Delta T(l)\myeq \Delta T$, respectively.
Therefore $\sigma(\delta)$ can be simplified as
\begin{equation}
	\begin{aligned}
		\sigma(\delta) = \sum_{l\myeq 1}^{N_c}\sigma_{l}(\delta),
	\end{aligned}
	\label{sigma2}
\end{equation}
with $\sigma_{l}(\delta) \myeq f(\delta)\mathcal{R}^{\left(l-1\right)}\mathcal{T}\mathrm{e}^{-\mathrm{i}(l-1)\delta \Delta T}$, the reflection coefficient $ \mathcal{R}\myeq\mathrm{e}^{-\frac{\Omega^2}{\Gamma}\uptau}$, and the transmission coefficient $\mathcal{T}\myeq 1-\mathrm{e}^{-\left(\frac{\Omega^2}{\Gamma}+\mathrm{i}\delta\right)\uptau}$.
Obviously, Eq.~(\ref{sigma2}) is analogous to the light transmission in a FP cavity, as shown in Fig.~\ref{experiment_pulses}(a).
According to Eq.~(\ref{sigma2}), constructive interferences occur at $\delta \Delta T \myeq  2m\pi $ $ (m \in \mathbb{Z})$, which exactly give the resonance peaks in our experimental spectra (see Fig.~\ref{fig2}).

To further show the power of our analytical results, we compare the experimental, numerical and analytical linewidths.
In analogy to the linewidth of FP cavity~\cite{siegman1986lasers}, the FWHM of spectrum can be given as $\Delta \nu \myeq  (2\Delta \nu_{FSR}/\pi)\arcsin\left[(1-\sqrt{\mathcal{R}})/(2\sqrt[4]{\mathcal{R}})\right]$ with $\Delta \nu_{FSR} \myeq  1/\Delta T$ corresponding to the FSR of FP cavity.
Accordingly, the linewidth will increase with the FSR which is proportional to the repeated frequency of inserted CPT pulses.
Fig.~\ref{experiment_pulses} clearly shows that the experimental results are well consistent with the analytical and numerical ones.
The linewidth increases with the pulse number $N$ for a given integration time $T$, while it will decrease with the integration time $T$ for a given pulse number $N$.
However, as labelled by the yellow arrows in Fig.~\ref{experiment_pulses}~(b,c), there appear some exotic dips in the experimental and numerical results.
These exotic dips are actually caused by a tiny contribution of magneto-sensitive transitions under the resonant condition $m/(B_z \Delta T)~\myeq~1.4~\textrm{MHz/G}$, see more details in Supplementary Information.
The linewidth increases with the pulse length $\uptau$ and the average Rabi frequency $\Omega$ when the other parameters are fixed, see Fig.~\ref{experiment_pulses}~(d,e).
\textcolor{black}{Here, the Rabi frequency is experimentally obtained by fitting CPT line shape with $f(\delta)$}.
\textcolor{black}{As the reflection coefficient $ \mathcal{R}\myeq\mathrm{e}^{-\frac{\Omega^2}{\Gamma}\uptau}$,}
this indicates that the linewidth decreases with the reflection coefficient.

\begin{figure}[!ht]
	\centering
	\includegraphics[width=1.0\linewidth]{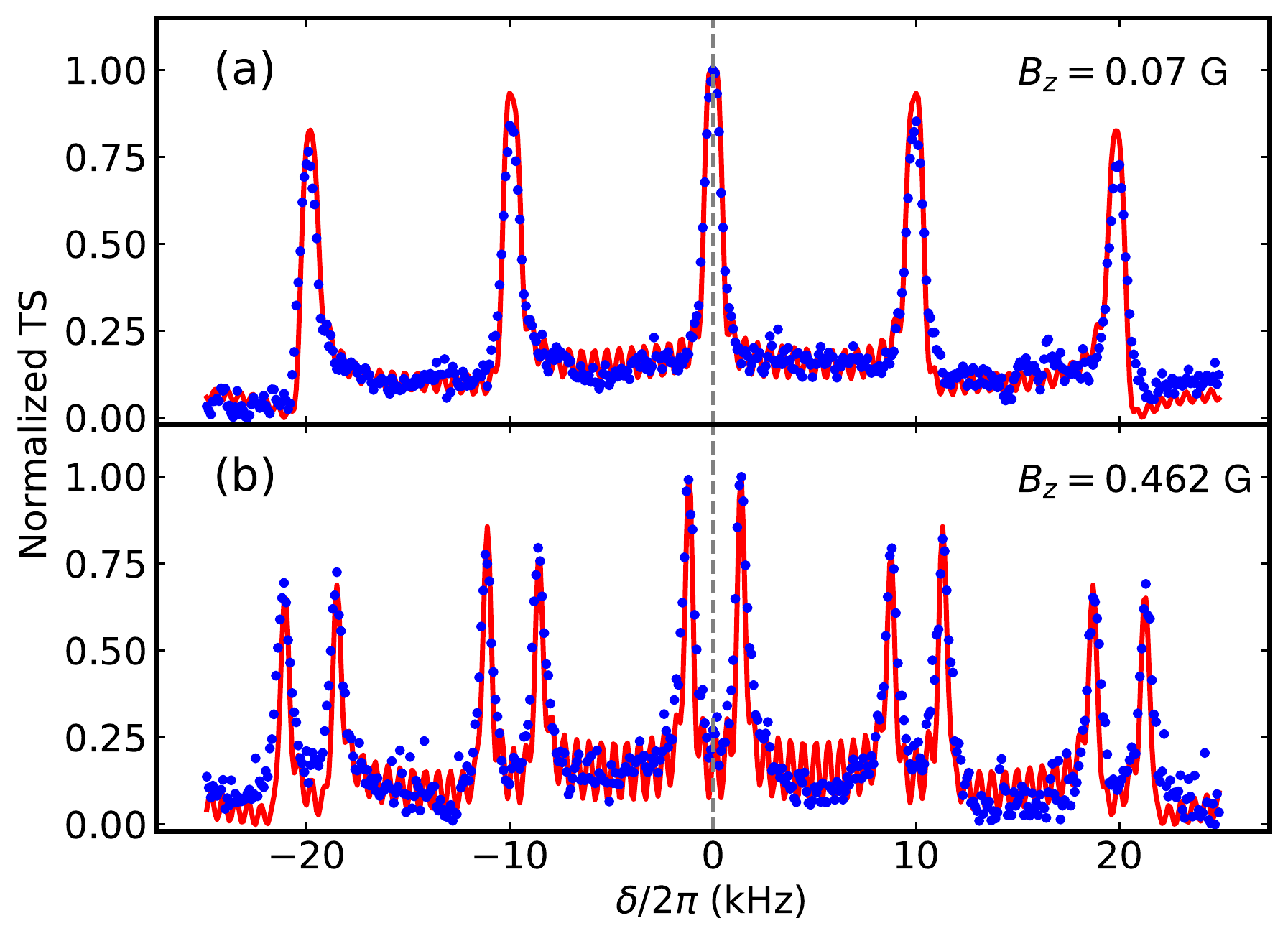}
	\caption{(Color online) Experimental TS (blue dots) under different bias magnetic fields $B_z$. In the experiments, 15 CPT pulses with length $\uptau=2~\mathrm{\mu s}$ are applied during the integration time $T=1.6$ ms. The numerical results (red lines) is fitted with $\Omega=1.6e6~\mathrm{s^{-1}}$. The vertical gray dash line labels $\delta=0$. (a) $B_z = 0.07~\mathrm{G}$. (b) $B_z = 0.462~\mathrm{G}$.}
	\label{length_repetition}
\end{figure}

Actually, the two involved Land\'{e} g factors $g_1\myeq -0.5017$ and $g_2\myeq 0.4997$ have small amplitude difference and this difference will bring a frequency shift $ 2|g_1+g_2|\frac{\mu_B}{2\pi\hbar} B_z = 5568 B_z $ Hz/G between the two magnetic-insensitive transitions $\left|1\right\rangle \leftrightarrow \left|4\right\rangle$ and $\left|2\right\rangle \leftrightarrow \left|3\right\rangle$.
However, due to the frequency shift between the two magnetic-insensitive transitions is too small compared with their resonant frequencies and the corresponding CPT resonance linewidths, it is difficult to be directly observed in conventional single-pulse CPT experiments or two-pulse CPT-Ramsey experiments~\cite{SM}.
In contrast, through applying the multi-pulse CPT sequence, the side peaks are suppressed compared with two-pulse CPT-Ramsey interference and the linewidth is decreased compared with single-pulse CPT spectra.
Therefore, when the magnetic field is not too weak, we directly observe the splitting.
As shown in Fig.~\ref{length_repetition}, each peak splits into two new ones when the bias magnetic field increases to a strength such that the frequency shift is larger than the linewidth.
This peak splitting is beneficial for not only eliminating linear Zeeman shifts by averaging each pair of peaks, but also measuring a magnetic field without involving magneto-sensitive transitions.

In conclusion, we have experimentally demonstrated a temporal spinwave FP interferometry via laser-cooled $^{87}$Rb atoms under the lin$||$lin configuration.
The transmission spectrum appears as a high-contrast comb, in which a sequence of equidistant resonant peaks and non-resonant plains are respectively due to constructive and destructive interferences.
We develop an analytical theory for the temporal spinwave FP interferometry based upon the five-level optical Bloch equations for our double-$\Lambda$ CPT system.
Beyond identifying the central CPT-Ramsey fringe, our scheme could be directly used to measure clock transition frequency~\cite{Vanier2005, Merimaa:03, PhysRevApplied.7.014018, PhysRevApplied.8.054001} and static magnetic field~\cite{PhysRevLett.69.1360, Nagel_1998, doi:10.1063/1.1839274, Tripathi2019}.
The temporal spinwave FP interferometry protocol could be also extended to other systems, such as, coherent storage of photons in EIT~\cite{RevModPhys.77.633} and coherent control of internal spin states in diamond defects~\cite{PhysRevLett.116.043603} or artificial atoms~\cite{PhysRevB.77.165312, donarini2019coherent, kelly2010direct}.

\acknowledgements{R. Fang, C. Han and X. Jiang contributed equally to this work. This work is supported by the Key-Area Research and Development Program of GuangDong Province (2019B030330001), the National Natural Science Foundation of China (12025509, 11874434), and the Science and Technology Program of Guangzhou (201904020024). J.H. is partially supported by the Guangzhou Science and Technology Projects (202002030459). B.L. is partially supported by the Guangdong Natural Science Foundation (2018A030313988) and the Guangzhou Science and Technology Projects (201804010497).}


%
\let\addcontentsline\oldaddcontentsline
\widetext
\pagebreak
\onecolumngrid
{\centering \textbf{\Large Supplementary Material}\\}
\setcounter{equation}{0}
\setcounter{figure}{0}
\newcounter{sfigure}
\setcounter{sfigure}{1}
\setcounter{table}{0}
\renewcommand{\theequation}{S\arabic{equation}}
\renewcommand\thefigure{S{\arabic{figure}}}
\renewcommand{\thesection}{S\arabic{section}}
\section{Experimental system and models}\label{MODEL}
\subsection{Experimental System}
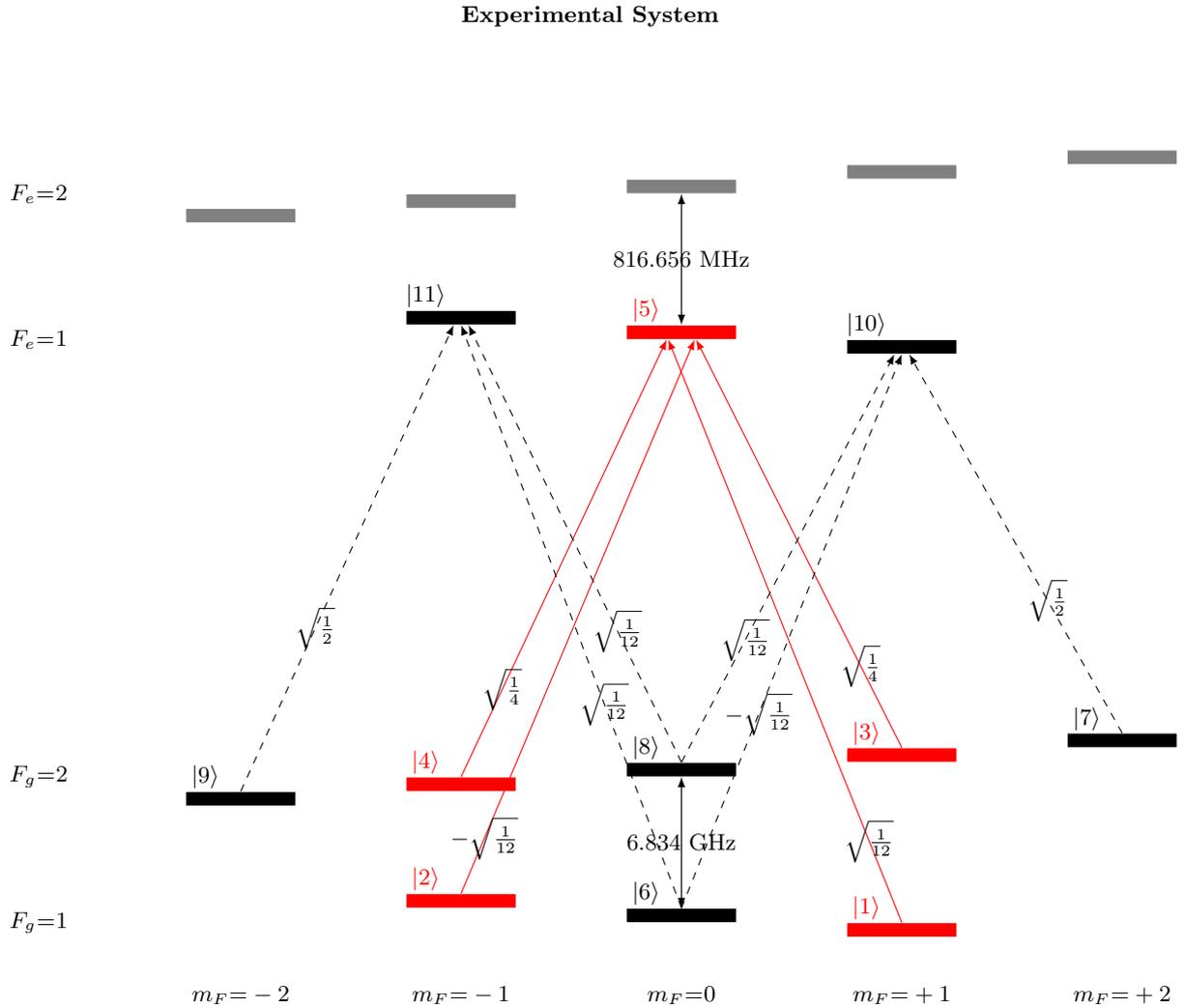
\begin{figure}[!ht]
	\begin{tikzpicture}%
		\path[draw,color=white,draw,fill=red] (5.0,-7.8) rectangle (6.5,-7.6);%
		\node at (5.27,-7.3999999999999995) {\textcolor{red}{$|2\rangle$}};%
		\path[draw,color=white,draw,fill=black] (8.0,-8.0) rectangle (9.5,-7.8);%
		\node at (8.27,-7.6) {\textcolor{black}{$|6\rangle$}};%
		\path[draw,color=white,draw,fill=red] (11.0,-8.2) rectangle (12.5,-8.0);%
		\node at (11.27,-7.799999999999999) {\textcolor{red}{$|1\rangle$}};%
		\path[draw,color=white,draw,fill=black] (2.0,-6.4) rectangle (3.5,-6.2);%
		\node at (2.27,-6.0) {\textcolor{black}{$|9\rangle$}};%
		\path[draw,color=white,draw,fill=red] (5.0,-6.2) rectangle (6.5,-6.0);%
		\node at (5.27,-5.8) {\textcolor{red}{$|4\rangle$}};%
		\path[draw,color=white,draw,fill=black] (8.0,-6.0) rectangle (9.5,-5.8);%
		\node at (8.27,-5.6) {\textcolor{black}{$|8\rangle$}};%
		\path[draw,color=white,draw,fill=red] (11.0,-5.8) rectangle (12.5,-5.6);%
		\node at (11.27,-5.3999999999999995) {\textcolor{red}{$|3\rangle$}};%
		\path[draw,color=white,draw,fill=black] (14.0,-5.6) rectangle (15.5,-5.4);%
		\node at (14.27,-5.199999999999999) {\textcolor{black}{$|7\rangle$}};%
		\path[draw,color=white,draw,fill=black] (5.0,0.2) rectangle (6.5,0.4);%
		\node at (5.27,0.6000000000000001) {\textcolor{black}{$|11\rangle$}};%
		\path[draw,color=white,draw,fill=red] (8.0,0.0) rectangle (9.5,0.2);%
		\node at (8.27,0.4) {\textcolor{red}{$|5\rangle$}};%
		\path[draw,color=white,draw,fill=black] (11.0,-0.2) rectangle (12.5,0.0);%
		\node at (11.27,0.2) {\textcolor{black}{$|10\rangle$}};%
		\path[draw,color=white,draw,fill=black!50] (2.0,1.6) rectangle (3.5,1.8);%
		\path[draw,color=white,draw,fill=black!50] (5.0,1.8) rectangle (6.5,2.0);%
		\path[draw,color=white,draw,fill=black!50] (8.0,2.0) rectangle (9.5,2.2);%
		\path[draw,color=white,draw,fill=black!50] (11.0,2.2) rectangle (12.5,2.4);%
		\path[draw,color=white,draw,fill=black!50] (14.0,2.4000000000000004) rectangle (15.5,2.6);%
		\path[solid,draw] (5.75,-7.6) edge[-latex,color=red] (8.93,0.0);%
		\node at (6.068,-6.84) {$-\sqrt{1\over12}$};%
		\path[solid,draw] (11.75,-8.0) edge[-latex,color=red] (8.57,0.0);%
		\node at (11.3048,-6.88) {$\sqrt{1\over12}$};%
		\path[solid,draw] (5.75,-6.0) edge[-latex,color=red] (8.55,0.0);%
		\node at (6.3100000000000005,-4.8) {$\sqrt{1\over4}$};%
		\path[solid,draw] (11.75,-5.6) edge[-latex,color=red] (8.95,0.0);%
		\node at (11.19,-4.4799999999999995) {$\sqrt{1\over4}$};%
		\path[dashed,draw] (2.75,-6.2) edge[-latex,color=black] (5.65,0.2);%
		\node at (3.765,-3.9600000000000004) {$\sqrt{1\over2}$};%
		\path[dashed,draw] (8.75,-5.8) edge[-latex,color=black] (5.85,0.2);%
		\node at (7.88,-4.0) {$\sqrt{1\over12}$};%
		\path[dashed,draw] (8.75,-7.8) edge[-latex,color=black] (5.75,0.2);%
		\node at (7.7,-5.0) {$\sqrt{1\over12}$};%
		\path[dashed,draw] (14.75,-5.3999999999999995) edge[-latex,color=black] (11.85,-0.2);%
		\node at (13.735,-3.58) {$\sqrt{1\over2}$};%
		\path[dashed,draw] (8.75,-5.8) edge[-latex,color=black] (11.65,-0.2);%
		\node at (9.620000000000001,-4.12) {$\sqrt{1\over12}$};%
		\path[dashed,draw] (8.75,-7.8) edge[-latex,color=black] (11.75,-0.2);%
		\node at (9.8,-5.140000000000001) {$-\sqrt{1\over12}$};%
		\path[solid,draw] (8.75,0.2) edge[latex-latex,color=black] (8.75,2.0);%
		\node at (8.75,1.1) {$816.656~\mathrm{MHz}$};%
		\path[solid,draw] (8.75,-7.8) edge[latex-latex,color=black] (8.75,-6.0);%
		\node at (8.75,-6.9) {$6.834~\mathrm{GHz}$};%
		\node at (2.75,-9.0) {$m_F\mkern0.5mu{=}\mkern0.5mu-2$};%
		\node at (5.75,-9.0) {$m_F\mkern0.5mu{=}\mkern0.5mu-1$};%
		\node at (8.75,-9.0) {$m_F\mkern0.5mu{=}\mkern0.5mu0$};%
		\node at (11.75,-9.0) {$m_F\mkern0.5mu{=}\mkern0.5mu +1$};%
		\node at (14.75,-9.0) {$m_F\mkern0.5mu{=}\mkern0.5mu +2$};%
		\node at (0.0,-8.0) {$F_g \mkern0.5mu{=}\mkern0.5mu 1$};%
		\node at (0.0,-6.0) {$F_g \mkern0.5mu{=}\mkern0.5mu 2$};%
		\node at (0.0,0.0) {$F_e \mkern0.5mu{=}\mkern0.5mu 1$};%
		\node at (0.0,2.0) {$F_e \mkern0.5mu{=}\mkern0.5mu 2$};%
	\end{tikzpicture}%
	\caption{\textbf{The manifold of $\mathrm{^{87}Rb}$ $\mathrm{D_1}$ transition.} The linearly polarized bichromatic light couples the $|F_g\myeq 1\rangle$ $|F_g\myeq 2\rangle$ and $|F_e\myeq 1\rangle$, the number of involved hyperfine levels is 11. For the convince of discussion, the involved Zeeman sublevels are labeled as $|1\rangle-|11\rangle$. In our experimental setup, the possible transitions are marked as dashed and solid lines. The coefficients marked at the lines represent the transition dipole moments(t.d.m) in units of $\langle J\myeq 1/2|e\boldsymbol{r}|J^\prime\myeq 1/2\rangle$.}
	\label{six_teen_levels}
\end{figure}
The laser light is locked at the resonance of $|F_g=2\rangle$ to $|F_e=1\rangle$ transition. A fiber-coupled electro-optic phase modulator(EOPM) modulates the laser light with $6.8~\mathrm{GHz}$ RF field, and generates sidebands. One of the sideband couples the $|F_g=1\rangle$ to $|F_e=2\rangle$. Since the applying magnetic field alone alone z direction $B_z<1~\mathrm{G}$ and the linewidth $\Gamma=2\pi\times 5.746~\mathrm{MHz}$, the frequency separation of $|F_e=1\rangle$ and $|F_e=2\rangle$ $\Delta\omega_e = 2\pi\times 816.656~\mathrm{MHz}$ is much larger than the linewidth $\Gamma$ and Zeeman shift. Therefore, we can exclude the $|F_e=2\rangle$ in our experimental setup and only 11 Zeeman sublevels are involved.
\subsection{Eleven-level Model}
The density matrix of 11 Zeeman sublevels is
\begin{equation}
	\begin{aligned}
		\rho &= \left(
		\begin{matrix}
			\rho_{5\times 5}&\rho_{5\times 6}\\
			\rho_{6\times 5}&\rho_{6\times 6}
		\end{matrix}
		\right),\\
		&=\left(\begin{array}{ccccccccccc}\rho_{1 1} & \rho_{1 2} & \rho_{1 3} & \rho_{1 4} & \rho_{1 5} & \rho_{1 6} & \rho_{1 7} & \rho_{1 8} & \rho_{1 9} & \rho_{1,10} & \rho_{1,11}\\\rho^*_{1 2} & \rho_{2 2} & \rho_{2 3} & \rho_{2 4} & \rho_{2 5} & \rho_{2 6} & \rho_{2 7} & \rho_{2 8} & \rho_{2 9} & \rho_{2,10} & \rho_{2,11}\\\rho^*_{1 3} & \rho^*_{2 3} & \rho_{3 3} & \rho_{3 4} & \rho_{3 5} & \rho_{3 6} & \rho_{3 7} & \rho_{3 8} & \rho_{3 9} & \rho_{3,10} & \rho_{3,11}\\\rho^*_{1 4} & \rho^*_{2 4} & \rho^*_{3 4} & \rho_{4 4} & \rho_{4 5} & \rho_{4 6} & \rho_{4 7} & \rho_{4 8} & \rho_{4 9} & \rho_{4,10} & \rho_{4,11}\\\rho^*_{1 5} & \rho^*_{2 5} & \rho^*_{3 5} & \rho^*_{4 5} & \rho_{5 5} & \rho_{5 6} & \rho_{5 7} & \rho_{5 8} & \rho_{5 9} & \rho_{5,10} & \rho_{5,11}\\\rho^*_{1 6} & \rho^*_{2 6} & \rho^*_{3 6} & \rho^*_{4 6} & \rho^*_{5 6} & \rho_{6 6} & \rho_{6 7} & \rho_{6 8} & \rho_{6 9} & \rho_{6,10} & \rho_{6,11}\\\rho^*_{1 7} & \rho^*_{2 7} & \rho^*_{3 7} & \rho^*_{4 7} & \rho^*_{5 7} & \rho^*_{6 7} & \rho_{7 7} & \rho_{7 8} & \rho_{7 9} & \rho_{7,10} & \rho_{7,11}\\\rho^*_{1 8} & \rho^*_{2 8} & \rho^*_{3 8} & \rho^*_{4 8} & \rho^*_{5 8} & \rho^*_{6 8} & \rho^*_{7 8} & \rho_{8 8} & \rho_{8 9} & \rho_{8,10} & \rho_{8,11}\\\rho^*_{1 9} & \rho^*_{2 9} & \rho^*_{3 9} & \rho^*_{4 9} & \rho^*_{5 9} & \rho^*_{6 9} & \rho^*_{7 9} & \rho^*_{8 9} & \rho_{9 9} & \rho_{9,10} & \rho_{9,11}\\\rho^*_{1,10} & \rho^*_{2,10} & \rho^*_{3,10} & \rho^*_{4,10} & \rho^*_{5,10} & \rho^*_{6,10} & \rho^*_{7,10} & \rho^*_{8,10} & \rho^*_{9,10} & \rho_{10,10} & \rho_{10,11}\\\rho^*_{1,11} & \rho^*_{2,11} & \rho^*_{3,11} & \rho^*_{4,11} & \rho^*_{5,11} & \rho^*_{6,11} & \rho^*_{7,11} & \rho^*_{8,11} & \rho^*_{9,11} & \rho^*_{10,11} & \rho_{11,11}\end{array}\right)
	\end{aligned}
\end{equation}
The time-evolution of system is governed by the Liouville equation for the density matrix~\cite{Shahriar2014},
\begin{equation}
	\frac{\partial}{\partial t}\rho = -\frac{\mathrm{i}}{\hbar}\left(\hat{H}\rho - \rho \hat{H}^{\dagger}\right) + \dot{\rho}_{src} + \dot{\rho}_{trans-decay},
	\label{liouville}
\end{equation}
$\hat{H}$ is the Hamiltonian, $\dot{\rho}_{src}$ is the source term, and the $\dot{\rho}_{trans-decay}$ is the dephasing term, We will discuss this in more detail below.
The Hamiltonian with 11 energy levels is
\begin{equation}
	\hat{H} =\left(
	\begin{matrix}
		\hat{H}_{5\times 5}&0\\
		0&\hat{H}_{6\times 6}
	\end{matrix}
	\right),
\end{equation}
where
\begin{equation*}
	\begin{aligned}
		\hat{H}_{5\times 5} &= \hbar\left(\begin{matrix}\frac{B_{z} \mu_B g_{1}}{\hbar} + \delta & 0 & 0 & 0 & \frac{\Omega_{1-5}}{2}\\0 & - \frac{B_{z} \mu_B g_{1}}{\hbar} + \delta & 0 & 0 & \frac{\Omega_{2-5}}{2}\\0 & 0 & \frac{B_{z} \mu_B g_{2}}{\hbar} & 0 & \frac{\Omega_{3-5}}{2}\\0 & 0 & 0 & - \frac{B_{z} \mu_B g_{2}}{\hbar} & \frac{\Omega_{4-5}}{2}\\\frac{\Omega_{1-5}}{2} & \frac{\Omega_{2-5}}{2} & \frac{\Omega_{3-5}}{2} & \frac{\Omega_{4-5}}{2} & - \frac{i \Gamma}{2}\end{matrix}\right),\\
		\hat{H}_{6\times 6} & = \hbar\left(\begin{matrix}\delta & 0 & 0 & 0 & \frac{\Omega_{6-10}}{2} & \frac{\Omega_{6-11}}{2}\\0 & \frac{2 B_{z} \mu_B g_{2}}{\hbar} & 0 & 0 & \frac{\Omega_{7-10}}{2} & 0\\0 & 0 & 0 & 0 & \frac{\Omega_{8-10}}{2} & \frac{\Omega_{8-11}}{2}\\0 & 0 & 0 & - \frac{2 B_{z} \mu_B g_{2}}{\hbar} & 0 & \frac{\Omega_{9-11}}{2}\\\frac{\Omega_{6-10}}{2} & \frac{\Omega_{7-10}}{2} & \frac{\Omega_{8-10}}{2} & 0 & \frac{B_{z} \mu_B g_{3}}{\hbar} - \frac{i \Gamma}{2} & 0\\\frac{\Omega_{6-11}}{2} & 0 & \frac{\Omega_{8-11}}{2} & \frac{\Omega_{9-11}}{2} & 0 & - \frac{B_{z} \mu_B g_{3}}{\hbar} - \frac{i \Gamma}{2}\end{matrix}\right).
	\end{aligned}
\end{equation*}
Here, $g_1\myeq -0.5017$, $g_2\myeq 0.4997$, and $g_3=-{1\over 6}$ are Land\'{e} g-factors, and $\mu_B\myeq 2\pi \hbar\times 1.4$ MHz/G is the Bohr magneton, $\Omega_{i-j}$ are the Rabi frequencies. The Hamiltonian $\hat{H}$ is the direct sum of $\hat{H}_{5\times 5}$ and $\hat{H}_{6\times 6}$, which are the Hamiltonian of five-level system and six-level system respectively.
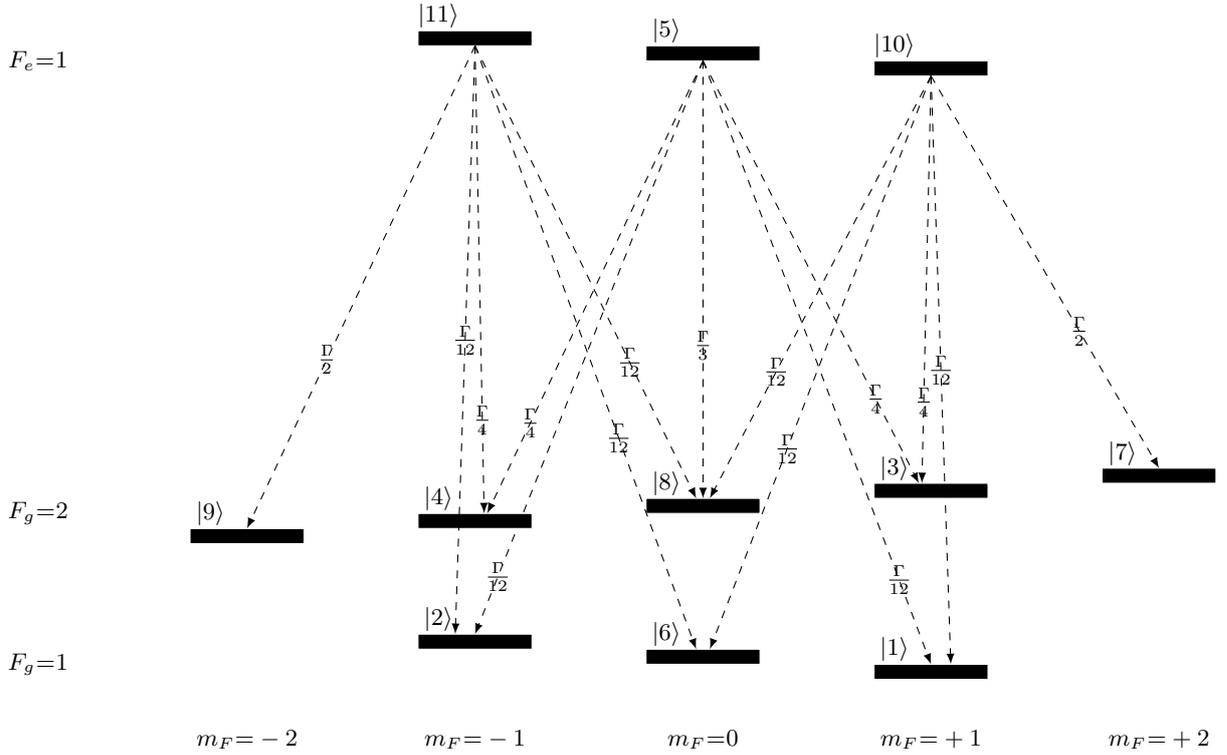
\begin{figure}
	\begin{tikzpicture}%
		\path[draw,color=white,draw,fill=black] (5.0,-7.8) rectangle (6.5,-7.6);%
		\node at (5.27,-7.3999999999999995) {\textcolor{black}{$|2\rangle$}};%
		\path[draw,color=white,draw,fill=black] (8.0,-8.0) rectangle (9.5,-7.8);%
		\node at (8.27,-7.6) {\textcolor{black}{$|6\rangle$}};%
		\path[draw,color=white,draw,fill=black] (11.0,-8.2) rectangle (12.5,-8.0);%
		\node at (11.27,-7.799999999999999) {\textcolor{black}{$|1\rangle$}};%
		\path[draw,color=white,draw,fill=black] (2.0,-6.4) rectangle (3.5,-6.2);%
		\node at (2.27,-6.0) {\textcolor{black}{$|9\rangle$}};%
		\path[draw,color=white,draw,fill=black] (5.0,-6.2) rectangle (6.5,-6.0);%
		\node at (5.27,-5.8) {\textcolor{black}{$|4\rangle$}};%
		\path[draw,color=white,draw,fill=black] (8.0,-6.0) rectangle (9.5,-5.8);%
		\node at (8.27,-5.6) {\textcolor{black}{$|8\rangle$}};%
		\path[draw,color=white,draw,fill=black] (11.0,-5.8) rectangle (12.5,-5.6);%
		\node at (11.27,-5.3999999999999995) {\textcolor{black}{$|3\rangle$}};%
		\path[draw,color=white,draw,fill=black] (14.0,-5.6) rectangle (15.5,-5.4);%
		\node at (14.27,-5.199999999999999) {\textcolor{black}{$|7\rangle$}};%
		\path[draw,color=white,draw,fill=black] (5.0,0.2) rectangle (6.5,0.4);%
		\node at (5.27,0.6000000000000001) {\textcolor{black}{$|11\rangle$}};%
		\path[draw,color=white,draw,fill=black] (8.0,0.0) rectangle (9.5,0.2);%
		\node at (8.27,0.4) {\textcolor{black}{$|5\rangle$}};%
		\path[draw,color=white,draw,fill=black] (11.0,-0.2) rectangle (12.5,0.0);%
		\node at (11.27,0.2) {\textcolor{black}{$|10\rangle$}};%
		\path[dashed,draw] (8.75,0.0) edge[-latex,color=black] (5.75,-7.6);%
		\node at (6.05,-6.84) {${\Gamma\over12}$};%
		\path[dashed,draw] (8.75,0.0) edge[-latex,color=black] (11.75,-8.0);%
		\node at (11.33,-6.88) {${\Gamma\over12}$};%
		\path[dashed,draw] (8.75,0.0) edge[-latex,color=black] (5.91,-6.0);%
		\node at (6.478,-4.8) {${\Gamma\over4}$};%
		\path[dashed,draw] (8.75,0.0) edge[-latex,color=black] (11.59,-5.6);%
		\node at (11.022,-4.4799999999999995) {${\Gamma\over4}$};%
		\path[dashed,draw] (5.75,0.2) edge[-latex,color=black] (5.87,-6.0);%
		\node at (5.846,-4.76) {${\Gamma\over4}$};%
		\path[dashed,draw] (11.75,-0.2) edge[-latex,color=black] (11.63,-5.6);%
		\node at (11.654,-4.52) {${\Gamma\over4}$};%
		\path[dashed,draw] (5.75,0.2) edge[-latex,color=black] (5.49,-7.6);%
		\node at (5.62,-3.6999999999999997) {${\Gamma\over12}$};%
		\path[dashed,draw] (11.75,-0.2) edge[-latex,color=black] (12.01,-8.0);%
		\node at (11.879999999999999,-4.1) {${\Gamma\over12}$};%
		\path[dashed,draw] (5.75,0.2) edge[-latex,color=black] (2.75,-6.2);%
		\node at (3.8,-3.9600000000000004) {${\Gamma\over2}$};%
		\path[dashed,draw] (5.75,0.2) edge[-latex,color=black] (8.66,-5.8);%
		\node at (7.787,-4.0) {${\Gamma\over12}$};%
		\path[dashed,draw] (5.75,0.2) edge[-latex,color=black] (8.66,-7.8);%
		\node at (7.641500000000001,-5.0) {${\Gamma\over12}$};%
		\path[dashed,draw] (11.75,-0.2) edge[-latex,color=black] (14.75,-5.3999999999999995);%
		\node at (13.7,-3.58) {${\Gamma\over2}$};%
		\path[dashed,draw] (11.75,-0.2) edge[-latex,color=black] (8.84,-5.8);%
		\node at (9.713,-4.12) {${\Gamma\over12}$};%
		\path[dashed,draw] (11.75,-0.2) edge[-latex,color=black] (8.84,-7.8);%
		\node at (9.8585,-5.140000000000001) {${\Gamma\over12}$};%
		\path[dashed,draw] (8.75,0.0) edge[-latex,color=black] (8.75,-5.8);%
		\node at (8.75,-3.77) {${\Gamma\over3}$};%
		\node at (2.75,-9.0) {$m_F\mkern0.5mu{=}\mkern0.5mu-2$};%
		\node at (5.75,-9.0) {$m_F\mkern0.5mu{=}\mkern0.5mu-1$};%
		\node at (8.75,-9.0) {$m_F\mkern0.5mu{=}\mkern0.5mu0$};%
		\node at (11.75,-9.0) {$m_F\mkern0.5mu{=}\mkern0.5mu +1$};%
		\node at (14.75,-9.0) {$m_F\mkern0.5mu{=}\mkern0.5mu +2$};%
		\node at (0.0,-8.0) {$F_g \mkern0.5mu{=}\mkern0.5mu 1$};%
		\node at (0.0,-6.0) {$F_g \mkern0.5mu{=}\mkern0.5mu 2$};%
		\node at (0.0,0.0) {$F_e \mkern0.5mu{=}\mkern0.5mu 1$};%
	\end{tikzpicture}%
	\caption{\textbf{Radiative decay channels of $F_e\myeq 1$ to $F_g\myeq 1$ and $F_g\myeq 2$.} The population of excited states will decay proportionately to ground states.}
	\label{six_teen_levels_decay}
\end{figure}
Considering the radiative decay channels as Fig.~\ref{six_teen_levels_decay} shows, the source term $\dot{\rho}_{src}$ is,
\begin{equation}
	\dot{\rho}_{src}=
	\left(
	\begin{array}{ccccccccccc}
		\dot{\rho}_{src11}&0&0&0&0&0&0&0&0&0&0\\
		0&\dot{\rho}_{src22}&0&0&0&0&0&0&0&0&0\\
		0&0&\dot{\rho}_{src33}&0&0&0&0&0&0&0&0\\
		0&0&0&\dot{\rho}_{src44}&0&0&0&0&0&0&0\\
		0&0&0&0&0&0&0&0&0&0&0\\
		0&0&0&0&0&\dot{\rho}_{src66}&0&0&0&0&0\\
		0&0&0&0&0&0&\dot{\rho}_{src77}&0&0&0&0\\
		0&0&0&0&0&0&0&\dot{\rho}_{src88}&0&0&0\\
		0&0&0&0&0&0&0&0&\dot{\rho}_{src99}&0&0\\
		0&0&0&0&0&0&0&0&0&0&0\\
		0&0&0&0&0&0&0&0&0&0&0
	\end{array}
	\right),
\end{equation}
where
\begin{equation}
	\begin{cases}
		\dot{\rho}_{src11} = {\Gamma\over 12} \rho_{55} + {\Gamma\over 12} \rho_{10,10}\\
		\dot{\rho}_{src22} = {\Gamma\over 12} \rho_{55} + {\Gamma\over 12} \rho_{11,11}\\
		\dot{\rho}_{src33} = {\Gamma\over 4} \rho_{55} + {\Gamma\over 4} \rho_{10,10}\\
		\dot{\rho}_{src44} = {\Gamma\over 4} \rho_{55} + {\Gamma\over 4} \rho_{11,11}\\
		\dot{\rho}_{src66} = {\Gamma\over 12} \rho_{10,10} + {\Gamma\over 12} \rho_{11,11}\\
		\dot{\rho}_{src77} = {\Gamma\over 2} \rho_{10,10}\\
		\dot{\rho}_{src88} = {\Gamma\over 12} \rho_{10,10} + {\Gamma\over 3} \rho_{55} + {\Gamma\over 12} \rho_{11,11}\\
		\dot{\rho}_{src99} = {\Gamma\over 2} \rho_{11,11}
	\end{cases}
	\label{rho_src_elements}
\end{equation}
The five-level system and six-level system couple with each other by the source term of Bloch equation. This coupling will result in the population exchange between five-level system and six-level system. However, for lin$||$lin configuration, the population of ground states change a little, therefore we can infer the population sourced from five-level system to six-level system is equal to that source from six-level system to five-level system, then we have
\begin{equation}
	{\Gamma\over 3}\rho_{55} \approxeq {\Gamma\over 3} (\rho_{10,10}+\rho_{11,11}).
	\label{rho_src_elements_approx}
\end{equation}
Since the $|10\rangle$ and $|11\rangle$ are symmetry, therefore $\rho_{10,10}\approxeq\rho_{11,11}$, then
\begin{equation}
	{\Gamma\over 3}\rho_{55} \approxeq {2\Gamma\over 3}\rho_{10,10}\approxeq {2\Gamma\over 3}\rho_{11,11},
	\label{rho_src_elements_approx2}
\end{equation}
and Eq.~\ref{rho_src_elements} will be
\begin{equation}
	\begin{cases}
		\dot{\rho}_{src11} = {\Gamma\over 8} \rho_{55}\\
		\dot{\rho}_{src22} = {\Gamma\over 8} \rho_{55}\\
		\dot{\rho}_{src33} = {3\Gamma\over 8} \rho_{55}\\
		\dot{\rho}_{src44} = {3\Gamma\over 8} \rho_{55}\\
		\dot{\rho}_{src66} = {\Gamma\over 12} \rho_{10,10} + {\Gamma\over 12} \rho_{11,11}\\
		\dot{\rho}_{src77} = {\Gamma\over 2} \rho_{10,10}\\
		\dot{\rho}_{src88} = {5\Gamma\over 12} \rho_{10,10} + {5\Gamma\over 12} \rho_{11,11}\\
		\dot{\rho}_{src99} = {\Gamma\over 2} \rho_{11,11}
	\end{cases}
	\label{rho_src_elements2}
\end{equation}
then $\dot{\rho}_{src}$ is the direct sum of $\dot{\rho}_{src5\times 5}$ and $\dot{\rho}_{src6\times 6}$
\begin{equation}
	\dot{\rho}_{src}=\left(
	\begin{matrix}
		\dot{\rho}_{src5\times 5} & 0 \\
		0 & \dot{\rho}_{src6\times 6}
	\end{matrix}
	\right)
\end{equation}
where
\begin{equation*}
	\begin{aligned}
		\dot{\rho}_{src5\times 5}&=\left(
		\begin{matrix}
			\frac{\Gamma_1}{2}\rho_{55}  & 0 & 0 & 0 & 0\\
			0 & \frac{\Gamma_1}{2}\rho_{55}  & 0 & 0 & 0\\
			0 & 0 & \frac{\Gamma_2}{2}\rho_{55}  & 0 & 0\\
			0 & 0 & 0 & \frac{\Gamma_2}{2}\rho_{55}  & 0\\
			0 & 0 & 0 & 0 & 0\\
		\end{matrix}\right)\\
		\dot{\rho}_{src6\times 6}&=\left(
		\begin{matrix}
			{\Gamma\over 12} \rho_{10,10} + {\Gamma\over 12} \rho_{11,11}  & 0 & 0 & 0 & 0 & 0\\
			0 & {\Gamma\over 2} \rho_{10,10}  & 0 & 0 & 0\\
			0 & 0 & {5\Gamma\over 12} \rho_{10,10} + {5\Gamma\over 12} \rho_{11,11}  & 0 & 0 & 0\\
			0 & 0 & 0 & {\Gamma\over 2} \rho_{11,11}  & 0 & 0\\
			0 & 0 & 0 & 0 & 0 & 0\\
			0 & 0 & 0 & 0 & 0 & 0
		\end{matrix}
		\right)
	\end{aligned}
\end{equation*}
here, $\Gamma_{1}={\Gamma\over 4}$ and $\Gamma_{1}={3\Gamma\over 4}$.

The term $\dot{\rho}_{trans-decay}$ accounts for the decay rate between ground states caused by dephasing and it is in form of
\begin{equation}
	\dot{\rho}_{trans-decay}=\left(
	\begin{matrix}
		\dot{\rho}_{trans-decay5\times 5}&\dot{\rho}_{trans-decay5\times 6}\\
		\dot{\rho}_{trans-decay6\times 5}&\dot{\rho}_{trans-decay6\times6}
	\end{matrix}
	\right)
\end{equation}
where
\begin{equation}
	\begin{aligned}
		\dot{\rho}_{trans-decay5\times 5} &= \left(\begin{matrix}0 & - \gamma_{12} \rho_{1 2} & - \gamma_{13} \rho_{1 3} & - \gamma_{14} \rho_{1 4} & 0\\- \gamma_{12} \rho^*_{1 2} & 0 & - \gamma_{23} \rho_{2 3} & - \gamma_{24} \rho_{2 4} & 0\\- \gamma_{13} \rho^*_{1 3} & - \gamma_{23} \rho^*_{2 3} & 0 & - \gamma_{34} \rho_{3 4} & 0\\- \gamma_{14} \rho^*_{1 4} & - \gamma_{24} \rho^*_{2 4} & - \gamma_{34} \rho^*_{3 4} & 0 & 0\\0 & 0 & 0 & 0 & 0\end{matrix}\right),\\
		\dot{\rho}_{trans-decay5\times 6} &= \left(\begin{matrix}- \gamma_{16} \rho_{1 6} & - \gamma_{17} \rho_{1 7} & - \gamma_{18} \rho_{1 8} & - \gamma_{19} \rho_{1 9} & 0 & 0\\- \gamma_{26} \rho_{2 6} & - \gamma_{27} \rho_{2 7} & - \gamma_{28} \rho_{2 8} & - \gamma_{29} \rho_{2 9} & 0 & 0\\- \gamma_{36} \rho_{3 6} & - \gamma_{37} \rho_{3 7} & - \gamma_{38} \rho_{3 8} & - \gamma_{39} \rho_{3 9} & 0 & 0\\- \gamma_{46} \rho_{4 6} & - \gamma_{47} \rho_{4 7} & - \gamma_{48} \rho_{4 8} & - \gamma_{49} \rho_{4 9} & 0 & 0\\0 & 0 & 0 & 0 & 0 & 0\end{matrix}\right),\\
		\dot{\rho}_{trans-decay6\times 5} &= \left(\begin{matrix}- \gamma_{16} \rho^*_{1 6} & - \gamma_{26} \rho^*_{2 6} & - \gamma_{36} \rho^*_{3 6} & - \gamma_{46} \rho^*_{4 6} & 0\\- \gamma_{17} \rho^*_{1 7} & - \gamma_{27} \rho^*_{2 7} & - \gamma_{37} \rho^*_{3 7} & - \gamma_{47} \rho^*_{4 7} & 0\\- \gamma_{18} \rho^*_{1 8} & - \gamma_{28} \rho^*_{2 8} & - \gamma_{38} \rho^*_{3 8} & - \gamma_{48} \rho^*_{4 8} & 0\\- \gamma_{19} \rho^*_{1 9} & - \gamma_{29} \rho^*_{2 9} & - \gamma_{39} \rho^*_{3 9} & - \gamma_{49} \rho^*_{4 9} & 0\\0 & 0 & 0 & 0 & 0\\0 & 0 & 0 & 0 & 0\end{matrix}\right),\\
		\dot{\rho}_{trans-decay6\times6} &= \left(\begin{matrix}0 & - \gamma_{67} \rho_{6 7} & - \gamma_{68} \rho_{6 8} & - \gamma_{69} \rho_{6 9} & 0 & 0\\- \gamma_{67} \rho^*_{6 7} & 0 & - \gamma_{78} \rho_{7 8} & - \gamma_{79} \rho_{7 9} & 0 & 0\\- \gamma_{68} \rho^*_{6 8} & - \gamma_{78} \rho^*_{7 8} & 0 & - \gamma_{89} \rho_{8 9} & 0 & 0\\- \gamma_{69} \rho^*_{6 9} & - \gamma_{79} \rho^*_{7 9} & - \gamma_{89} \rho^*_{8 9} & 0 & 0 & 0\\0 & 0 & 0 & 0 & 0 & 0\\0 & 0 & 0 & 0 & 0 & 0\end{matrix}\right),
	\end{aligned}
\end{equation}
here $\gamma_{ij}$ is the dephasing rate. In the $\dot{\rho}_{trans-decay5\times 5}$, the parameters $\{\gamma_{12},\gamma_{13},\gamma_{24},\gamma_{34}\}$ describe the dephasing rate of magneto-sensitive transitions and so that they are sensitive to the fluctuations of magnetic field~\cite{Baumgart2016}. The $\{\gamma_{14},\gamma_{23}\}$ describe the dephasing of magneto-insensitive transitions and can be regarded as zero.
Hence, the Eq.~\ref{bloch_eq} can be decomposed into two independent Bloch equation of five-level system and six-level system.
\begin{equation}
	\frac{\partial}{\partial t}\rho_{5\times 5} = -\frac{\mathrm{i}}{\hbar}\left(\hat{H}_{5\times 5}\rho_{5\times 5} - \rho_{5\times 5} \hat{H}^{\dagger}_{5\times 5}\right) + \dot{\rho}_{src5\times 5} + \dot{\rho}_{trans-decay5\times 5},
\end{equation}
and
\begin{equation}
	\frac{\partial}{\partial t}\rho_{6\times 6} = -\frac{\mathrm{i}}{\hbar}\left(\hat{H}_{6\times 6}\rho_{6\times 6} - \rho_{6\times 6} \hat{H}^{\dagger}_{6\times 6}\right) + \dot{\rho}_{src6\times 6} + \dot{\rho}_{trans-decay6\times 6}.
	\label{bloch6}
\end{equation}
The five-level system and six-level system are independent with each other. The absorption signal is proportional to $\rho_{55}+\rho_{10,10}+\rho_{11,11}$. The $\rho_{55}$ is determined by five-level system and $\rho_{10,10}+\rho_{11,11}$ is determined by six-level system.
For six-level system, if all the Rabi frequencies are real, and $\dot{\rho}_{trans-decay}=0$, expanding the Eq.~\ref{bloch6} we have
\begin{equation}
	\begin{cases}
		{\partial \over \partial t}\rho_{66} = \frac{\Gamma \rho_{10,10}}{12} + \frac{\Gamma \rho_{11,11}}{12} - \frac{i \Omega_{6-10} \rho^*_{6,10}}{2} + \frac{i \Omega_{6-10} \rho_{6,10}}{2} - \frac{i \Omega_{6-11} \rho^*_{6,11}}{2} + \frac{i \Omega_{6-11} \rho_{6,11}}{2},\\
		{\partial \over \partial t}\rho_{67} = - \frac{i \Omega_{6-10} \rho^*_{7,10}}{2} - \frac{i \Omega_{6-11} \rho^*_{7,11}}{2} + \frac{i \Omega_{7-10} \rho_{6,10}}{2},\\
		{\partial \over \partial t}\rho_{68} = - \frac{i \Omega_{6-10} \rho^*_{8,10}}{2} - \frac{i \Omega_{6-11} \rho^*_{8,11}}{2} + \frac{i \Omega_{8-10} \rho_{6,10}}{2} + \frac{i \Omega_{8-11} \rho_{6,11}}{2} - i \delta \rho_{6 8},\\
		{\partial \over \partial t}\rho_{69} = - \frac{i \Omega_{6-10} \rho^*_{9,10}}{2} - \frac{i \Omega_{6-11} \rho^*_{9,11}}{2} + \frac{i \Omega_{9-11} \rho_{6,11}}{2},\\
		{\partial \over \partial t}\rho_{6,10} = \frac{i B_{z} \mu_B \rho_{6,10} g_{3}}{\hbar} - \frac{\Gamma \rho_{6,10}}{2} - \frac{i \Omega_{6-10} \rho_{10,10}}{2} + \frac{i \Omega_{6-10} \rho_{6 6}}{2} + \frac{i \Omega_{8-10} \rho_{6 8}}{2} - i \delta \rho_{6,10},\\
		{\partial \over \partial t}\rho_{6,11} = - \frac{i B_{z} \mu_B \rho_{6,11} g_{3}}{\hbar} - \frac{\Gamma \rho_{6,11}}{2} - \frac{i \Omega_{6-11} \rho_{11,11}}{2} + \frac{i \Omega_{6-11} \rho_{6 6}}{2} + \frac{i \Omega_{8-11} \rho_{6 8}}{2} - i \delta \rho_{6,11},\\
		{\partial \over \partial t}\rho_{77} = \frac{\Gamma \rho_{10,10}}{2} - \frac{i \Omega_{7-10} \rho^*_{7,10}}{2} + \frac{i \Omega_{7-10} \rho_{7,10}}{2},\\
		{\partial \over \partial t}\rho_{78} = - \frac{i \Omega_{7-10} \rho^*_{8,10}}{2} + \frac{i \Omega_{8-10} \rho_{7,10}}{2} + \frac{i \Omega_{8-11} \rho_{7,11}}{2},\\
		{\partial \over \partial t}\rho_{79} = - \frac{4 i B_{z} \mu_B \rho_{7 9} g_{2}}{\hbar} - \frac{i \Omega_{7-10} \rho^*_{9,10}}{2} + \frac{i \Omega_{9-11} \rho_{7,11}}{2},\\
		{\partial \over \partial t}\rho_{7,10} = - \frac{2 i B_{z} \mu_B \rho_{7,10} g_{2}}{\hbar} + \frac{i B_{z} \mu_B \rho_{7,10} g_{3}}{\hbar} - \frac{\Gamma \rho_{7,10}}{2} - \frac{i \Omega_{7-10} \rho_{10,10}}{2} + \frac{i \Omega_{7-10} \rho_{7 7}}{2},\\
		{\partial \over \partial t}\rho_{7,11} = - \frac{2 i B_{z} \mu_B \rho_{7,11} g_{2}}{\hbar} - \frac{i B_{z} \mu_B \rho_{7,11} g_{3}}{\hbar} - \frac{\Gamma \rho_{7,11}}{2} + \frac{i \Omega_{9-11} \rho_{7 9}}{2},\\
		{\partial \over \partial t}\rho_{88} = \frac{5 \Gamma \rho_{10,10}}{12} + \frac{5 \Gamma \rho_{11,11}}{12} - \frac{i \Omega_{8-10} \rho^*_{8,10}}{2} + \frac{i \Omega_{8-10} \rho_{8,10}}{2} - \frac{i \Omega_{8-11} \rho^*_{8,11}}{2} + \frac{i \Omega_{8-11} \rho_{8,11}}{2},\\
		{\partial \over \partial t}\rho_{89} = - \frac{i \Omega_{8-10} \rho^*_{9,10}}{2} - \frac{i \Omega_{8-11} \rho^*_{9,11}}{2} + \frac{i \Omega_{9-11} \rho_{8,11}}{2},\\
		{\partial \over \partial t}\rho_{8,10} = \frac{i B_{z} \mu_B \rho_{8,10} g_{3}}{\hbar} - \frac{\Gamma \rho_{8,10}}{2} + \frac{i \Omega_{6-10} \rho^*_{6 8}}{2} - \frac{i \Omega_{8-10} \rho_{10,10}}{2} + \frac{i \Omega_{8-10} \rho_{8 8}}{2},\\
		{\partial \over \partial t}\rho_{8,11} = - \frac{i B_{z} \mu_B \rho_{8,11} g_{3}}{\hbar} - \frac{\Gamma \rho_{8,11}}{2} + \frac{i \Omega_{6-11} \rho^*_{6 8}}{2} - \frac{i \Omega_{8-11} \rho_{11,11}}{2} + \frac{i \Omega_{8-11} \rho_{8 8}}{2},\\
		{\partial \over \partial t}\rho_{99} = \frac{\Gamma \rho_{11,11}}{2} - \frac{i \Omega_{9-11} \rho^*_{9,11}}{2} + \frac{i \Omega_{9-11} \rho_{9,11}}{2},\\
		{\partial \over \partial t}\rho_{9,10} = \frac{2 i B_{z} \mu_B \rho_{9,10} g_{2}}{\hbar} + \frac{i B_{z} \mu_B \rho_{9,10} g_{3}}{\hbar} - \frac{\Gamma \rho_{9,10}}{2} + \frac{i \Omega_{7-10} \rho^*_{7 9}}{2},\\
		{\partial \over \partial t}\rho_{9,11} = \frac{2 i B_{z} \mu_B \rho_{9,11} g_{2}}{\hbar} - \frac{i B_{z} \mu_B \rho_{9,11} g_{3}}{\hbar} - \frac{\Gamma \rho_{9,11}}{2} - \frac{i \Omega_{9-11} \rho_{11,11}}{2} + \frac{i \Omega_{9-11} \rho_{9 9}}{2},\\
		{\partial \over \partial t}\rho_{10,10} = - \Gamma \rho_{10,10} + \frac{i \Omega_{6-10} \rho^*_{6,10}}{2} - \frac{i \Omega_{6-10} \rho_{6,10}}{2} + \frac{i \Omega_{7-10} \rho^*_{7,10}}{2} - \frac{i \Omega_{7-10} \rho_{7,10}}{2} + \frac{i \Omega_{8-10} \rho^*_{8,10}}{2} - \frac{i \Omega_{8-10} \rho_{8,10}}{2},\\
		{\partial \over \partial t}\rho_{10,11} = - \frac{i \Omega_{6-10} \rho_{6,11}}{2} + \frac{i \Omega_{6-11} \rho^*_{6,10}}{2} - \frac{i \Omega_{7-10} \rho_{7,11}}{2} - \frac{i \Omega_{8-10} \rho_{8,11}}{2}+ \frac{i \Omega_{8-11} \rho^*_{8,10}}{2} + \frac{i \Omega_{9-11} \rho^*_{9,10}}{2},\\
		{\partial \over \partial t}\rho_{11,11} = - \Gamma \rho_{11,11} + \frac{i \Omega_{6-11} \rho^*_{6,11}}{2} - \frac{i \Omega_{6-11} \rho_{6,11}}{2} + \frac{i \Omega_{8-11} \rho^*_{8,11}}{2} - \frac{i \Omega_{8-11} \rho_{8,11}}{2} + \frac{i \Omega_{9-11} \rho^*_{9,11}}{2} - \frac{i \Omega_{9-11} \rho_{9,11}}{2},\\
	\end{cases}
\end{equation}
The rapid evolution or decay elements of density matrix is small. Therefore near the magneto-insensitive two-photon resonance, in the presence of magnetic field $B_z$, density matrix elements except $\{\rho_{66},\rho_{77},\rho_{88},\rho_{99},\rho_{68},\rho_{68}^*\}$ can be regarded as small perturbations. The ground state can be regarded as invariant and $\rho_{66}=\rho_{77}=\rho_{88}=\rho_{99}=0.25$.
Using the adiabatically eliminating (which implies consideration of the system at times $t\mygg \frac{1}{\Gamma}$)~\cite{Chuchelov2019},
\begin{equation}
	\begin{cases}
		\rho_{6,10}=\frac{i \left(0.25 \Omega_{6-10} + \Omega_{8-10} \rho_{6 8}\right)}{\Gamma},\\
		\rho_{6,11}=\frac{i \left(0.25 \Omega_{6-11} + \Omega_{8-11} \rho_{6 8}\right)}{\Gamma},\\
		\rho_{8,10}^*=- \frac{i \left(\Omega_{6-10} \rho_{6 8} + 0.25 \Omega_{8-10}\right)}{\Gamma},\\
		\rho_{8,11}^*=- \frac{i \left(\Omega_{6-11} \rho_{6 8} + 0.25 \Omega_{8-11}\right)}{\Gamma}.
	\end{cases}
\end{equation}
Then the evolution of $\rho_{68}$ is
\begin{equation}
	\begin{aligned}
		{\partial \over \partial t} \rho_{68}=&- i \delta \rho_{6 8} - \frac{\Omega_{6-10}^{2} \rho_{6 8}}{2 \Gamma} - \frac{0.25 \Omega_{6-10} \Omega_{8-10}}{\Gamma} - \frac{\Omega_{6-11}^{2} \rho_{6 8}}{2 \Gamma}\\
		& - \frac{0.25 \Omega_{6-11} \Omega_{8-11}}{\Gamma} - \frac{\Omega_{8-10}^{2} \rho_{6 8}}{2 \Gamma} - \frac{\Omega_{8-11}^{2} \rho_{6 8}}{2 \Gamma}
	\end{aligned}
	\label{rho68}
\end{equation}
As the Fig.~\ref{six_teen_levels} shows, the t.d.m of $|6\rangle$ to $|10\rangle$ and $|11\rangle$ are opposite, and the t.d.m of $|8\rangle$ to $|10\rangle$ and $|11\rangle$ are equal. In lin||lin configuration, the $\sigma^+$ and $\sigma^-$ components of each linear polarization light have equal intensity and zero phase shift, hence $\Omega_{6-10}=-\Omega_{6-11}$ and $\Omega_{8-10}=\Omega_{8-11}$.
Then the Eq.~\ref{rho68} will be
\begin{equation}
	{\partial \over \partial t} \rho_{68}=- i \delta \rho_{6 8} - \frac{\Omega_{6-10}^{2} \rho_{6 8}}{\Gamma}-\frac{\Omega_{8-10}^{2} \rho_{6 8}}{\Gamma}.
\end{equation}
The steady-state solution is $\rho_{68}=0$, $|F_g\myeq1,m_F\myeq0\rangle-|F_g\myeq2,m_F\myeq0\rangle$ resonance is absent. Using the adiabatically eliminating we can easy obtain that
\begin{equation}
	\begin{cases}
		\rho_{10,10}={1\over4\Gamma}\left(\Omega_{6-10}^2+\Omega_{7-10}^2+\Omega_{8-10}^2\right),\\
		\rho_{11,11}={1\over4\Gamma}\left(\Omega_{6-11}^2+\Omega_{7-11}^2+\Omega_{8-11}^2\right).
	\end{cases}
\end{equation}
Because the $\rho_{10,10}$ and $\rho_{11,11}$ do not vary with $\delta$ when near the magneto-insensitive resonance. Therefore, when near the magneto-insensitive resonance the absorption signal is proportional to $\rho_{5, 5}+C$, where $C=\rho_{10,10}+\rho_{11,11}$ is a constant. Hence we can consider the five-level system only. An intuitive physical image is that due to the destructive interference of two-photon transitions in the case of lin$||$lin configuration, the $|F_g\myeq1,m_F\myeq0\rangle-|F_g\myeq2,m_F\myeq0\rangle$ resonance is absent\cite{Taichenachev2005a}.
\subsection{Five-level Model}
\begin{figure}[!h]
	\centering
	\includegraphics[width=0.6\linewidth]{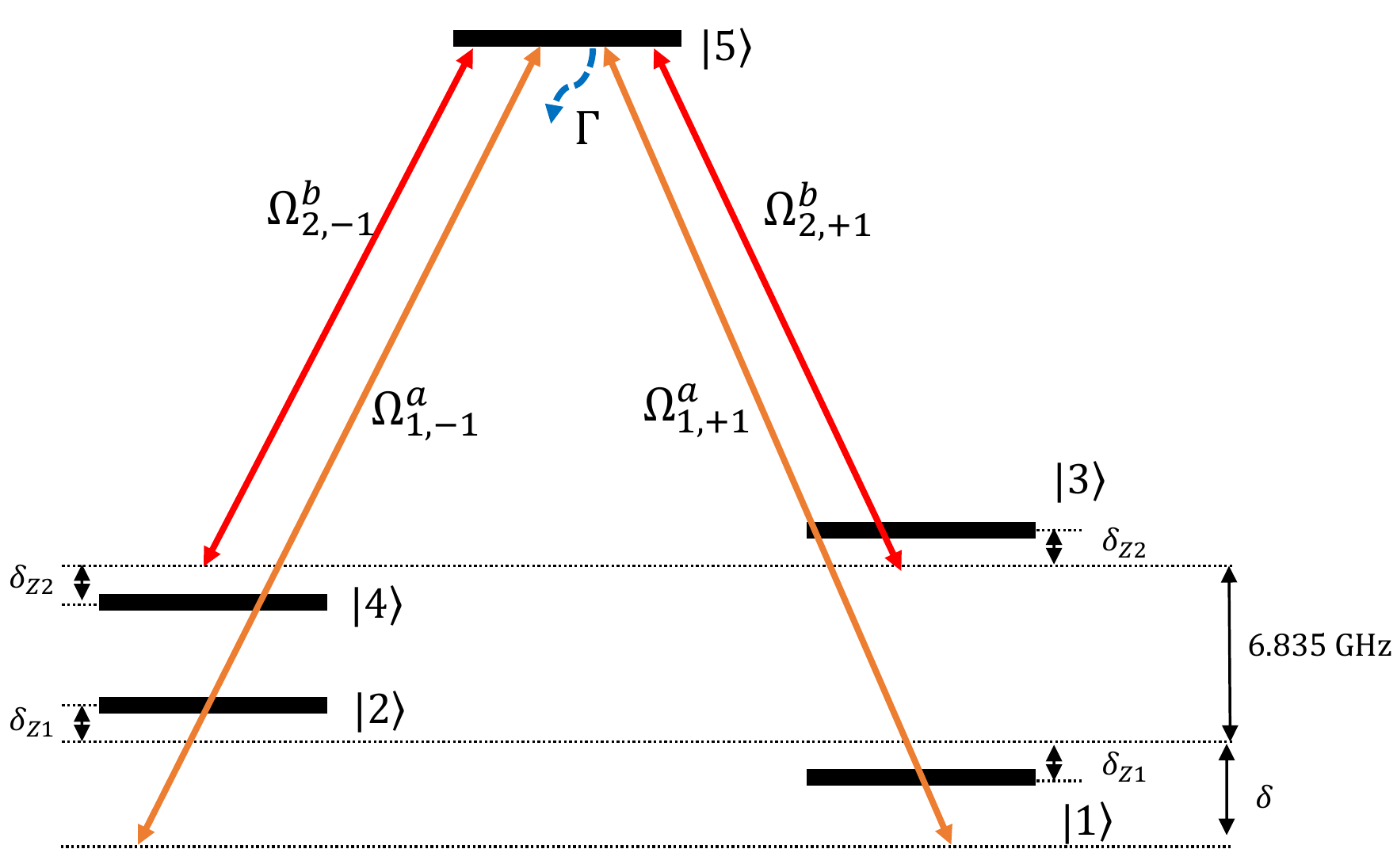}
	\caption{\textbf{Schematic of the five-level model}. The double-$\Lambda$ configuration of a linearly polarized bichromatic light can be described by a five-level model.
		In the absence of the external magnetic field, the transition frequencies for $|1\rangle \to |4\rangle$ and $|2\rangle \to |3\rangle$ are both $6.835$ GHz.
		When a small magnetic field $B_z$ along z-axis is applied and the second-order Zeeman effect is ignored, these Zeeman sublevels of the hyperfine ground state experience linear Zeeman shift as $\delta_{Z1}\myeq g_1\mu_B B_z/\hbar$ and $\delta_{Z2}\myeq g_2\mu_B B_z/\hbar$.
		The $\Omega^{a}_{1,\pm 1}$ and $\Omega^{b}_{2,\pm 1}$ are Rabi frequencies.}
	\label{five_level}
\end{figure}

The five-level system contains double $\Lambda$ models as (see Fig.~\ref{five_level}) shows.
%

As shown in Fig.~\ref{five_level}, a bichromatic field with frequencies $\omega_a$ and $\omega_b$ couples the five levels. If the Rabi frequency $\Omega^{a}_{1,\pm 1}$ and $\Omega^{b}_{2,\pm 1}$ are real, the Hamiltonian $\hat{H}_{5\times 5}$ reads,
\begin{equation}
	\hat{H}_{5\times 5}=\hbar
	\left(\begin{matrix}
		\Delta_1 & 0 & 0 & 0 & \frac{\Omega^{a}_{1,+1}}{2}\\
		0 & \Delta_2& 0 & 0 & \frac{\Omega^{a}_{1,-1}}{2}\\
		0 & 0 & \Delta_3& 0 & \frac{\Omega^{b}_{2,+1}}{2}\\
		0 & 0 & 0 & \Delta_4 & \frac{\Omega^{b}_{2,-1}}{2}\\
		\frac{\Omega^{a}_{1,+1}}{2}&\frac{\Omega^{a}_{1,-1}}{2}&\frac{\Omega^{b}_{2,+1}}{2}&\frac{\Omega^{b}_{2,-1}}{2}&  - \frac{\mathrm{i}\Gamma}{2}
	\end{matrix}\right).
	\label{hamilton}
\end{equation}
In the presence of magnetic field, the diagonal terms of the $ \hat{H}_{5\times 5}$ matrix are defined in terms of the $\delta$ and linear Zeeman shift ($\delta_{Z1}$ and $\delta_{Z2}$),
\begin{equation}
	\label{define_ham}
	\begin{aligned}
		\Delta_1 &= \delta + \delta_{Z1},\\
		\Delta_2 &= \delta - \delta_{Z1} ,\\
		\Delta_3 &= \delta_{Z2},\\
		\Delta_4 &= -\delta_{Z2}.
	\end{aligned}
\end{equation}
\newpage

\section{Analytical Results from Five-level Model}\label{theory}
\subsection{Simplification and Derivation}
The five-level model Bloch equation reads
\begin{equation}\
	\begin{cases}\
		\frac{\partial}{\partial t}\rho_{11}=\frac{\Gamma_1}{2}\rho_{55} + \frac{\mathrm{i}\Omega_{1,+1}^{a}}{2}\rho_{15} - \frac{\mathrm{i}\Omega_{1,+1}^{a}}{2}\rho_{15}^{*},\\\
		\frac{\partial}{\partial t}\rho_{12}=-\frac{\mathrm{i}\Omega_{1,+1}^{a}}{2}\rho_{25}^{*} + \frac{\mathrm{i}\Omega_{1,-1}^{a}}{2}\rho_{15} - \gamma_{12}\rho_{12} - \mathrm{i}\rho_{12}\left(\Delta_1-\Delta_2\right),\\\
		\frac{\partial}{\partial t}\rho_{13}=-\frac{\mathrm{i}\Omega_{1,+1}^{a}}{2}\rho_{35}^{*} + \frac{\mathrm{i}\Omega_{2,+1}^{b}}{2}\rho_{15} - \gamma_{13}\rho_{13} - \mathrm{i}\rho_{13}(\Delta_1 - \Delta_{3})\\\
		\frac{\partial}{\partial t}\rho_{14}=-\frac{\mathrm{i}\Omega_{1,+1}^{a}}{2}\rho_{45}^{*} + \frac{\mathrm{i}\Omega_{2,-1}^{b}}{2}\rho_{15} - \gamma_{14}\rho_{14} - \mathrm{i}\rho_{14}(\Delta_1 - \Delta_4),\\\
		\frac{\partial}{\partial t}\rho_{15}=-\frac{\Gamma}{2}\rho_{15} + \frac{\mathrm{i}\Omega_{1,+1}^{a}}{2}\rho_{11} - \frac{\mathrm{i}\Omega_{1,+1}^{a}}{2}\rho_{55} + \frac{\mathrm{i}\Omega_{1,-1}^{a}}{2}\rho_{12} + \frac{\mathrm{i}\Omega_{2,+1}^{b}}{2}\rho_{13} + \frac{\mathrm{i}\Omega_{2,-1}^{b}}{2}\rho_{14} - \mathrm{i}\rho_{15}\Delta_1,\\\
		\frac{\partial}{\partial t}\rho_{22}=\frac{\Gamma_1}{2}\rho_{55} + \frac{\mathrm{i}\Omega_{1,-1}^{a}}{2}\rho_{25} - \frac{\mathrm{i}\Omega_{1,-1}^{a}}{2}\rho_{25}^{*},\\\
		\frac{\partial}{\partial t}\rho_{23}=-\frac{\mathrm{i}\Omega_{1,-1}^{a}}{2}\rho_{35}^{*} + \frac{\mathrm{i}\Omega_{2,+1}^{b}}{2}\rho_{25} - \gamma_{23}\rho_{23} - \mathrm{i}\rho_{23}(\Delta_2 - \Delta_3),\\\
		\frac{\partial}{\partial t}\rho_{24}=-\frac{\mathrm{i}\Omega_{1,-1}^{a}}{2}\rho_{45}^{*} + \frac{\mathrm{i}\Omega_{2,-1}^{b}}{2}\rho_{25} - \gamma_{24}\rho_{24} - \mathrm{i}\rho_{24}(\Delta_2 - \Delta_4),\\\
		\frac{\partial}{\partial t}\rho_{25}=-\frac{\Gamma}{2}\rho_{25} + \frac{\mathrm{i}\Omega_{1,+1}^{a}}{2}\rho_{12}^{*} + \frac{\mathrm{i}\Omega_{1,-1}^{a}}{2}\rho_{22} - \frac{\mathrm{i}\Omega_{1,-1}^{a}}{2}\rho_{55} + \frac{\mathrm{i}\Omega_{2,+1}^{b}}{2}\rho_{23} + \frac{\mathrm{i}\Omega_{2,-1}^{b}}{2}\rho_{24} - \mathrm{i}\rho_{25}\Delta_2,\\\
		\frac{\partial}{\partial t}\rho_{33}=\frac{\Gamma_2}{2}\rho_{55} + \frac{\mathrm{i}\Omega_{2,+1}^{b}}{2}\rho_{35} - \frac{\mathrm{i}\Omega_{2,+1}^{b}}{2}\rho_{35}^{*},\\\
		\frac{\partial}{\partial t}\rho_{34}=-\frac{\mathrm{i}\Omega_{2,+1}^{b}}{2}\rho_{45}^{*} + \frac{\mathrm{i}\Omega_{2,-1}^{b}}{2}\rho_{35} - \gamma_{34}\rho_{34} - \mathrm{i}\rho_{34}(\Delta_3 - \Delta_4),\\\
		\frac{\partial}{\partial t}\rho_{35}=-\frac{\Gamma}{2}\rho_{35} + \frac{\mathrm{i}\Omega_{1,+1}^{a}}{2}\rho_{13}^{*} + \frac{\mathrm{i}\Omega_{1,-1}^{a}}{2}\rho_{23}^{*} + \frac{\mathrm{i}\Omega_{2,+1}^{b}}{2}\rho_{33} - \frac{\mathrm{i}\Omega_{2,+1}^{b}}{2}\rho_{55} + \frac{\mathrm{i}\Omega_{2,-1}^{b}}{2}\rho_{34} - \mathrm{i}\rho_{35}\Delta_3,\\\
		\frac{\partial}{\partial t}\rho_{44}=\frac{\Gamma_2}{2}\rho_{55} + \frac{\mathrm{i}\Omega_{2,-1}^{b}}{2}\rho_{45} - \frac{\mathrm{i}\Omega_{2,-1}^{b}}{2}\rho_{45}^{*},\\\
		\frac{\partial}{\partial t}\rho_{45}=-\frac{\Gamma}{2}\rho_{45} + \frac{\mathrm{i}\Omega_{1,+1}^{a}}{2}\rho_{14}^{*} + \frac{\mathrm{i}\Omega_{1,-1}^{a}}{2}\rho_{24}^{*} + \frac{\mathrm{i}\Omega_{2,+1}^{b}}{2}\rho_{34}^{*} + \frac{\mathrm{i}\Omega_{2,-1}^{b}}{2}\rho_{44} - \frac{\mathrm{i}\Omega_{2,-1}^{b}}{2}\rho_{55} - \mathrm{i}\rho_{45}\Delta_4,\\\
		\frac{\partial}{\partial t}\rho_{55}=-\Gamma\rho_{55} - \frac{\mathrm{i}\Omega_{1,+1}^{a}}{2}\rho_{15} + \frac{\mathrm{i}\Omega_{1,+1}^{a}}{2}\rho_{15}^{*} - \frac{\mathrm{i}\Omega_{1,-1}^{a}}{2}\rho_{25} + \frac{\mathrm{i}\Omega_{1,-1}^{a}}{2}\rho_{25}^{*} - \frac{\mathrm{i}\Omega_{2,+1}^{b}}{2}\rho_{35}\\
		\quad \quad \quad \quad  +\frac{\mathrm{i}\Omega_{2,+1}^{b}}{2}\rho_{35}^{*} - \frac{\mathrm{i}\Omega_{2,-1}^{b}}{2}\rho_{45} + \frac{\mathrm{i}\Omega_{2,-1}^{b}}{2}\rho_{45}^{*}.\\
	\end{cases}
	\label{bloch_eq}
\end{equation}
%
%
%
The decaying from excited states to the ground states $|F\myeq 1\rangle$ and $|F\myeq 2\rangle$ satisfy $\Gamma_1:\Gamma_2\myeq 1:3$.
%
%
In our experiment, the intensities of bichromatic light field are equal, so the four Rabi frequencies satisfy~\cite{steck2001rubidium, Shahriar2014},
\begin{equation}
	\label{rabi_ratio}
	\Omega_{1,+1}^{a}:\Omega_{1,-1}^{a}:\Omega_{2,+1}^{b}:\Omega_{2,-1}^{b}=1:-1:\sqrt{3}:\sqrt{3}.
\end{equation}
%
%
%
%
For simplicity, in theoretical analysis we do not consider degenerate Zeeman
sublevels~\cite{Chen_PRA_2000}, and set all four Rabi frequencies as the average Rabi frequency and the damping rates $\Gamma_1 \myeq \Gamma_2 \myeq \Gamma$.
To compare with the experimental observation, the average Rabi frequency can be given as $\Omega\myeq\sqrt{\left[(\Omega_{1,\pm 1}^{a})^2+(\Omega_{2,\pm 1}^{b})^2\right]/2}$~\cite{ Hemmer_1989}.
%
Using the adiabatically eliminating (which implies consideration of the model at times $t\mygg \frac{1}{\Gamma}$ and $\frac{\partial}{\partial t} \rho_{55}\myap 0$)~\cite{Chuchelov2019}, solving Eq.~\ref{bloch_eq}, the excited state population can be given
\begin{equation}
	\rho_{55} \approx \frac{\Omega}{\Gamma} \left(\mathrm{Im}\left(\rho_{15}\right)+\mathrm{Im}\left(\rho_{25}\right)+
	\mathrm{Im}\left(\rho_{35}\right)+\mathrm{Im}\left(\rho_{45}\right)\right),
	\label{rho55}
\end{equation}
which is relevant to the light absorption~\cite{PhysRevA.79.063837}.
For $^{87}$Rb atoms, the excited-state decaying rate of D1 line is $\Gamma \myeq2\pi\times 5.75$ MHz, and so that the accumulated excited-state population $\rho_{55}$ is small comparing with that of ground states.
Thus, in the near resonant region of $\delta \ll \Gamma$, we can simplify the time-evolution as
\begin{equation}
	\begin{cases}\
		\frac{\partial}{\partial t}\rho_{15}\approx -\frac{\Gamma}{2}\rho_{15}  + \frac{\mathrm{i}\Omega}{2}\rho_{11} + \frac{\mathrm{i}\Omega}{2}\left(\rho_{12}+\rho_{13}+\rho_{14}\right),\\\
		\frac{\partial}{\partial t}\rho_{25}\approx-\frac{\Gamma}{2}\rho_{25} + \frac{\mathrm{i}\Omega}{2}\rho_{22} + \frac{\mathrm{i}\Omega}{2}\left(\rho_{12}^*+\rho_{23}+\rho_{24}\right),\\\
		\frac{\partial}{\partial t}\rho_{35}\approx-\frac{\Gamma}{2}\rho_{35} + \frac{\mathrm{i}\Omega}{2}\rho_{33} + \frac{\mathrm{i}\Omega}{2}\left(\rho_{13}^*+\rho_{23}^*+\rho_{34}\right),\\\
		\frac{\partial}{\partial t}\rho_{45}\approx-\frac{\Gamma}{2}\rho_{45} + \frac{\mathrm{i}\Omega}{2}\rho_{44} +  \frac{\mathrm{i}\Omega}{2}\left(\rho_{14}^*+\rho_{24}^*+\rho_{34}^*\right).
	\end{cases}\
	\label{rho15s}
\end{equation}
%
%
Under adiabatically eliminating, solving Eq.~\ref{rho15s}, we can get
\begin{equation}
	\begin{cases}\
		\rho_{15}\approx \frac{\mathrm{i}\Omega}{\Gamma} \left(\rho_{11} + \rho_{12}+\rho_{13}+\rho_{14}\right),\\\
		\rho_{25}\approx \frac{\mathrm{i}\Omega}{\Gamma} \left(\rho_{22} + \rho_{12}^*+\rho_{23}+\rho_{24}\right),\\\
		\rho_{35}\approx \frac{\mathrm{i}\Omega}{\Gamma} \left(\rho_{33} + \rho_{13}^*+\rho_{23}^*+\rho_{34}\right),\\\
		\rho_{45}\approx \frac{\mathrm{i}\Omega}{\Gamma} \left(\rho_{44} + \rho_{14}^*+\rho_{24}^*+\rho_{34}^*\right).
	\end{cases}\
	\label{rho15s_approx}
\end{equation}
%
%
The population of excited-state $\rho_{55} \ll 1$, and the population of ground state approximately $\rho_{11}=\rho_{22}=\rho_{33}=\rho_{44}= 1/4$.
Thus Eq.~\ref{rho55} reads
\begin{equation}
	\begin{aligned}
		\rho_{55} =&\frac{\Omega^2}{\Gamma^2}\left(\rho_{11}+\rho_{22}+\rho_{33}+\rho_{44} + 2\mathrm{Re}\left(\rho_{12}\right)+2\mathrm{Re}\left(\rho_{34}\right)\right.\\
		&\left.+2\mathrm{Re}\left(\rho_{13}\right)+2\mathrm{Re}\left(\rho_{24}\right)+2\mathrm{Re}\left(\rho_{14}\right)+2\mathrm{Re}\left(\rho_{23}\right)\right)\\
		=&\frac{\Omega^2}{\Gamma^2}\left(1 + 2\mathrm{Re}\left(\rho_{12}\right)+2\mathrm{Re}\left(\rho_{34}\right)+2\mathrm{Re}\left(\rho_{13}\right)\right.\\
		&\left.+2\mathrm{Re}\left(\rho_{24}\right)+2\mathrm{Re}\left(\rho_{14}\right)+2\mathrm{Re}\left(\rho_{23}\right)\right).
	\end{aligned}
	\label{rho55_all}
\end{equation}
This means that the excited-state population $\rho_{55}$ is relevant to the real parts of six density-matrix elements $\{\rho_{12}, \rho_{34}, \rho_{13}, \rho_{24}, \rho_{14}, \rho_{23}\}$~\cite{Chuchelov2019}.
Combining Eq.~\ref{bloch_eq} and Eq.~\ref{rho15s_approx}, the time-evolution can be described by
\begin{equation}
	\begin{cases}
		\frac{\partial}{\partial t}\rho_{12}= -\frac{\Omega^2}{2\Gamma}\left(\frac{1}{2}+2\rho_{12}+\rho_{13}+\rho_{14}+\rho_{23}^*+\rho_{24}^*\right) - \mathrm{i}\rho_{12}\Delta_{12}, \\
		\frac{\partial}{\partial t}\rho_{34}= -\frac{\Omega^2}{2\Gamma}\left(\frac{1}{2}+2\rho_{34}+\rho_{13}^*+\rho_{23}^*+\rho_{14}+\rho_{24}^*\right) - \mathrm{i}\rho_{23}\Delta_{34},\\
		\frac{\partial}{\partial t}\rho_{13}= -\frac{\Omega^2}{2\Gamma}\left(\frac{1}{2}+2\rho_{13}+\rho_{12}+\rho_{14}+\rho_{23}+\rho_{34}^*\right) - \mathrm{i}\rho_{13}\Delta_{13}, \\
		\frac{\partial}{\partial t}\rho_{24}= -\frac{\Omega^2}{2\Gamma}\left(\frac{1}{2}+2\rho_{24}+\rho_{12}^*+\rho_{23}+\rho_{14}+\rho_{34}\right) - \mathrm{i}\rho_{24}\Delta_{24},\\
		\frac{\partial}{\partial t}\rho_{14}= -\frac{\Omega^2}{2\Gamma}\left(\frac{1}{2}+2\rho_{14}+\rho_{12}+\rho_{13}+\rho_{24}+\rho_{34}\right) - \mathrm{i}\rho_{14}\Delta_{14}, \\
		\frac{\partial}{\partial t}\rho_{23}= -\frac{\Omega^2}{2\Gamma}\left(\frac{1}{2}+2\rho_{23}+\rho_{12}^*+\rho_{24}+\rho_{13}+\rho_{34}^*\right) - \mathrm{i}\rho_{23}\Delta_{23}.
	\end{cases}
	\label{rho14s_1}
\end{equation}
Where
\begin{equation}
	\begin{cases}
		\Delta_{12}=\Delta_1-\Delta_2-\mathrm{i}\gamma_{12}, \\
		\Delta_{34}=\Delta_3-\Delta_4-\mathrm{i}\gamma_{34},\\
		\Delta_{13}=\Delta_1-\Delta_3-\mathrm{i}\gamma_{13}, \\
		\Delta_{24}=\Delta_2-\Delta_4-\mathrm{i}\gamma_{24},\\
		\Delta_{14}=\Delta_1-\Delta_4-\mathrm{i}\gamma_{14}, \\
		\Delta_{23}=\Delta_2-\Delta_3-\mathrm{i}\gamma_{23}.
	\end{cases}
	\label{define_Deltaij}
\end{equation}
%
%
Subjected to bias magnetic field, the resonances of magneto-sensitive transitions are sufficiently separated in frequency
to not overlap with the magneto-insensitive transitions, Eq.~\ref{rho14s_1} can be further simplified as
\begin{equation}
	\begin{cases}
		\frac{\partial}{\partial t}\rho_{12}\approx -\frac{\Omega^2}{2\Gamma}\left(\frac{1}{2}+2\rho_{12}\right) - \mathrm{i}\rho_{12}\Delta_{12}, \\
		\frac{\partial}{\partial t}\rho_{34}\approx -\frac{\Omega^2}{2\Gamma}\left(\frac{1}{2}+2\rho_{34}\right) - \mathrm{i}\rho_{34}\Delta_{34},\\
		\frac{\partial}{\partial t}\rho_{13}\approx-\frac{\Omega^2}{2\Gamma}\left(\frac{1}{2}+2\rho_{13}\right) - \mathrm{i}\rho_{13}\Delta_{13}, \\
		\frac{\partial}{\partial t}\rho_{24}\approx -\frac{\Omega^2}{2\Gamma}\left(\frac{1}{2}+2\rho_{24}\right) - \mathrm{i}\rho_{24}\Delta_{24},\\
		\frac{\partial}{\partial t}\rho_{14}\approx -\frac{\Omega^2}{2\Gamma}\left(\frac{1}{2}+2\rho_{14}\right) - \mathrm{i}\rho_{14}\Delta_{14}, \\
		\frac{\partial}{\partial t}\rho_{23}\approx -\frac{\Omega^2}{2\Gamma}\left(\frac{1}{2}+2\rho_{23}\right) - \mathrm{i}\rho_{23}\Delta_{23}.\\
	\end{cases}
	\label{rho14s_2}
\end{equation}
%
%
If $\Omega$ remains unchanged for a period of time, by solving Eq.~\ref{rho14s_2}, the solution of $ \rho_{14} $ is
\begin{equation}
	\rho_{14}\left(t_0+t\right)=f(\Delta_{14})\left(1-\mathrm{e}^{-\left(\mathrm{i}\Delta_{14}+\frac{\Omega^2}{\Gamma}\right)t}\right) + \rho_{14}\left(t_0\right)\mathrm{e}^{-\left(\mathrm{i}\Delta_{14}+\frac{\Omega^2}{\Gamma}\right)t}
	\label{rho14s_solution}
\end{equation}
with
\begin{equation}
	f(x) = -\frac{\Omega^2}{4\Gamma\left(\mathrm{i}x + \frac{\Omega^2}{\Gamma}\right)},
	\label{fx_def}
\end{equation}
and $ t_0 $ is the initial time and $ t $ is the evolution time.

\begin{figure}[!h]
	\centering
	\includegraphics[width=0.6\linewidth]{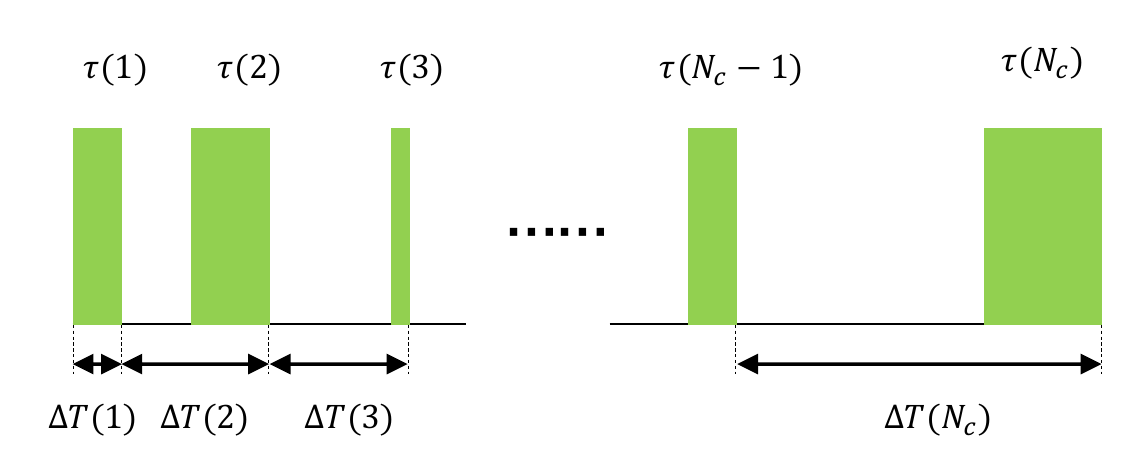}
	\caption{\textbf{Time sequence of multi-pulses.} Each pulse interval $\uptau \left(k\right)$ and pulse interval $\Delta T \left(k\right)$ $(k \in \mathbb{N})$ are variable.}
	\label{fig_s2}
\end{figure}	

When a sequence of pulses is applied, $\rho_{14} $ depends on the pulse length $\uptau(k)$ and the pulse interval $\Delta T(k)$ as shown in Fig.~\ref{fig_s2}.
Given $\rho_{14}(0)\myeq 0$, according to Eq.\ref{rho14s_solution}, after the first pulse, we have
\begin{equation}
	\begin{aligned}
		\rho_{14}^1&= f(\Delta_{14})\left(1-\mathrm{e}^{-\left(\frac{\Omega^2}{\Gamma}+\mathrm{i}\Delta_{14}\right)\uptau(1)}\right)\\
		&=f(\Delta_{14})\sum_{l=1}^{1}\mathrm{e}^{-\frac{\Omega^2}{\Gamma}\sum_{k=l+1}^{1}\uptau(k)}\left[1-\mathrm{e}^{-\left(\frac{\Omega^2}{\Gamma}+\mathrm{i}\Delta_{14}\right)\uptau(l)}\right] \mathrm{e}^{-\mathrm{i}\Delta_{14} \sum_{k=l+1}^{1}\Delta T(k)}.\\
	\end{aligned}
	\label{rho14_1}
\end{equation}
%
Similarly, assume that after the $N_t$-th $(N_t\in \mathbb{N})$ pulse, we have
\begin{equation}
	\begin{aligned}
		\rho_{14}^{N_t}
		&=f(\Delta_{14})\sum_{l=1}^{N_t}\mathrm{e}^{-\frac{\Omega^2}{\Gamma}\sum_{k=l+1}^{N_t}\uptau(k)}\left[1-\mathrm{e}^{-\left(\frac{\Omega^2}{\Gamma}+\mathrm{i}\Delta_{14}\right)\uptau(l)}\right] \mathrm{e}^{-\mathrm{i}\Delta_{14} \sum_{k=l+1}^{N_t}\Delta T(k)}.
	\end{aligned}
\end{equation}
Before applying the $(N_t+1)$-th pulse $\uptau(N_t + 1)$, there is a time duration of $\Delta T(N_t+1)-\uptau(N_t+1)$ without CPT light, thus we have
\begin{equation}
	\begin{aligned}
		\rho_{14}^{N_t + 1}&= f(\Delta_{14})\left(1-\mathrm{e}^{-\left(\frac{\Omega^2}{\Gamma}+\mathrm{i}\Delta_{14}\right)\uptau(N_t+1)}\right) + \rho_{14}^{N_t}\mathrm{e}^{-\mathrm{i}\Delta_{14} \left(\Delta T(N_t+1)-\uptau(N_t+1)\right)}\times \mathrm{e}^{-\left(\frac{\Omega^2}{\Gamma}+\mathrm{i}\Delta_{14}\right)\uptau(N_t+1)}\\
		&=f(\Delta_{14})\sum_{l=1}^{N_t+1}\mathrm{e}^{-\frac{\Omega^2}{\Gamma}\sum_{k=l+1}^{N_t+1}\uptau(k)}\left[1-\mathrm{e}^{-\left(\frac{\Omega^2}{\Gamma}+\mathrm{i}\Delta_{14}\right)\uptau(l)}\right] \mathrm{e}^{-\mathrm{i}\Delta_{14} \sum_{k=l+1}^{N_t+1}\Delta T(k)}.\\
	\end{aligned}
\end{equation}
Note that if $l+1 \textgreater N_t$, $\sum_{k=l+1}^{N_t}(\cdots)\myeq 0$. Then when $N_c$ pulses are applied, we have
\textcolor{black}{
	\begin{equation}
		\rho_{14}(N_c \Delta T )=\sigma\left(\Delta_{14}\right),
		\label{rho_14_solu}
	\end{equation}
}
where
\begin{equation}
	\sigma(x)=f(x)\sum_{l=1}^{N_c}\mathrm{e}^{-\frac{\Omega^2}{\Gamma}\sum_{k=l+1}^{N_c}\uptau(k)}\left[1-\mathrm{e}^{-\left(\frac{\Omega^2}{\Gamma}+\mathrm{i}x\right)\uptau(l)}\right] \mathrm{e}^{-\mathrm{i}x \sum_{k=l+1}^{N_c}\Delta T(k)}.
	\label{sigma2}
\end{equation}
%
%
Similar to the derivation of $\rho_{14}$, one can easily obtain
\begin{equation}
	\rho_{ij}(N_c \Delta T) = \sigma\left(\Delta_{ij}\right) \quad (i,j\in \{1, 2, 3, 4\}, i\neq j).
	\label{fp23}
\end{equation}

\subsection{Analog to the FP Cavity}

According to Eq.~\ref{rho55_all} and Eq.~\ref{fp23}, $ \rho_{55} $ can be expressed as
\begin{equation}
	\begin{aligned}
		\rho_{55} =& \frac{\Omega^2}{\Gamma^2}\Biggr\{1 + 2\mathrm{Re}\left[\sigma(\Delta_{12})\right]
		+2\mathrm{Re}\left[\sigma(\Delta_{34})\right]
		+ 2\mathrm{Re}\left[\sigma(\Delta_{13})\right]\\
		&+2\mathrm{Re}\left[\sigma(\Delta_{24})\right]+ 2\mathrm{Re}\left[\sigma(\Delta_{14})\right]
		+2\mathrm{Re}\left[\sigma(\Delta_{23})\right]\Biggr\}.
	\end{aligned}
	\label{rho55_all_term}
\end{equation}
Since the dephasing as well as the detuning near center resonance in magneto-sensitive coupling are large, so that $\{\rho_{12},\rho_{34},\rho_{13},\rho_{24}\}$ can be neglected around $ \delta\myeq 0 $, that is,
\begin{equation}
	\rho_{55} = \frac{\Omega^2}{\Gamma^2}\Biggr\{1 + 2\mathrm{Re}\left[\sigma(\Delta_{14})\right]
	+2\mathrm{Re}\left[\sigma(\Delta_{23})\right]\Biggr\}.
	\label{rho55_fp}
\end{equation}
Considering the weak magnetic field such that $|g_1+g_2|\mu_B Bz/(2\pi\hbar) \ll \mathrm{FWHM}$ (where FWHM is linewidth of conventional CPT spectrum) and $\gamma_{14}\myeq \gamma_{23}\myeq 0$, we have
\begin{equation}
	\Delta_{14} \approx \Delta_{23} \approx \delta.
\end{equation}
Thus Eq.~\ref{rho55_fp} can be further simplified as
\begin{equation}
	\rho_{55} = \frac{\Omega^2}{\Gamma^2}\Biggr\{1 + 4\mathrm{Re}\left[\sigma(\delta)\right]\Biggr\}.
	\label{rho55_simplify}
\end{equation}
Fixing the pulse length $\uptau(k)\myeq \uptau$ and the pulse interval $\Delta T(k)\myeq \Delta T$, according to Eq.~\ref{sigma2}, we have
\begin{equation}
	\begin{aligned}
		\sigma(\delta)&=f(\delta)\sum_{l=1}^{N_c}\mathrm{e}^{-\frac{\Omega^2}{\Gamma}(N_c-l)\uptau}\left[1-\mathrm{e}^{-\left(\frac{\Omega^2}{\Gamma}+\mathrm{i}\delta\right)\uptau}\right]
		\mathrm{e}^{-\mathrm{i}\delta (N_c-l)\Delta T}\\
		&=f(\delta)\sum_{l\myeq 1}^{N_c}\mathcal{R}^{\left(N_c-l\right)}\mathcal{T}\mathrm{e}^{-\mathrm{i}(N_c-l)\delta \Delta T}\\
		&=f(\delta)\sum_{l\myeq 1}^{N_c}\mathcal{R}^{\left(l-1\right)}\mathcal{T}\mathrm{e}^{-\mathrm{i}(l-1)\delta \Delta T}\\
		&=\sum_{l\myeq 1}^{N_c}\sigma_{l}(\delta),
	\end{aligned}
	\label{sigma3}
\end{equation}
where $\sigma_{l}(\delta)\myeq f(\delta)\mathcal{R}^{\left(l-1\right)}\mathcal{T}\mathrm{e}^{-\mathrm{i}(l-1)\delta \Delta T}$.
\begin{figure}[!ht]
	\centering
	\includegraphics[width=0.5\linewidth]{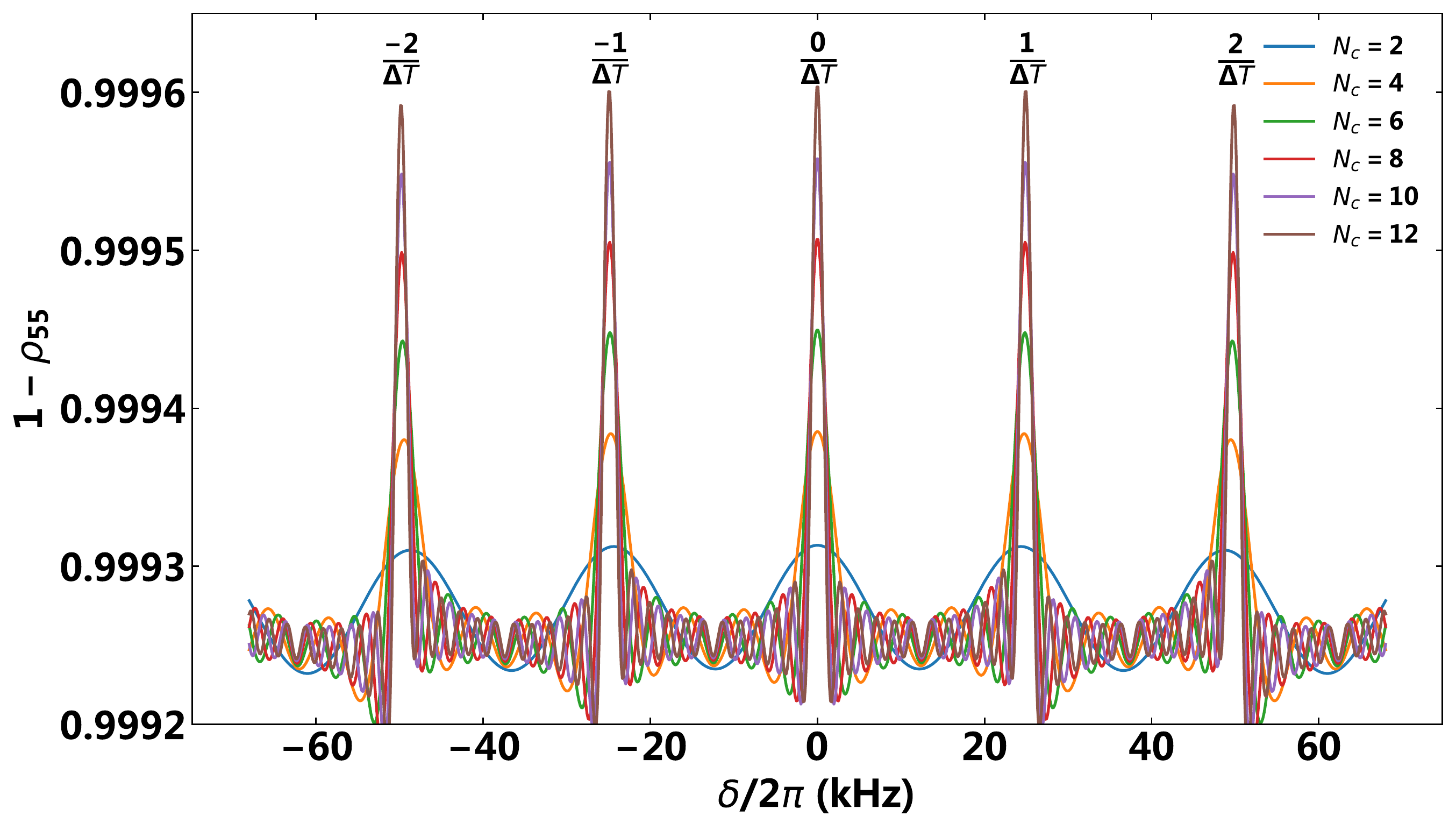}
	\caption{\textbf{Analytical results of different pulse number $ N $.} We set $\Omega\myeq 1e6~\mathrm{s^{-1}}$, $\Delta T \myeq 0.04 $ ms, $\uptau\myeq 0.002$ ms. As $ N $ increases, the peaks become sharper at repeated frequency $\{\frac{n}{\Delta T} \mid n\in \mathbb{Z}\}$.}
	\label{fp_rho55_1}
\end{figure}
Obviously, Eq.~\ref{sigma3} is analogous to the transmission of light in a Fabry-P\'{e}rot (FP) cavity.
Here, the effective reflection coefficient $\mathcal{R}\equiv \mathrm{e}^{-\frac{\Omega^2}{\Gamma}\uptau}$, the effective transmission coefficient $\mathcal{T}\myeq 1-\mathrm{e}^{-\left(\frac{\Omega^2}{\Gamma}+\mathrm{i}\delta\right)\uptau}$ and the effective free spectral range (FSR) $\Delta\nu_{FSR}\myeq \frac{1}{\Delta T}$.
The constructive interference occurs when $\delta\Delta T\myeq 2m\pi  \quad (m\in \mathbb{N})$, otherwise destructive interference occurs.
In FIG.~\ref{fp_rho55_1}, we show the signals of $1-\rho_{55}$ that are analogous to the transmission signal of FP cavity, where the constructive interference appears as comb-like resonant peaks at the frequencies: $\{\frac{n}{\Delta T} \mid n\in \mathbb{Z}\}$.

\begin{figure}[!ht]
	\centering
	\includegraphics[width=0.8\linewidth]{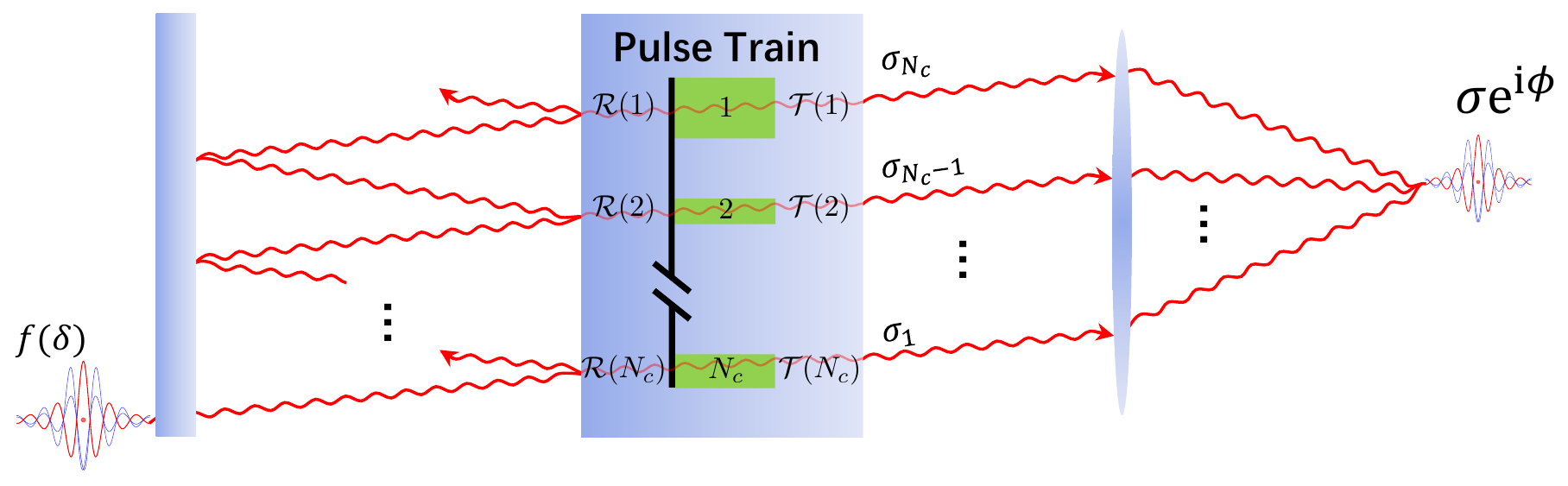}
	\caption{The mapping between each pulse of pulse train and each FP reflection event. The mapping sequence is inverted. The $l$-th pulse maps to $(N_c-l+1)$-th reflection event with reflection coefficient $\mathcal{R}(l)\myeqv \mathrm{e}^{-{\Omega^2\over \Gamma}\uptau(l)}$, transmission coefficient $\mathcal{T}(l)\myeqv 1-\e^{-\left({\Omega^2\over \Gamma}+\mi\delta\right)\uptau(l)}$, and free spectrum range $\Delta\nu_{FSR}(l)\myeqv 1/\Delta T(l)$.}
	\label{mapping2b1}
\end{figure}

Here, if reflection coefficient $\mathcal{R}(l)\myeqv \mathrm{e}^{-{\Omega^2\over \Gamma}\uptau(l)}$ and transmission coefficient $\mathcal{T}(l)\myeqv 1-\e^{-\left({\Omega^2\over \Gamma}+\mi\delta\right)\uptau(l)}$ is variable, then
\begin{equation}
	\label{eqR2_1b}
	\begin{aligned}
		\sigma(\delta)&=f(\delta)\sum_{l=1}^{N_c}\e^{-{\Omega^2\over \Gamma}\sum_{k=l+1}^{N_c}\uptau(k)}\left[1-\e^{-\left({\Omega^2\over \Gamma}+\mi\delta\right)\uptau(l)}\right]\e^{-\mi \delta\sum_{k=l+1}^{N_c}\Delta T(k)}\\
		=&f(\delta)\sum_{l=1}^{N_c}\left[\prod_{k=l+1}^{N_c}\mathcal{R}(k)\right]\mathcal{T}(l)\e^{-\mi \delta\sum_{k=l+1}^{N_c}\Delta T(k)}.
	\end{aligned}
\end{equation}
Refering to Eq. \ref{eqR2_1b}, the temporal spinwave FP  is not just a simple analog, but also a fully mapping. The $l$-th CPT pulse maps to $(N_c-l+1)$-th reflection event of FP FP interferometry. For example, the first CPT pulse of pulse train maps to the last reflection event and the second maps to the last but one et. al as shown in Fig. \ref{mapping2b1}.

Each pulse length $\uptau(l)$ and pulse period $\Delta T(l)$ of $l$-th pulse can be adjusted independently. Accordingly, the reflection coefficient $\mathcal{R}(l)\myeqv \mathrm{e}^{-{\Omega^2\over \Gamma}\uptau(l)}$, transmission coefficient $\mathcal{R}(l)\myeqv \mathrm{e}^{-{\Omega^2\over \Gamma}\uptau(l)}$, and free spectrum range $\Delta\nu_{FSR}(l)\myeqv 1/\Delta T(l)$ can be adjusted too.

\subsection{Connection with Fourier Transform}\label{Fourier}
FP interferometry can be degraded into a result of Fourier transform of pulses train if three conditions can be fixed.
\begin{itemize}
	\item[A] Each pulse length $\uptau$ and pulse period $\Delta T$ of are the small.
	\item[B] Rabi-frequency is much smaller than $\Gamma$, $\Omega\ll \Gamma$.
	\item[C] Pulse length is much smaller than the pulse period $\uptau\ll \Delta T$.
\end{itemize}
Now, we do a simple derivation to find the connection between FP interferometry and Fourier transform. According to our theory, when the CPT pulses are identical the transmission spectrum can be analytically explained by Eq.~\ref{rho55_simplify}.
The $\sigma(\delta)$ can be reformed as
\begin{equation}
	\label{eqR3}
	\begin{aligned}
		\sigma(\delta)&=f(\delta)\sum_{l=1}^{N_c}\left[1-\e^{-\left({\Omega^2\over\Gamma}+\mi \delta \right)\uptau}\right]\e^{-{\Omega^2\over \Gamma}(N_c-l)\uptau}\e^{-\mi \delta(N_c-l)\Delta T},\\
		&=f(\delta)\left[1-\e^{-\left({\Omega^2\over\Gamma}+\mi \delta \right)\uptau}\right]\e^{-\left({\Omega^2\uptau\over \Gamma \Delta T}+\mi \delta \right)N_c\Delta T}\sum_{l=1}^{N_c}\e^{\left({\Omega^2\uptau\over \Gamma \Delta T}+\mi \delta \right)l\Delta T},\\
		&=f(\delta)\left[1-\e^{-({\Omega^2\over \Gamma}+\mi \delta)\uptau}\right]\e^{-\left({\Omega^2\uptau\over \Gamma \Delta T}+\mi \delta \right)T}{{\Omega^2\uptau\over \Gamma \Delta T}+\mi \delta \over 1-\e^{-\left({\Omega^2\uptau\over \Gamma \Delta T}+\mi \delta \right)\uptau}}\\
		&\times \sum_{l=1}^{N_c}\int_{l\Delta T-\uptau}^{l\Delta T}\e^{\left({\Omega^2\uptau\over \Gamma \Delta T}+\mi\delta\right)t}dt,\\
		&=f(\delta)\left[1-\e^{-({\Omega^2\over \Gamma}+\mi \delta)\uptau}\right]{{\Omega^2\uptau\over \Gamma \Delta T}+\mi \delta \over 1-\e^{-\left({\Omega^2\uptau\over \Gamma \Delta T}+\mi \delta \right)\uptau}} \int_{0}^{T} p(t) \e^{-\left({\Omega^2\uptau\over \Gamma \Delta T}+\mi\delta\right)(T-t)}dt,\\
	\end{aligned}
\end{equation}
$p(t)$ is a pulse train, $f(\delta)=-{\Omega^2\over 4\Gamma (\mi \delta + {\Omega^2\over \Gamma})}$,
\begin{equation}
	p(t)=
	\begin{cases}
		1~~t=[l\Delta T-\uptau, l\Delta T],~l\in [1,\dots,N_c],\\
		0~~t=\mathrm{other}.
	\end{cases}
\end{equation}
If the pulse length $\uptau\ll \Delta T$,
\begin{equation}
	\label{laplace}
	\begin{aligned}
		\sigma(\delta)&=f(\delta)\left({\Omega^2\over \Gamma} + \mi \delta\right)\uptau{{\Omega^2\uptau\over \Gamma \Delta T}+\mi \delta \over \left({\Omega^2\uptau\over \Gamma \Delta T}+\mi \delta \right)\uptau}\int_{0}^{T} p(t) \cdot \e^{-\left({\Omega^2\uptau\over \Gamma \Delta T}+\mi\delta\right)(T-t)}dt,\\
		&=-{\Omega^2\over 4\Gamma}\int_{0}^{T}p(t) \e^{-\left({\Omega^2\uptau\over \Gamma \Delta T}+\mi\delta\right)(T-t)}dt.
	\end{aligned}
\end{equation}
Eq. \ref{laplace} is a convolution of a pulse train. Let $t^\prime=T-t$, then Eq. \ref{laplace} can be derived as
\begin{equation}
	\label{laplace2}
	\begin{aligned}
		\sigma(\delta)&=-{\Omega^2\over 4\Gamma}\int_{0}^{T}p(T-t^\prime) \e^{-\left({\Omega^2\uptau\over \Gamma \Delta T}+\mi\delta\right)t^\prime}dt^\prime.\\
		&=-{\Omega^2\over 4\Gamma}\int_{0}^{T}p(T-t) \e^{-\left({\Omega^2\uptau\over \Gamma \Delta T}+\mi\delta\right)t}dt.
	\end{aligned}
\end{equation}
When the Rabi frequency $\Omega\ll \Gamma$ and pulse length $\uptau\ll \Delta T$, the ${\Omega^2\tau \over\Gamma \Delta T}$ can be neglected and Eq. \ref{laplace2} degrades into a Fourier transform.

Despite the FP interferometry could be mathematically degraded into a form of Fourier Transform at some extreme conditions, our experiment is a temporal spinwave FP interferometry, rather than Fourier Transform of pulse train. The conditions for Fourier Transform such as the small pulse length is also not easy to realized experimentally.

\subsection{Multi-Pulse CPT-Ramsey Spectrum in a Non-weak Bias Magnetic Field}\label{general_case}

\begin{figure}[!ht]
	\centering
	\includegraphics[width=0.6\linewidth]{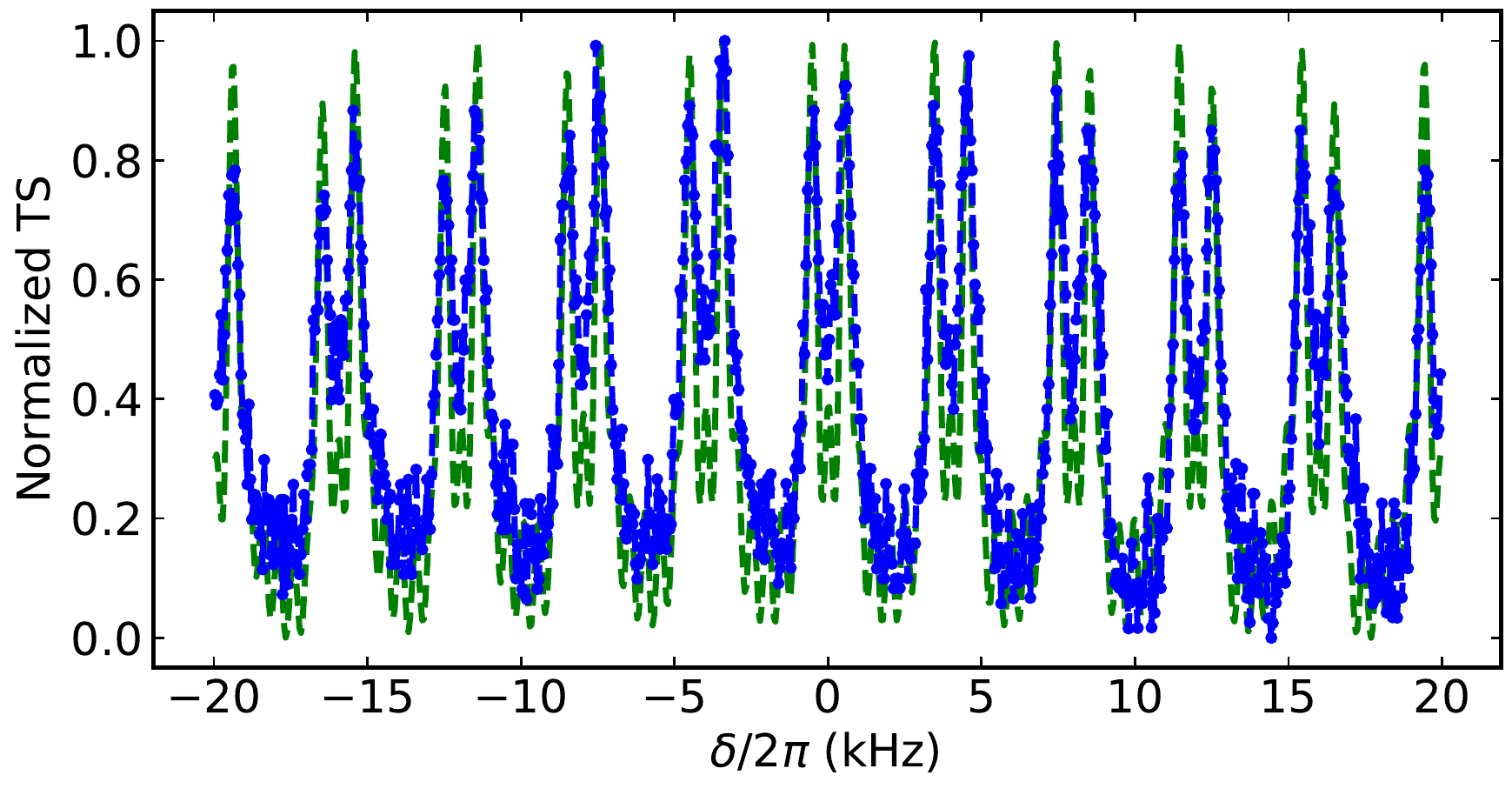}
	\caption{\textbf{Multi-pulse CPT-Ramsey spectrum in presence of a non-weak bias magnetic field.}  Each peak splits into two new ones. Green line is the analytical result of normalized $1-\rho_{55}$, and blue dash-dotted line is the experimental data. The parameters are chosen as $\Omega \myeq 1.6e6~\mathrm{s^{-1}}$, $\Delta T \myeq  0.25$ ms, $\uptau\myeq 2$ $ \mu $s, $B_z\myeq 0.2$ G and $ N \myeq$7.}
	\label{fp_rho55}
\end{figure}

The above analyses are performed under the condition of $|g_1+g_2|\mu_B Bz/(2\pi\hbar) \ll \mathrm{FWHM}$ and so that the assumption of $ \Delta_{14} \approx \Delta_{23} \approx \delta $ is valid.
Actually, due to the small difference in two involved Land\'{e} g factors ($g_1\myeq -0.5017$ and $g_2\myeq 0.4997$), there is a frequency shift $2|g_1+g_2|\mu_B B_z/(2\pi\hbar) \myeq5568 B_z $ Hz between the two transitions $\left\{|1\rangle \to |4\rangle, |2\rangle \to |3\rangle\right\}$ when a magnetic field $ B_z $ is applied.
For a weak magnetic field, it only broadens the linewidth.
But if the bias magnetic field is not too weak, the approximation $\Delta_{14} \approx \Delta_{23}$ becomes invalid, one has to use Eq.~\ref{rho55_fp} instead of Eq.~\ref{rho55_simplify}.
Therefore the two-photon resonances occur at $ \Delta_{14}\myeq 0$ or $ \Delta_{23}\myeq 0$ and each peak splits into two new ones.
In Fig.~\ref{fp_rho55}, we show that the analytical results (green line) are well consistent with the experimental signals (blue dash-dotted line).
On one hand, we can eliminate the effects of Land\'{e} g factors on the clock transition frequency by averaging the frequencies of two peaks.
On the other hand, the frequency difference between two peaks can be used to measure magnetic field without involving magneto-sensitive transitions.

\begin{figure}[!htb]
	\centering
	\includegraphics[width=1\linewidth]{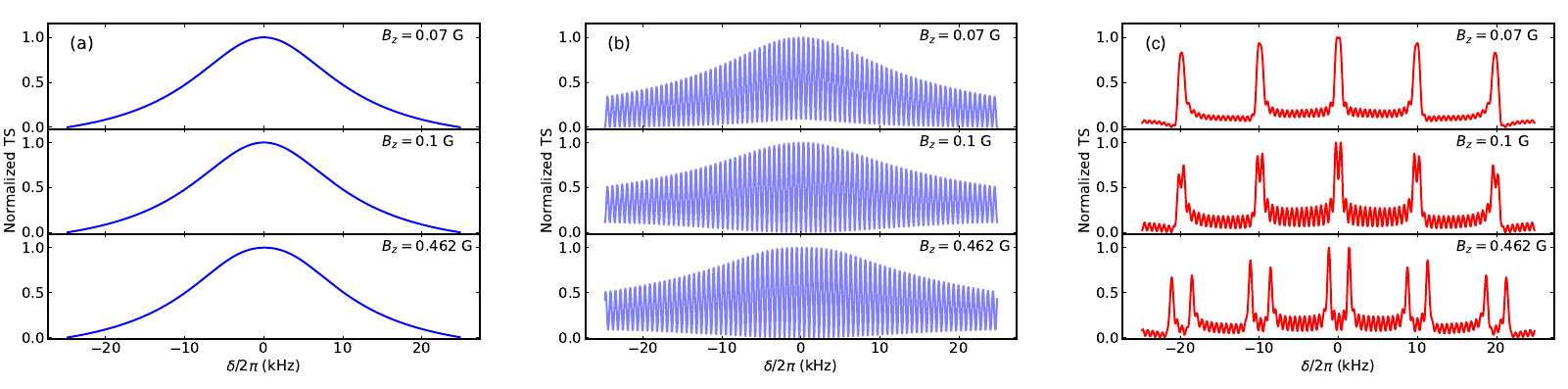}
	\caption{The spectra under different magnetic field. (a) CPT spectrum, (b) two-pulse Ramsey spectrum and (c) multi-pulse spectrum.}
	\label{cpt_split}
\end{figure}

Since the difference of nuclear g-factor is small, this kind of splitting is usually observed when magnetic field is large enough. For example, in the condition of $B_z=0.462$ G (the condition used in our experiment), this kind of splitting cannot be observed via the traditional CPT or two-pulse Ramsey interference.As shown in Fig.~\ref{cpt_split}~(a), the linewidth of the CPT spectrum is so broad that no splitting is caused. As shown in Fig.~\ref{cpt_split}~(b), since the Ramsey fringes are too dense, the two-pulse Ramsey interference also cannot resolve the splitting.

However, when applying the multi-pulse CPT sequence, the side peaks is suppressed compared with two-pulse Ramsey interference and the linewidth is decreased compared with one-pulse CPT. Therefore, even when the magnetic field is weak, we can directly observe the splitting caused by the nuclear g-factor, see Fig.~\ref{cpt_split}~(c).

\section{Numerical Simulation via Five-level Model}\label{simulation}
\subsection{Method of Simulation}

Generally, one has to numerically solve the optical Bloch equation~\ref{bloch_eq}.
And we set $\gamma_{12}\myeq \gamma_{13}\myeq \gamma_{24}\myeq \gamma_{34}\myeq \gamma_c$ for describing the dephasing induced by magnetic-field fluctuation.
%
%
In our simulation, we assume the population exchange rate from each ground state $\gamma_g\myeq 0$ and the Rabi frequencies are set according to Eq.~\ref{rabi_ratio}.

According to Eq.~\ref{rho55_simplify}, $(1-\rho_{55}) \mid _{\uptau \rightarrow \infty} \sim A_{1}f(\delta) + A_{2}$, where $A_{1}$ and $A_{2}$ are adjustable coefficients.
When the magnetic field $B_z$ is weak, the average Rabi frequency $\Omega$ can be obtained by fitting the central peak of experimental CPT spectrum using formula $A_{1}f(\delta) + A_{2}$.
%
%
Eq.~\ref{bloch_eq} can be constructed as
\begin{equation}
	\frac{\partial}{\partial t}\boldsymbol{V}(t) = \mathbf{M}(t)\boldsymbol{V}(t),
\end{equation}
where $\mathbf{M}$ is the time-dependent coefficient matrix, $\boldsymbol{V}$ corresponds to the density matrix $\rho$ being reshaped into one-dimensional vector.
In a short time step $t_1$, $\mathbf{M}$ can be regarded as a constant matrix, thus we have
\begin{equation}
	\boldsymbol{V}(t_0+t_1)=\exp(\mathbf{M} t_1)\boldsymbol{V}(t_0).
\end{equation}

\subsection{Comparison between Numerical and Experimental Results}\label{sim_exp}

\begin{figure}[!ht]
	\centering
	\includegraphics[width=0.58\linewidth]{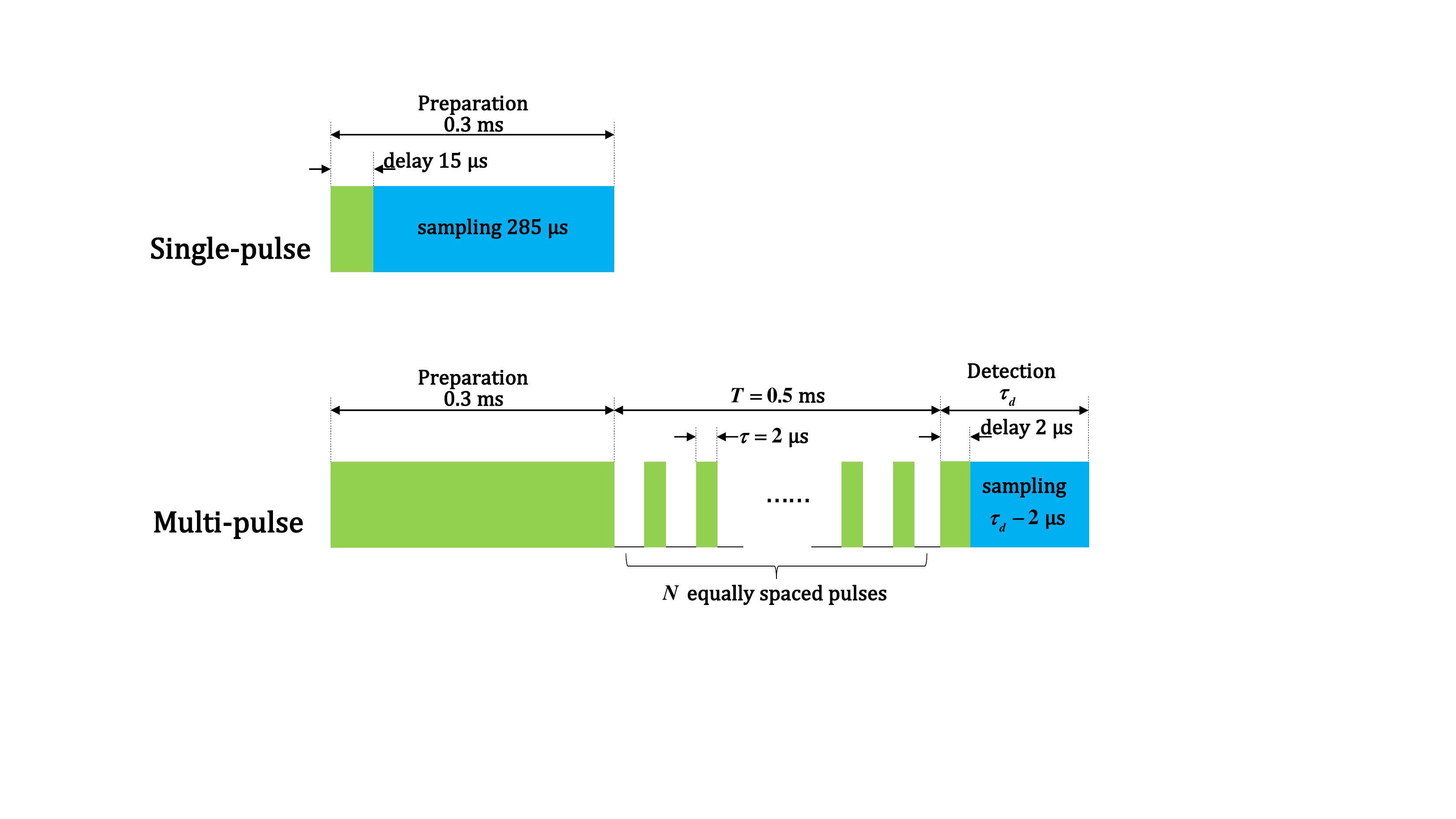}
	\caption{\textbf{Timeing sequence of single-pulse and multi-pulse CPT laser.}The preparation pulse is 0.3 ms, integration time $T\myeq0.5$ ms and the time of detection pulse $\uptau_d$. For single-pulse CPT, the signal is collected by sampling and averaging the rear $285~\mu $s of preparation pulse. For multi-pulse interference, the equally spaced N pulses are implemented between preparation pulse and detection pulse. The pulse length $ \uptau \myeq 2  \mu $s and the total integration time $T\myeq 0.5$ ms. When the detection pulse starts to be emitted. After a delay of 2 $\mu s$, the signal is obtained by sampling and averaging the remaining detection pulse. The sampling rate is 1 MHz.}
	\label{sequence}
\end{figure}

\begin{figure}[!ht]
	\centering
	\includegraphics[width=0.58\linewidth]{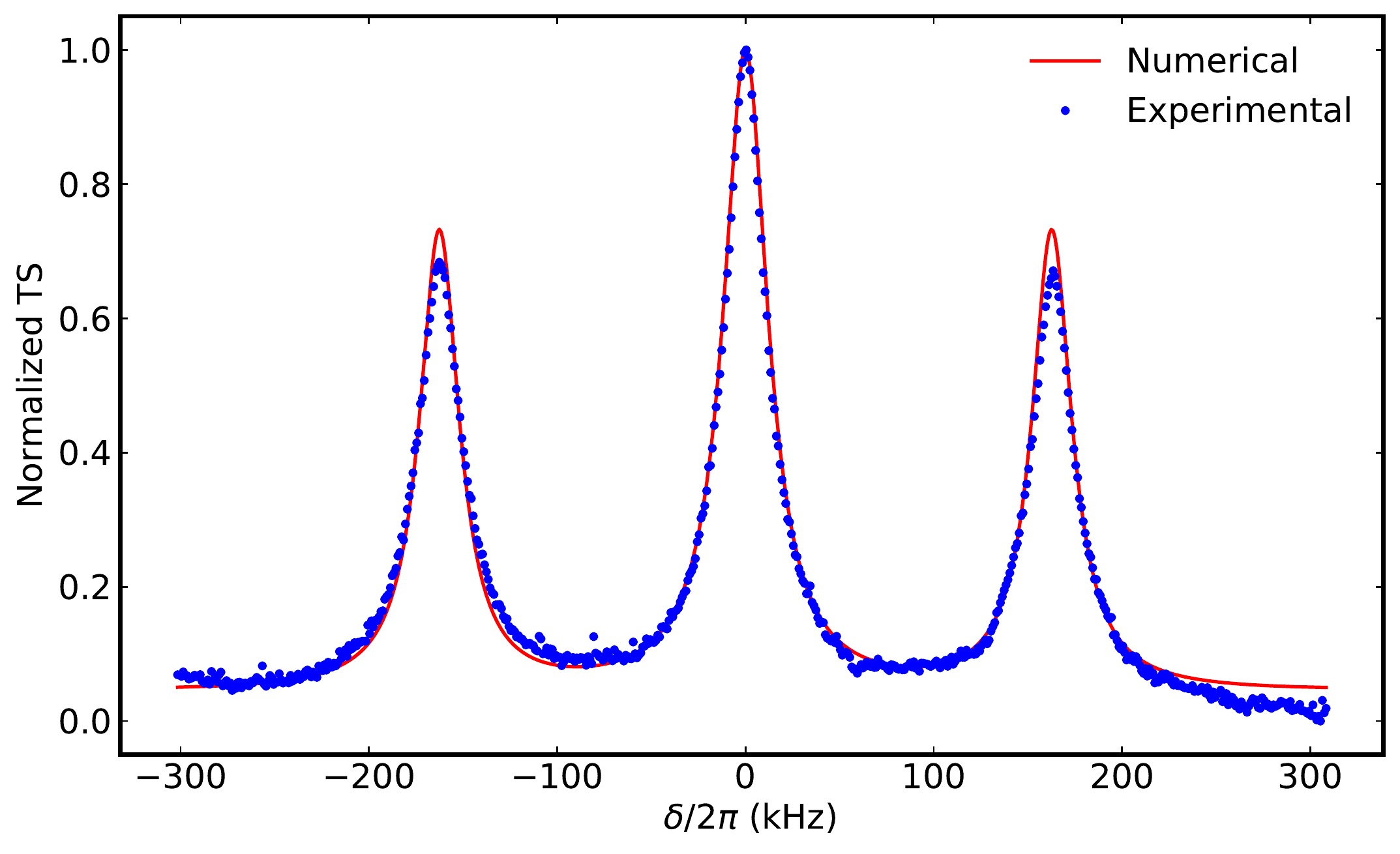}
	\caption{\textbf{The normalized single-pulse CPT spectrum of experimental signal (blue dots) and the numerical result (red line) of the normalized $1-\rho_{55}$}. The parameters used in numerical simulation are set as $\pm\Omega_{1,+ 1}^a\myeq \frac{1}{\sqrt{3}}\Omega_{2,\pm 1}^b\myeq 1.25e6~\mathrm{s^{-1}}$, $B_z\myeq 0.116$ G and $\gamma_c\myeq 1.2e4~\mathrm{s^{-1}}$. }
	\label{fitcpt}
\end{figure}

\begin{figure}[!t]
	\centering
	\includegraphics[width=0.6\linewidth]{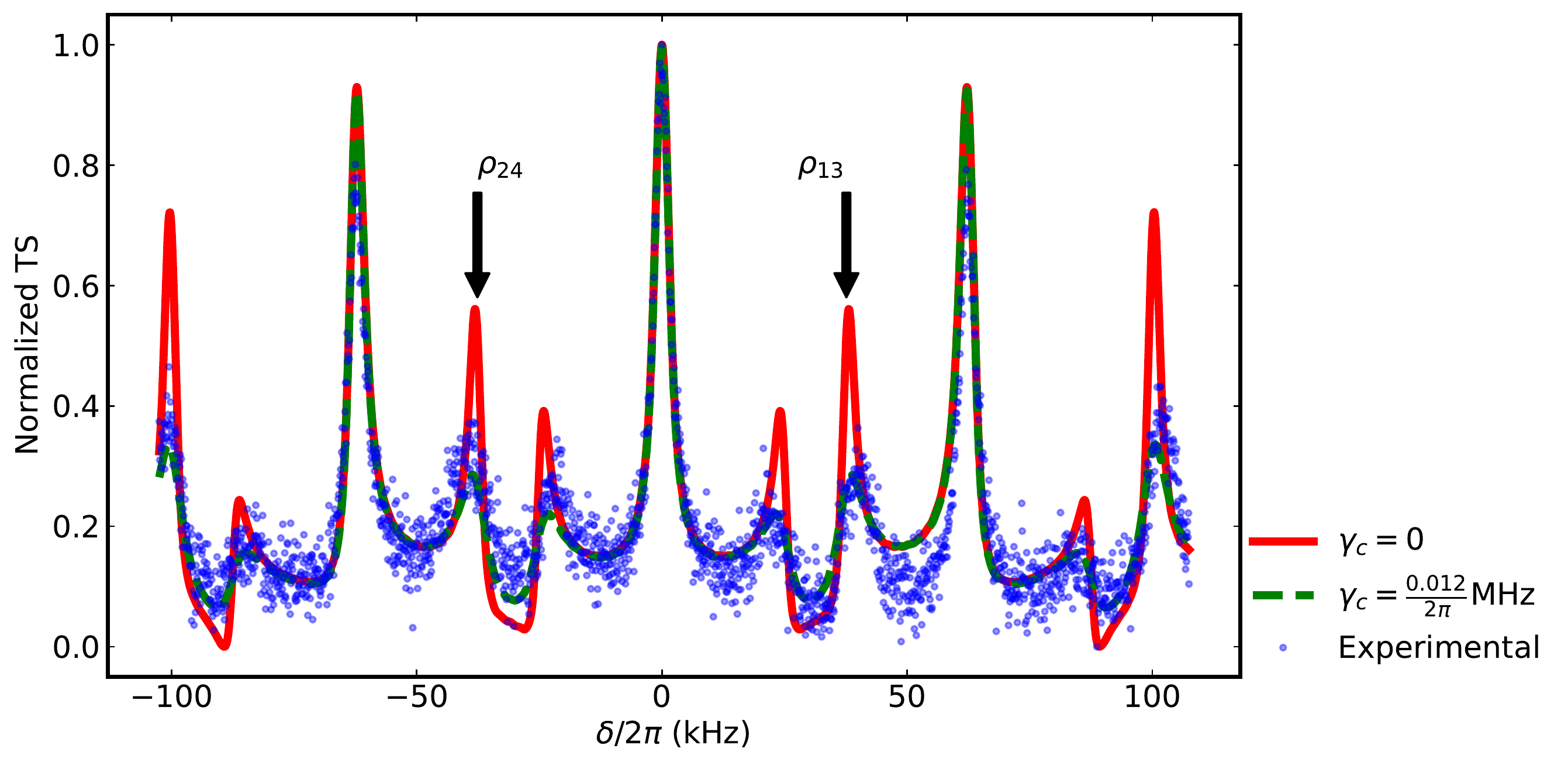}
	\caption{\textbf{Numerical result (green line for $\gamma_c\myeq 0$ and red line for $\gamma_c\myeq 1.2e4~\mathrm{s^{-1}}$) and experiment (blue dots).} The preparation time is 0.3 ms, total free-evolution time $T\myeq 0.5$ ms and 31 pulses with the length $\uptau \myeq$ 2 $ \mu $s is inserted. The length of detection pulse $\uptau_d\myeq 3\mu \text{s}$. The parameters used in numerical simulation are set as $\pm\Omega_{1,+ 1}^a\myeq \frac{1}{\sqrt{3}}\Omega_{2,\pm 1}^b\myeq 1.25e6~\mathrm{s^{-1}}$, $B_z\myeq 0.116$ G. The constructive peaks of $\rho_{13}$ and $\rho_{24}$ are annotated.}
	\label{discuss_gamma_c}
\end{figure}

In the experiment, the single-pulse CPT spectrum is collected by sampling and averaging the rear 285 $\mu$s of the preparation pulse of CPT laser, as shown in Fig.~\ref{sequence}.
%
%
%
Fig.~\ref{fitcpt} shows the numerical result is consistent with experimental CPT spectrum.

The multi-pulse interference are implemented by inserting corresponding CPT pulses between CPT-Ramsey pulses respectively called as preparation and detection.
%
%
Then one detection pulse starts to be emitted, and the length of the probe light is $\uptau_{d}$. The signal is obtained by sampling and averaging this pulse with $2~\mu\text{s}$ delay.
The sampling rate is 1 MHz of both single-pulse and multi-pulse spectrum.
%

%
%
To discuss the $\gamma_c$, we perform the multi-pulse experimental and numerical result.
%
As the Fig.~\ref{discuss_gamma_c} shown, by comparing the experimental and numerical result, we find they agree better when setting $\gamma_c\myeq 1.2e4~\mathrm{s^{-1}}$.

In order to obtain a better experimental signal, we generally set the detection pulse $\uptau_d\myeq 8~\mu\text{s}$ and average this pulse after a delay of $2~\mu\text{s}$. However, near the $\delta\myeq 0$ the $\rho_{13}$ and $\rho_{24}$ can be considered to be rapidly oscillating. These items will be averaged out if the detection time is long. Therefore the detection pulse is set to $\uptau_{d}\myeq 3~\mu\text{s}$ in order to observe the peaks of these terms.
\\
\\
\subsection{Comparison between Numerical and Analytical Results}
%
%
\begin{figure}[!th]
	\centering
	\includegraphics[width=0.8\linewidth]{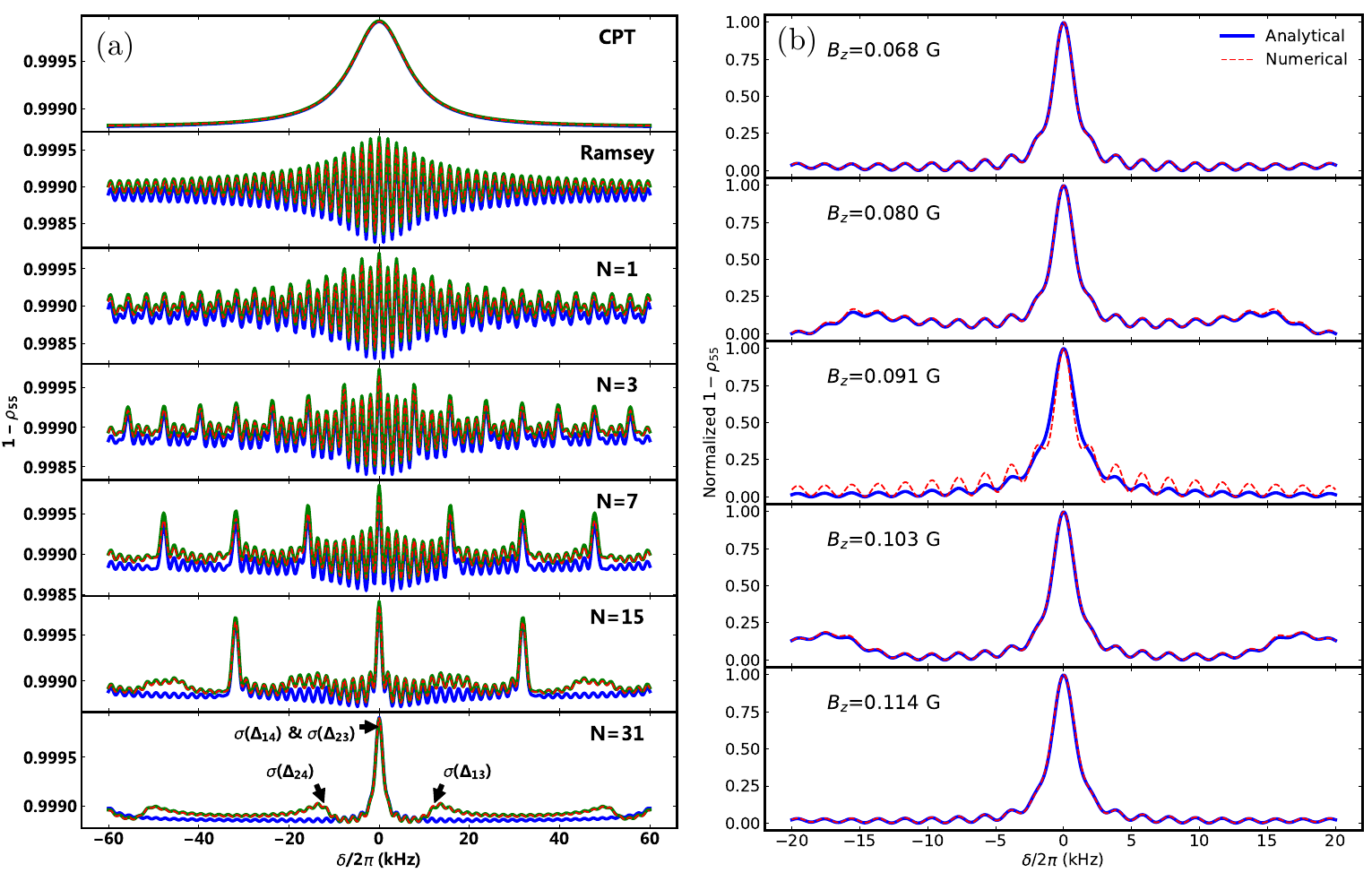}
	\caption{\textbf{Numerical (red dash) and analytical result (green line for Eq.~\ref{rho55_all_term} and blue line for Eq.~\ref{rho55_fp}).} The preparation time is 0.3 ms, total free-evolution time $T\myeq 0.5$ ms and $ N $ pulses with the length  $\uptau\myeq 2~\mu $s is inserted. Bias magnetic field is set as $B_z\myeq 0.1$ G. During detection pulse, the second sample of $1-\rho_{55}$ is collected as the signal. (a)The signals of different pulse number, when $N\myeq 31$, the peaks corresponding to the different components are annotated. (b)The signals of different magnetic field, when $B_z\myeq0.091$ G, the analytical result has a little difference with numerical result.}
	\label{them_theo}
\end{figure}
Both numerical and analytical parameters are set as $\Omega_{1,\pm 1}^{a}\myeq \Omega_{2,\pm 1}^{b}\myeq \Omega\myeq 1.25e6~\mathrm{s^{-1}}$, $\gamma_{14}\myeq \gamma_{23}\myeq 0$, $\Gamma_1\myeq \Gamma_2\myeq \frac{1}{2} \Gamma$, $\gamma_c\myeq 1.2e4~\mathrm{s^{-1}}$ and detection time $\uptau_d\myeq3~\mu \text{s}$.
The analytical result is referring to Eq.~\ref{rho55_all_term} and Eq.~\ref{rho55_fp}.
Fig.~\ref{them_theo}(a) shows that the numerical results agree with Eq.~\ref{rho55_all_term} as well as the Eq.~\ref{rho55_fp}, which means the neglect of $\rho_{12}$, $\rho_{34}$, $\rho_{13}$ and $\rho_{24}$ is feasible when $ \gamma_c $ is relatively large.

However, there are some critical cases, in which $\{|g_1-g_2|\mu_B/(2\pi\hbar)\myeq m \frac{1}{\Delta T} \mid m\in\mathbb{N}\}$, the constructive peaks of $\sigma(\Delta_{13})$ and $\sigma(\Delta_{24})$ will overlay with that of $\sigma(\Delta_{14})$ and $\sigma(\Delta_{23})$ at $\delta\myeq 0$.
In this specific situation, the approximation of Eq.~\ref{rho14s_2} is not so precise that analytical result has a slight difference with simulated result (the 3rd subplot) in Fig.~\ref{them_theo}(b).
%
%

\section{Numerical Simulation via Eleven-level Model}\label{simVSana}
\begin{figure}[!ht]
	\centering
	\includegraphics[width=0.6\linewidth]{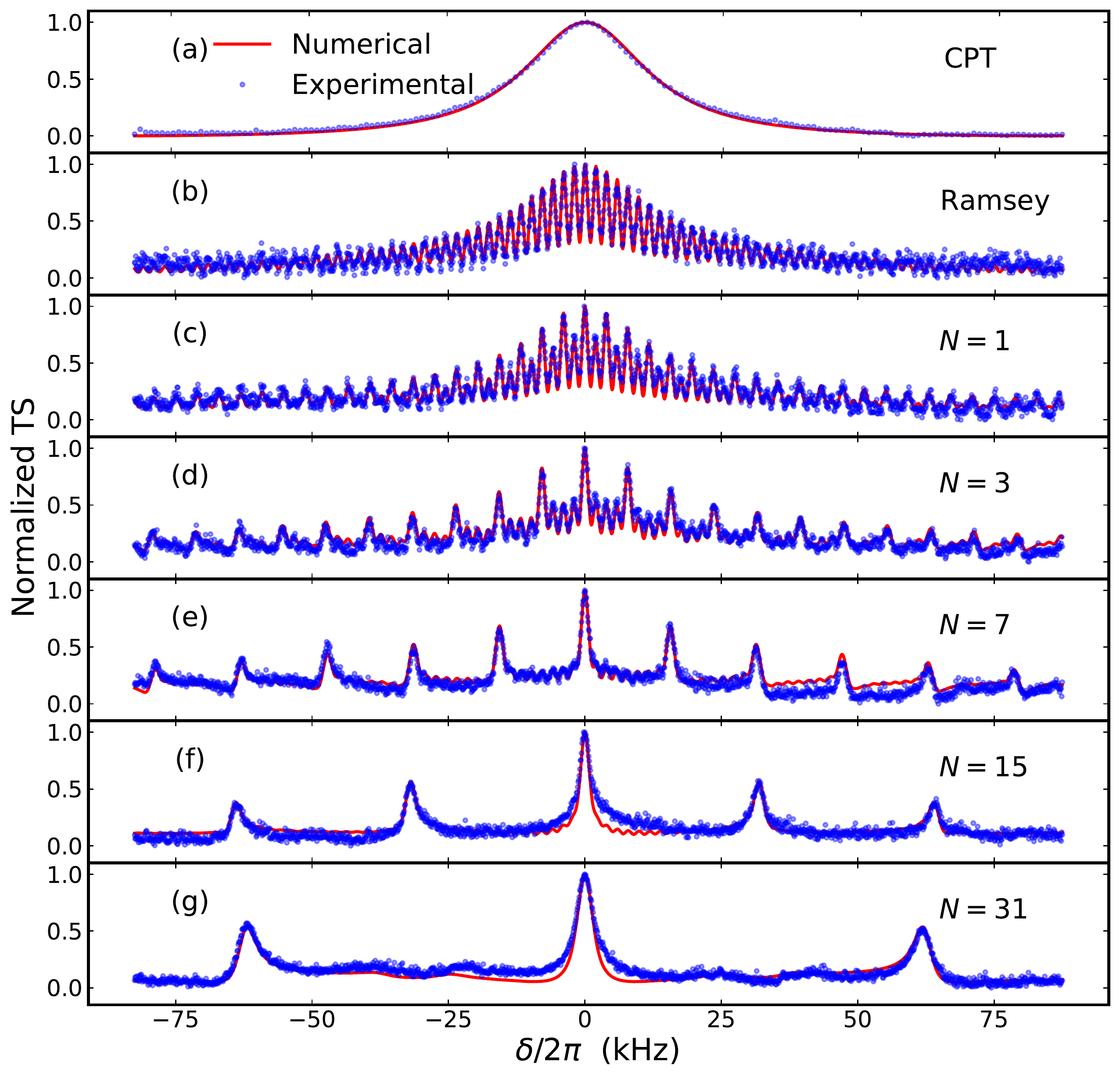}
	\caption{\textbf{Numerical (red line) and experimental (red dots) signals of eleven-level model in different conditions.} (a) Numerical and experimental result of single-pulse CPT spectrum. (b) Two-pulse CPT-Ramsey spectrum is obtained with a integration time of 0.5 ms. (c)-(g) Multi-pulse CPT-Ramsey spectra via temporal spinwave FP interferometry of $N$ equidistant pulses with a length $\uptau \myeq $2 $\mu$s into the integration time of 0.5 ms. }
	\label{unclose}
\end{figure}

As the Eq.~\ref{rho_src_elements_approx} shows, we can approximately consider the population sourced from five-level model to six-level model is equal to that source from six-level model to five-level model. However, the population exchange is a little unbalanced and population will flow into or outflow from five-level model, that cause the numerical contrasts deviate from experimental data slightly. The eleven level numerical simulation can fix the experimental well as the Fig.~\ref{unclose} shows.




%
%

\begin{thebibliography}{X}%
	\makeatletter
	\providecommand \@ifxundefined [1]{%
		\@ifx{#1\undefined}
	}%
	\providecommand \@ifnum [1]{%
		\ifnum #1\expandafter \@firstoftwo
		\else \expandafter \@secondoftwo
		\fi
	}%
	\providecommand \@ifx [1]{%
		\ifx #1\expandafter \@firstoftwo
		\else \expandafter \@secondoftwo
		\fi
	}%
	\providecommand \natexlab [1]{#1}%
	\providecommand \enquote  [1]{``#1''}%
	\providecommand \bibnamefont  [1]{#1}%
	\providecommand \bibfnamefont [1]{#1}%
	\providecommand \citenamefont [1]{#1}%
	\providecommand \href@noop [0]{\@secondoftwo}%
	\providecommand \href [0]{\begingroup \@sanitize@url \@href}%
	\providecommand \@href[1]{\@@startlink{#1}\@@href}%
	\providecommand \@@href[1]{\endgroup#1\@@endlink}%
	\providecommand \@sanitize@url [0]{\catcode `\\12\catcode `\$12\catcode
		`\&12\catcode `\#12\catcode `\^12\catcode `\_12\catcode `\%12\relax}%
	\providecommand \@@startlink[1]{}%
	\providecommand \@@endlink[0]{}%
	\providecommand \url  [0]{\begingroup\@sanitize@url \@url }%
	\providecommand \@url [1]{\endgroup\@href {#1}{\urlprefix }}%
	\providecommand \urlprefix  [0]{URL }%
	\providecommand \Eprint [0]{\href }%
	\providecommand \doibase [0]{http://dx.doi.org/}%
	\providecommand \selectlanguage [0]{\@gobble}%
	\providecommand \bibinfo  [0]{\@secondoftwo}%
	\providecommand \bibfield  [0]{\@secondoftwo}%
	\providecommand \translation [1]{[#1]}%
	\providecommand \BibitemOpen [0]{}%
	\providecommand \bibitemStop [0]{}%
	\providecommand \bibitemNoStop [0]{.\EOS\space}%
	\providecommand \EOS [0]{\spacefactor3000\relax}%
	\providecommand \BibitemShut  [1]{\csname bibitem#1\endcsname}%
	\let\auto@bib@innerbib\@empty
	\bibitem [{\citenamefont {Gray}\ \emph {et~al.}(1978)\citenamefont {Gray},
		\citenamefont {Whitley},\ and\ \citenamefont {Stroud}}]{Gray:78}%
	\BibitemOpen
	\bibfield  {author} {\bibinfo {author} {\bibfnamefont {H.~R.}\ \bibnamefont
			{Gray}}, \bibinfo {author} {\bibfnamefont {R.~M.}\ \bibnamefont {Whitley}}, \
		and\ \bibinfo {author} {\bibfnamefont {C.~R.}\ \bibnamefont {Stroud}},\
	}\href {\doibase 10.1364/OL.3.000218} {\bibfield  {journal} {\bibinfo
			{journal} {Opt. Lett.}\ }\textbf {\bibinfo {volume} {3}},\ \bibinfo {pages}
		{218} (\bibinfo {year} {1978})}\BibitemShut {NoStop}%
	\bibitem [{\citenamefont {Rogers}\ \emph {et~al.}(2014)\citenamefont {Rogers},
		\citenamefont {Jahnke}, \citenamefont {Metsch}, \citenamefont {Sipahigil},
		\citenamefont {Binder}, \citenamefont {Teraji}, \citenamefont {Sumiya},
		\citenamefont {Isoya}, \citenamefont {Lukin}, \citenamefont {Hemmer},\ and\
		\citenamefont {Jelezko}}]{PhysRevLett.113.263602}%
	\BibitemOpen
	\bibfield  {author} {\bibinfo {author} {\bibfnamefont {L.~J.}\ \bibnamefont
			{Rogers}}, \bibinfo {author} {\bibfnamefont {K.~D.}\ \bibnamefont {Jahnke}},
		\bibinfo {author} {\bibfnamefont {M.~H.}\ \bibnamefont {Metsch}}, \bibinfo
		{author} {\bibfnamefont {A.}~\bibnamefont {Sipahigil}}, \bibinfo {author}
		{\bibfnamefont {J.~M.}\ \bibnamefont {Binder}}, \bibinfo {author}
		{\bibfnamefont {T.}~\bibnamefont {Teraji}}, \bibinfo {author} {\bibfnamefont
			{H.}~\bibnamefont {Sumiya}}, \bibinfo {author} {\bibfnamefont
			{J.}~\bibnamefont {Isoya}}, \bibinfo {author} {\bibfnamefont {M.~D.}\
			\bibnamefont {Lukin}}, \bibinfo {author} {\bibfnamefont {P.}~\bibnamefont
			{Hemmer}}, \ and\ \bibinfo {author} {\bibfnamefont {F.}~\bibnamefont
			{Jelezko}},\ }\href {\doibase 10.1103/PhysRevLett.113.263602} {\bibfield
		{journal} {\bibinfo  {journal} {Phys. Rev. Lett.}\ }\textbf {\bibinfo
			{volume} {113}},\ \bibinfo {pages} {263602} (\bibinfo {year}
		{2014})}\BibitemShut {NoStop}%
	\bibitem [{\citenamefont {Das}\ \emph {et~al.}(2018)\citenamefont {Das},
		\citenamefont {Liu}, \citenamefont {Gr\'emaud},\ and\ \citenamefont
		{Mukherjee}}]{PhysRevA.97.033838}%
	\BibitemOpen
	\bibfield  {author} {\bibinfo {author} {\bibfnamefont {S.}~\bibnamefont
			{Das}}, \bibinfo {author} {\bibfnamefont {P.}~\bibnamefont {Liu}}, \bibinfo
		{author} {\bibfnamefont {B.}~\bibnamefont {Gr\'emaud}}, \ and\ \bibinfo
		{author} {\bibfnamefont {M.}~\bibnamefont {Mukherjee}},\ }\href {\doibase
		10.1103/PhysRevA.97.033838} {\bibfield  {journal} {\bibinfo  {journal} {Phys.
				Rev. A}\ }\textbf {\bibinfo {volume} {97}},\ \bibinfo {pages} {033838}
		(\bibinfo {year} {2018})}\BibitemShut {NoStop}%
	\bibitem [{\citenamefont {Xia}\ \emph {et~al.}(2015)\citenamefont {Xia},
		\citenamefont {Kolesov}, \citenamefont {Wang}, \citenamefont {Siyushev},
		\citenamefont {Reuter}, \citenamefont {Kornher}, \citenamefont {Kukharchyk},
		\citenamefont {Wieck}, \citenamefont {Villa}, \citenamefont {Yang},\ and\
		\citenamefont {Wrachtrup}}]{PhysRevLett.115.093602}%
	\BibitemOpen
	\bibfield  {author} {\bibinfo {author} {\bibfnamefont {K.}~\bibnamefont
			{Xia}}, \bibinfo {author} {\bibfnamefont {R.}~\bibnamefont {Kolesov}},
		\bibinfo {author} {\bibfnamefont {Y.}~\bibnamefont {Wang}}, \bibinfo {author}
		{\bibfnamefont {P.}~\bibnamefont {Siyushev}}, \bibinfo {author}
		{\bibfnamefont {R.}~\bibnamefont {Reuter}}, \bibinfo {author} {\bibfnamefont
			{T.}~\bibnamefont {Kornher}}, \bibinfo {author} {\bibfnamefont
			{N.}~\bibnamefont {Kukharchyk}}, \bibinfo {author} {\bibfnamefont {A.~D.}\
			\bibnamefont {Wieck}}, \bibinfo {author} {\bibfnamefont {B.}~\bibnamefont
			{Villa}}, \bibinfo {author} {\bibfnamefont {S.}~\bibnamefont {Yang}}, \ and\
		\bibinfo {author} {\bibfnamefont {J.}~\bibnamefont {Wrachtrup}},\ }\href
	{\doibase 10.1103/PhysRevLett.115.093602} {\bibfield  {journal} {\bibinfo
			{journal} {Phys. Rev. Lett.}\ }\textbf {\bibinfo {volume} {115}},\ \bibinfo
		{pages} {093602} (\bibinfo {year} {2015})}\BibitemShut {NoStop}%
	\bibitem [{\citenamefont {Santori}\ \emph {et~al.}(2006)\citenamefont
		{Santori}, \citenamefont {Tamarat}, \citenamefont {Neumann}, \citenamefont
		{Wrachtrup}, \citenamefont {Fattal}, \citenamefont {Beausoleil},
		\citenamefont {Rabeau}, \citenamefont {Olivero}, \citenamefont {Greentree},
		\citenamefont {Prawer}, \citenamefont {Jelezko},\ and\ \citenamefont
		{Hemmer}}]{PhysRevLett.97.247401}%
	\BibitemOpen
	\bibfield  {author} {\bibinfo {author} {\bibfnamefont {C.}~\bibnamefont
			{Santori}}, \bibinfo {author} {\bibfnamefont {P.}~\bibnamefont {Tamarat}},
		\bibinfo {author} {\bibfnamefont {P.}~\bibnamefont {Neumann}}, \bibinfo
		{author} {\bibfnamefont {J.}~\bibnamefont {Wrachtrup}}, \bibinfo {author}
		{\bibfnamefont {D.}~\bibnamefont {Fattal}}, \bibinfo {author} {\bibfnamefont
			{R.~G.}\ \bibnamefont {Beausoleil}}, \bibinfo {author} {\bibfnamefont
			{J.}~\bibnamefont {Rabeau}}, \bibinfo {author} {\bibfnamefont
			{P.}~\bibnamefont {Olivero}}, \bibinfo {author} {\bibfnamefont {A.~D.}\
			\bibnamefont {Greentree}}, \bibinfo {author} {\bibfnamefont {S.}~\bibnamefont
			{Prawer}}, \bibinfo {author} {\bibfnamefont {F.}~\bibnamefont {Jelezko}}, \
		and\ \bibinfo {author} {\bibfnamefont {P.}~\bibnamefont {Hemmer}},\ }\href
	{\doibase 10.1103/PhysRevLett.97.247401} {\bibfield  {journal} {\bibinfo
			{journal} {Phys. Rev. Lett.}\ }\textbf {\bibinfo {volume} {97}},\ \bibinfo
		{pages} {247401} (\bibinfo {year} {2006})}\BibitemShut {NoStop}%
	\bibitem [{\citenamefont {Jamonneau}\ \emph {et~al.}(2016)\citenamefont
		{Jamonneau}, \citenamefont {H\'etet}, \citenamefont {Dr\'eau}, \citenamefont
		{Roch},\ and\ \citenamefont {Jacques}}]{PhysRevLett.116.043603}%
	\BibitemOpen
	\bibfield  {author} {\bibinfo {author} {\bibfnamefont {P.}~\bibnamefont
			{Jamonneau}}, \bibinfo {author} {\bibfnamefont {G.}~\bibnamefont {H\'etet}},
		\bibinfo {author} {\bibfnamefont {A.}~\bibnamefont {Dr\'eau}}, \bibinfo
		{author} {\bibfnamefont {J.-F.}\ \bibnamefont {Roch}}, \ and\ \bibinfo
		{author} {\bibfnamefont {V.}~\bibnamefont {Jacques}},\ }\href {\doibase
		10.1103/PhysRevLett.116.043603} {\bibfield  {journal} {\bibinfo  {journal}
			{Phys. Rev. Lett.}\ }\textbf {\bibinfo {volume} {116}},\ \bibinfo {pages}
		{043603} (\bibinfo {year} {2016})}\BibitemShut {NoStop}%
	\bibitem [{\citenamefont {Ni}\ \emph {et~al.}(2008)\citenamefont {Ni},
		\citenamefont {Ospelkaus}, \citenamefont {de~Miranda}, \citenamefont
		{Pe{\textquoteright}er}, \citenamefont {Neyenhuis}, \citenamefont {Zirbel},
		\citenamefont {Kotochigova}, \citenamefont {Julienne}, \citenamefont {Jin},\
		and\ \citenamefont {Ye}}]{Ni231}%
	\BibitemOpen
	\bibfield  {author} {\bibinfo {author} {\bibfnamefont {K.-K.}\ \bibnamefont
			{Ni}}, \bibinfo {author} {\bibfnamefont {S.}~\bibnamefont {Ospelkaus}},
		\bibinfo {author} {\bibfnamefont {M.~H.~G.}\ \bibnamefont {de~Miranda}},
		\bibinfo {author} {\bibfnamefont {A.}~\bibnamefont {Pe{\textquoteright}er}},
		\bibinfo {author} {\bibfnamefont {B.}~\bibnamefont {Neyenhuis}}, \bibinfo
		{author} {\bibfnamefont {J.~J.}\ \bibnamefont {Zirbel}}, \bibinfo {author}
		{\bibfnamefont {S.}~\bibnamefont {Kotochigova}}, \bibinfo {author}
		{\bibfnamefont {P.~S.}\ \bibnamefont {Julienne}}, \bibinfo {author}
		{\bibfnamefont {D.~S.}\ \bibnamefont {Jin}}, \ and\ \bibinfo {author}
		{\bibfnamefont {J.}~\bibnamefont {Ye}},\ }\href {\doibase
		10.1126/science.1163861} {\bibfield  {journal} {\bibinfo  {journal}
			{Science}\ }\textbf {\bibinfo {volume} {322}},\ \bibinfo {pages} {231}
		(\bibinfo {year} {2008})}\BibitemShut {NoStop}%
	\bibitem [{\citenamefont {Aspect}\ \emph {et~al.}(1988)\citenamefont {Aspect},
		\citenamefont {Arimondo}, \citenamefont {Kaiser}, \citenamefont
		{Vansteenkiste},\ and\ \citenamefont {Cohen-Tannoudji}}]{PhysRevLett.61.826}%
	\BibitemOpen
	\bibfield  {author} {\bibinfo {author} {\bibfnamefont {A.}~\bibnamefont
			{Aspect}}, \bibinfo {author} {\bibfnamefont {E.}~\bibnamefont {Arimondo}},
		\bibinfo {author} {\bibfnamefont {R.}~\bibnamefont {Kaiser}}, \bibinfo
		{author} {\bibfnamefont {N.}~\bibnamefont {Vansteenkiste}}, \ and\ \bibinfo
		{author} {\bibfnamefont {C.}~\bibnamefont {Cohen-Tannoudji}},\ }\href
	{\doibase 10.1103/PhysRevLett.61.826} {\bibfield  {journal} {\bibinfo
			{journal} {Phys. Rev. Lett.}\ }\textbf {\bibinfo {volume} {61}},\ \bibinfo
		{pages} {826} (\bibinfo {year} {1988})}\BibitemShut {NoStop}%
	\bibitem [{\citenamefont {Vanier}(2005)}]{Vanier2005}%
	\BibitemOpen
	\bibfield  {author} {\bibinfo {author} {\bibfnamefont {J.}~\bibnamefont
			{Vanier}},\ }\href {\doibase 10.1007/s00340-005-1905-3} {\bibfield  {journal}
		{\bibinfo  {journal} {Appl. Phys. B}\ }\textbf {\bibinfo {volume} {81}},\
		\bibinfo {pages} {421} (\bibinfo {year} {2005})}\BibitemShut {NoStop}%
	\bibitem [{\citenamefont {Merimaa}\ \emph {et~al.}(2003)\citenamefont
		{Merimaa}, \citenamefont {Lindvall}, \citenamefont {Tittonen},\ and\
		\citenamefont {Ikonen}}]{Merimaa:03}%
	\BibitemOpen
	\bibfield  {author} {\bibinfo {author} {\bibfnamefont {M.}~\bibnamefont
			{Merimaa}}, \bibinfo {author} {\bibfnamefont {T.}~\bibnamefont {Lindvall}},
		\bibinfo {author} {\bibfnamefont {I.}~\bibnamefont {Tittonen}}, \ and\
		\bibinfo {author} {\bibfnamefont {E.}~\bibnamefont {Ikonen}},\ }\href
	{\doibase 10.1364/JOSAB.20.000273} {\bibfield  {journal} {\bibinfo  {journal}
			{J. Opt. Soc. Am. B}\ }\textbf {\bibinfo {volume} {20}},\ \bibinfo {pages}
		{273} (\bibinfo {year} {2003})}\BibitemShut {NoStop}%
	\bibitem [{\citenamefont {Yun}\ \emph {et~al.}(2017)\citenamefont {Yun},
		\citenamefont {Tricot}, \citenamefont {Calosso}, \citenamefont {Micalizio},
		\citenamefont {Fran\ifmmode~\mbox{\c{c}}\else \c{c}\fi{}ois}, \citenamefont
		{Boudot}, \citenamefont {Gu\'erandel},\ and\ \citenamefont
		{de~Clercq}}]{PhysRevApplied.7.014018}%
	\BibitemOpen
	\bibfield  {author} {\bibinfo {author} {\bibfnamefont {P.}~\bibnamefont
			{Yun}}, \bibinfo {author} {\bibfnamefont {F.}~\bibnamefont {Tricot}},
		\bibinfo {author} {\bibfnamefont {C.~E.}\ \bibnamefont {Calosso}}, \bibinfo
		{author} {\bibfnamefont {S.}~\bibnamefont {Micalizio}}, \bibinfo {author}
		{\bibfnamefont {B.}~\bibnamefont {Fran\ifmmode~\mbox{\c{c}}\else
				\c{c}\fi{}ois}}, \bibinfo {author} {\bibfnamefont {R.}~\bibnamefont
			{Boudot}}, \bibinfo {author} {\bibfnamefont {S.}~\bibnamefont {Gu\'erandel}},
		\ and\ \bibinfo {author} {\bibfnamefont {E.}~\bibnamefont {de~Clercq}},\
	}\href {\doibase 10.1103/PhysRevApplied.7.014018} {\bibfield  {journal}
		{\bibinfo  {journal} {Phys. Rev. Applied}\ }\textbf {\bibinfo {volume} {7}},\
		\bibinfo {pages} {014018} (\bibinfo {year} {2017})}\BibitemShut {NoStop}%
	\bibitem [{\citenamefont {Liu}\ \emph {et~al.}(2017{\natexlab{a}})\citenamefont
		{Liu}, \citenamefont {Ivanov}, \citenamefont {Yudin}, \citenamefont
		{Kitching},\ and\ \citenamefont {Donley}}]{PhysRevApplied.8.054001}%
	\BibitemOpen
	\bibfield  {author} {\bibinfo {author} {\bibfnamefont {X.}~\bibnamefont
			{Liu}}, \bibinfo {author} {\bibfnamefont {E.}~\bibnamefont {Ivanov}},
		\bibinfo {author} {\bibfnamefont {V.~I.}\ \bibnamefont {Yudin}}, \bibinfo
		{author} {\bibfnamefont {J.}~\bibnamefont {Kitching}}, \ and\ \bibinfo
		{author} {\bibfnamefont {E.~A.}\ \bibnamefont {Donley}},\ }\href {\doibase
		10.1103/PhysRevApplied.8.054001} {\bibfield  {journal} {\bibinfo  {journal}
			{Phys. Rev. Applied}\ }\textbf {\bibinfo {volume} {8}},\ \bibinfo {pages}
		{054001} (\bibinfo {year} {2017}{\natexlab{a}})}\BibitemShut {NoStop}%
	\bibitem [{\citenamefont {Scully}\ and\ \citenamefont
		{Fleischhauer}(1992)}]{PhysRevLett.69.1360}%
	\BibitemOpen
	\bibfield  {author} {\bibinfo {author} {\bibfnamefont {M.~O.}\ \bibnamefont
			{Scully}}\ and\ \bibinfo {author} {\bibfnamefont {M.}~\bibnamefont
			{Fleischhauer}},\ }\href {\doibase 10.1103/PhysRevLett.69.1360} {\bibfield
		{journal} {\bibinfo  {journal} {Phys. Rev. Lett.}\ }\textbf {\bibinfo
			{volume} {69}},\ \bibinfo {pages} {1360} (\bibinfo {year}
		{1992})}\BibitemShut {NoStop}%
	\bibitem [{\citenamefont {Nagel}\ \emph {et~al.}(1998)\citenamefont {Nagel},
		\citenamefont {Graf}, \citenamefont {Naumov}, \citenamefont {Mariotti},
		\citenamefont {Biancalana}, \citenamefont {Meschede},\ and\ \citenamefont
		{Wynands}}]{Nagel_1998}%
	\BibitemOpen
	\bibfield  {author} {\bibinfo {author} {\bibfnamefont {A.}~\bibnamefont
			{Nagel}}, \bibinfo {author} {\bibfnamefont {L.}~\bibnamefont {Graf}},
		\bibinfo {author} {\bibfnamefont {A.}~\bibnamefont {Naumov}}, \bibinfo
		{author} {\bibfnamefont {E.}~\bibnamefont {Mariotti}}, \bibinfo {author}
		{\bibfnamefont {V.}~\bibnamefont {Biancalana}}, \bibinfo {author}
		{\bibfnamefont {D.}~\bibnamefont {Meschede}}, \ and\ \bibinfo {author}
		{\bibfnamefont {R.}~\bibnamefont {Wynands}},\ }\href {\doibase
		10.1209/epl/i1998-00430-0} {\bibfield  {journal} {\bibinfo  {journal}
			{Europhys. Lett.}\ }\textbf {\bibinfo {volume} {44}},\ \bibinfo {pages} {31}
		(\bibinfo {year} {1998})}\BibitemShut {NoStop}%
	\bibitem [{\citenamefont {Schwindt}\ \emph {et~al.}(2004)\citenamefont
		{Schwindt}, \citenamefont {Knappe}, \citenamefont {Shah}, \citenamefont
		{Hollberg}, \citenamefont {Kitching}, \citenamefont {Liew},\ and\
		\citenamefont {Moreland}}]{doi:10.1063/1.1839274}%
	\BibitemOpen
	\bibfield  {author} {\bibinfo {author} {\bibfnamefont {P.~D.~D.}\
			\bibnamefont {Schwindt}}, \bibinfo {author} {\bibfnamefont {S.}~\bibnamefont
			{Knappe}}, \bibinfo {author} {\bibfnamefont {V.}~\bibnamefont {Shah}},
		\bibinfo {author} {\bibfnamefont {L.}~\bibnamefont {Hollberg}}, \bibinfo
		{author} {\bibfnamefont {J.}~\bibnamefont {Kitching}}, \bibinfo {author}
		{\bibfnamefont {L.-A.}\ \bibnamefont {Liew}}, \ and\ \bibinfo {author}
		{\bibfnamefont {J.}~\bibnamefont {Moreland}},\ }\href {\doibase
		10.1063/1.1839274} {\bibfield  {journal} {\bibinfo  {journal} {Appl. Phys.
				Lett.}\ }\textbf {\bibinfo {volume} {85}},\ \bibinfo {pages} {6409} (\bibinfo
		{year} {2004})}\BibitemShut {NoStop}%
	\bibitem [{\citenamefont {Tripathi}\ and\ \citenamefont
		{Pati}(2019)}]{Tripathi2019}%
	\BibitemOpen
	\bibfield  {author} {\bibinfo {author} {\bibfnamefont {R.}~\bibnamefont
			{Tripathi}}\ and\ \bibinfo {author} {\bibfnamefont {G.~S.}\ \bibnamefont
			{Pati}},\ }\href {\doibase 10.1109/jphot.2019.2922831} {\bibfield  {journal}
		{\bibinfo  {journal} {{IEEE} Photonics Journal}\ }\textbf {\bibinfo {volume}
			{11}},\ \bibinfo {pages} {1} (\bibinfo {year} {2019})}\BibitemShut {NoStop}%
	\bibitem [{\citenamefont {Zanon}\ \emph {et~al.}(2005)\citenamefont {Zanon},
		\citenamefont {Guerandel}, \citenamefont {de~Clercq}, \citenamefont
		{Holleville}, \citenamefont {Dimarcq},\ and\ \citenamefont
		{Clairon}}]{PhysRevLett.94.193002}%
	\BibitemOpen
	\bibfield  {author} {\bibinfo {author} {\bibfnamefont {T.}~\bibnamefont
			{Zanon}}, \bibinfo {author} {\bibfnamefont {S.}~\bibnamefont {Guerandel}},
		\bibinfo {author} {\bibfnamefont {E.}~\bibnamefont {de~Clercq}}, \bibinfo
		{author} {\bibfnamefont {D.}~\bibnamefont {Holleville}}, \bibinfo {author}
		{\bibfnamefont {N.}~\bibnamefont {Dimarcq}}, \ and\ \bibinfo {author}
		{\bibfnamefont {A.}~\bibnamefont {Clairon}},\ }\href {\doibase
		10.1103/PhysRevLett.94.193002} {\bibfield  {journal} {\bibinfo  {journal}
			{Phys. Rev. Lett.}\ }\textbf {\bibinfo {volume} {94}},\ \bibinfo {pages}
		{193002} (\bibinfo {year} {2005})}\BibitemShut {NoStop}%
	\bibitem [{\citenamefont {Vanier}\ \emph {et~al.}(2003)\citenamefont {Vanier},
		\citenamefont {Levine}, \citenamefont {Janssen},\ and\ \citenamefont
		{Delaney}}]{PhysRevA.67.065801}%
	\BibitemOpen
	\bibfield  {author} {\bibinfo {author} {\bibfnamefont {J.}~\bibnamefont
			{Vanier}}, \bibinfo {author} {\bibfnamefont {M.~W.}\ \bibnamefont {Levine}},
		\bibinfo {author} {\bibfnamefont {D.}~\bibnamefont {Janssen}}, \ and\
		\bibinfo {author} {\bibfnamefont {M.}~\bibnamefont {Delaney}},\ }\href
	{\doibase 10.1103/PhysRevA.67.065801} {\bibfield  {journal} {\bibinfo
			{journal} {Phys. Rev. A}\ }\textbf {\bibinfo {volume} {67}},\ \bibinfo
		{pages} {065801} (\bibinfo {year} {2003})}\BibitemShut {NoStop}%
	\bibitem [{\citenamefont {Liu}\ \emph {et~al.}(2017{\natexlab{b}})\citenamefont
		{Liu}, \citenamefont {Yudin}, \citenamefont {Taichenachev}, \citenamefont
		{Kitching},\ and\ \citenamefont {Donley}}]{doi:10.1063/1.5001179}%
	\BibitemOpen
	\bibfield  {author} {\bibinfo {author} {\bibfnamefont {X.}~\bibnamefont
			{Liu}}, \bibinfo {author} {\bibfnamefont {V.~I.}\ \bibnamefont {Yudin}},
		\bibinfo {author} {\bibfnamefont {A.~V.}\ \bibnamefont {Taichenachev}},
		\bibinfo {author} {\bibfnamefont {J.}~\bibnamefont {Kitching}}, \ and\
		\bibinfo {author} {\bibfnamefont {E.~A.}\ \bibnamefont {Donley}},\ }\href
	{\doibase 10.1063/1.5001179} {\bibfield  {journal} {\bibinfo  {journal}
			{Appl. Phys. Lett.}\ }\textbf {\bibinfo {volume} {111}},\ \bibinfo {pages}
		{224102} (\bibinfo {year} {2017}{\natexlab{b}})}\BibitemShut {NoStop}%
	\bibitem [{\citenamefont {Warren}\ \emph {et~al.}(2018)\citenamefont {Warren},
		\citenamefont {Shahriar}, \citenamefont {Tripathi},\ and\ \citenamefont
		{Pati}}]{Warren2018}%
	\BibitemOpen
	\bibfield  {author} {\bibinfo {author} {\bibfnamefont {Z.}~\bibnamefont
			{Warren}}, \bibinfo {author} {\bibfnamefont {M.~S.}\ \bibnamefont
			{Shahriar}}, \bibinfo {author} {\bibfnamefont {R.}~\bibnamefont {Tripathi}},
		\ and\ \bibinfo {author} {\bibfnamefont {G.~S.}\ \bibnamefont {Pati}},\
	}\href {\doibase 10.1063/1.5008402} {\bibfield  {journal} {\bibinfo
			{journal} {J. Appl. Phys.}\ }\textbf {\bibinfo {volume} {123}},\ \bibinfo
		{pages} {053101} (\bibinfo {year} {2018})}\BibitemShut {NoStop}%
	\bibitem [{\citenamefont {{Guerandel}}\ \emph {et~al.}(2007)\citenamefont
		{{Guerandel}}, \citenamefont {{Zanon}}, \citenamefont {{Castagna}},
		\citenamefont {{Dahes}}, \citenamefont {{de Clercq}}, \citenamefont
		{{Dimarcq}},\ and\ \citenamefont {{Clairon}}}]{4126869}%
	\BibitemOpen
	\bibfield  {author} {\bibinfo {author} {\bibfnamefont {S.}~\bibnamefont
			{{Guerandel}}}, \bibinfo {author} {\bibfnamefont {T.}~\bibnamefont
			{{Zanon}}}, \bibinfo {author} {\bibfnamefont {N.}~\bibnamefont {{Castagna}}},
		\bibinfo {author} {\bibfnamefont {F.}~\bibnamefont {{Dahes}}}, \bibinfo
		{author} {\bibfnamefont {E.}~\bibnamefont {{de Clercq}}}, \bibinfo {author}
		{\bibfnamefont {N.}~\bibnamefont {{Dimarcq}}}, \ and\ \bibinfo {author}
		{\bibfnamefont {A.}~\bibnamefont {{Clairon}}},\ }\href@noop {} {\bibfield
		{journal} {\bibinfo  {journal} {IEEE Trans. Instrum. Meas.}\ }\textbf
		{\bibinfo {volume} {56}},\ \bibinfo {pages} {383} (\bibinfo {year}
		{2007})}\BibitemShut {NoStop}%
	\bibitem [{\citenamefont {Yun}\ \emph {et~al.}(2012)\citenamefont {Yun},
		\citenamefont {Zhang}, \citenamefont {Liu}, \citenamefont {Deng},
		\citenamefont {You},\ and\ \citenamefont {Gu}}]{Yun_2012}%
	\BibitemOpen
	\bibfield  {author} {\bibinfo {author} {\bibfnamefont {P.}~\bibnamefont
			{Yun}}, \bibinfo {author} {\bibfnamefont {Y.}~\bibnamefont {Zhang}}, \bibinfo
		{author} {\bibfnamefont {G.}~\bibnamefont {Liu}}, \bibinfo {author}
		{\bibfnamefont {W.}~\bibnamefont {Deng}}, \bibinfo {author} {\bibfnamefont
			{L.}~\bibnamefont {You}}, \ and\ \bibinfo {author} {\bibfnamefont
			{S.}~\bibnamefont {Gu}},\ }\href {\doibase 10.1209/0295-5075/97/63004}
	{\bibfield  {journal} {\bibinfo  {journal} {Europhys. Lett.}\ }\textbf
		{\bibinfo {volume} {97}},\ \bibinfo {pages} {63004} (\bibinfo {year}
		{2012})}\BibitemShut {NoStop}%
	\bibitem [{\citenamefont {Yang}\ \emph {et~al.}(2018)\citenamefont {Yang},
		\citenamefont {Rui}, \citenamefont {Dai}, \citenamefont {Jin}, \citenamefont
		{Chen},\ and\ \citenamefont {Pan}}]{PhysRevA.98.033802}%
	\BibitemOpen
	\bibfield  {author} {\bibinfo {author} {\bibfnamefont {S.-J.}\ \bibnamefont
			{Yang}}, \bibinfo {author} {\bibfnamefont {J.}~\bibnamefont {Rui}}, \bibinfo
		{author} {\bibfnamefont {H.-N.}\ \bibnamefont {Dai}}, \bibinfo {author}
		{\bibfnamefont {X.-M.}\ \bibnamefont {Jin}}, \bibinfo {author} {\bibfnamefont
			{S.}~\bibnamefont {Chen}}, \ and\ \bibinfo {author} {\bibfnamefont {J.-W.}\
			\bibnamefont {Pan}},\ }\href {\doibase 10.1103/PhysRevA.98.033802} {\bibfield
		{journal} {\bibinfo  {journal} {Phys. Rev. A}\ }\textbf {\bibinfo {volume}
			{98}},\ \bibinfo {pages} {033802} (\bibinfo {year} {2018})}\BibitemShut
	{NoStop}%
	\bibitem [{\citenamefont {Nicolas}\ \emph {et~al.}(2018)\citenamefont
		{Nicolas}, \citenamefont {Delord}, \citenamefont {Jamonneau}, \citenamefont
		{Coto}, \citenamefont {Maze}, \citenamefont {Jacques},\ and\ \citenamefont
		{H{\'{e}}tet}}]{Nicolas2018}%
	\BibitemOpen
	\bibfield  {author} {\bibinfo {author} {\bibfnamefont {L.}~\bibnamefont
			{Nicolas}}, \bibinfo {author} {\bibfnamefont {T.}~\bibnamefont {Delord}},
		\bibinfo {author} {\bibfnamefont {P.}~\bibnamefont {Jamonneau}}, \bibinfo
		{author} {\bibfnamefont {R.}~\bibnamefont {Coto}}, \bibinfo {author}
		{\bibfnamefont {J.}~\bibnamefont {Maze}}, \bibinfo {author} {\bibfnamefont
			{V.}~\bibnamefont {Jacques}}, \ and\ \bibinfo {author} {\bibfnamefont
			{G.}~\bibnamefont {H{\'{e}}tet}},\ }\href {\doibase 10.1088/1367-2630/aab574}
	{\bibfield  {journal} {\bibinfo  {journal} {New J. Phys.}\ }\textbf {\bibinfo
			{volume} {20}},\ \bibinfo {pages} {033007} (\bibinfo {year}
		{2018})}\BibitemShut {NoStop}%
	\bibitem [{\citenamefont {Taichenachev}\ \emph {et~al.}(2005)\citenamefont
		{Taichenachev}, \citenamefont {Yudin}, \citenamefont {Velichansky},\ and\
		\citenamefont {Zibrov}}]{Taichenachev2005a}%
	\BibitemOpen
	\bibfield  {author} {\bibinfo {author} {\bibfnamefont {A.~V.}\ \bibnamefont
			{Taichenachev}}, \bibinfo {author} {\bibfnamefont {V.~I.}\ \bibnamefont
			{Yudin}}, \bibinfo {author} {\bibfnamefont {V.~L.}\ \bibnamefont
			{Velichansky}}, \ and\ \bibinfo {author} {\bibfnamefont {S.~A.}\ \bibnamefont
			{Zibrov}},\ }\href {\doibase 10.1134/1.2142864} {\bibfield  {journal}
		{\bibinfo  {journal} {Jetp Lett.}\ }\textbf {\bibinfo {volume} {82}},\
		\bibinfo {pages} {398} (\bibinfo {year} {2005})}\BibitemShut {NoStop}%
	\bibitem [{\citenamefont {Breschi}\ \emph {et~al.}(2009)\citenamefont
		{Breschi}, \citenamefont {Kazakov}, \citenamefont {Lammegger}, \citenamefont
		{Mileti}, \citenamefont {Matisov},\ and\ \citenamefont
		{Windholz}}]{PhysRevA.79.063837}%
	\BibitemOpen
	\bibfield  {author} {\bibinfo {author} {\bibfnamefont {E.}~\bibnamefont
			{Breschi}}, \bibinfo {author} {\bibfnamefont {G.}~\bibnamefont {Kazakov}},
		\bibinfo {author} {\bibfnamefont {R.}~\bibnamefont {Lammegger}}, \bibinfo
		{author} {\bibfnamefont {G.}~\bibnamefont {Mileti}}, \bibinfo {author}
		{\bibfnamefont {B.}~\bibnamefont {Matisov}}, \ and\ \bibinfo {author}
		{\bibfnamefont {L.}~\bibnamefont {Windholz}},\ }\href {\doibase
		10.1103/PhysRevA.79.063837} {\bibfield  {journal} {\bibinfo  {journal} {Phys.
				Rev. A}\ }\textbf {\bibinfo {volume} {79}},\ \bibinfo {pages} {063837}
		(\bibinfo {year} {2009})}\BibitemShut {NoStop}%
	\bibitem [{\citenamefont {Zibrov}\ \emph {et~al.}(2010)\citenamefont {Zibrov},
		\citenamefont {Novikova}, \citenamefont {Phillips}, \citenamefont
		{Walsworth}, \citenamefont {Zibrov}, \citenamefont {Velichansky},
		\citenamefont {Taichenachev},\ and\ \citenamefont
		{Yudin}}]{PhysRevA.81.013833}%
	\BibitemOpen
	\bibfield  {author} {\bibinfo {author} {\bibfnamefont {S.~A.}\ \bibnamefont
			{Zibrov}}, \bibinfo {author} {\bibfnamefont {I.}~\bibnamefont {Novikova}},
		\bibinfo {author} {\bibfnamefont {D.~F.}\ \bibnamefont {Phillips}}, \bibinfo
		{author} {\bibfnamefont {R.~L.}\ \bibnamefont {Walsworth}}, \bibinfo {author}
		{\bibfnamefont {A.~S.}\ \bibnamefont {Zibrov}}, \bibinfo {author}
		{\bibfnamefont {V.~L.}\ \bibnamefont {Velichansky}}, \bibinfo {author}
		{\bibfnamefont {A.~V.}\ \bibnamefont {Taichenachev}}, \ and\ \bibinfo
		{author} {\bibfnamefont {V.~I.}\ \bibnamefont {Yudin}},\ }\href {\doibase
		10.1103/PhysRevA.81.013833} {\bibfield  {journal} {\bibinfo  {journal} {Phys.
				Rev. A}\ }\textbf {\bibinfo {volume} {81}},\ \bibinfo {pages} {013833}
		(\bibinfo {year} {2010})}\BibitemShut {NoStop}%
	\bibitem [{\citenamefont {Mikhailov}\ \emph {et~al.}(2010)\citenamefont
		{Mikhailov}, \citenamefont {Horrom}, \citenamefont {Belcher},\ and\
		\citenamefont {Novikova}}]{Mikhailov:10}%
	\BibitemOpen
	\bibfield  {author} {\bibinfo {author} {\bibfnamefont {E.~E.}\ \bibnamefont
			{Mikhailov}}, \bibinfo {author} {\bibfnamefont {T.}~\bibnamefont {Horrom}},
		\bibinfo {author} {\bibfnamefont {N.}~\bibnamefont {Belcher}}, \ and\
		\bibinfo {author} {\bibfnamefont {I.}~\bibnamefont {Novikova}},\ }\href
	{\doibase 10.1364/JOSAB.27.000417} {\bibfield  {journal} {\bibinfo  {journal}
			{J. Opt. Soc. Am. B}\ }\textbf {\bibinfo {volume} {27}},\ \bibinfo {pages}
		{417} (\bibinfo {year} {2010})}\BibitemShut {NoStop}%
	\bibitem [{\citenamefont {Esnault}\ \emph {et~al.}(2013)\citenamefont
		{Esnault}, \citenamefont {Blanshan}, \citenamefont {Ivanov}, \citenamefont
		{Scholten}, \citenamefont {Kitching},\ and\ \citenamefont
		{Donley}}]{PhysRevA.88.042120}%
	\BibitemOpen
	\bibfield  {author} {\bibinfo {author} {\bibfnamefont {F.-X.}\ \bibnamefont
			{Esnault}}, \bibinfo {author} {\bibfnamefont {E.}~\bibnamefont {Blanshan}},
		\bibinfo {author} {\bibfnamefont {E.~N.}\ \bibnamefont {Ivanov}}, \bibinfo
		{author} {\bibfnamefont {R.~E.}\ \bibnamefont {Scholten}}, \bibinfo {author}
		{\bibfnamefont {J.}~\bibnamefont {Kitching}}, \ and\ \bibinfo {author}
		{\bibfnamefont {E.~A.}\ \bibnamefont {Donley}},\ }\href {\doibase
		10.1103/PhysRevA.88.042120} {\bibfield  {journal} {\bibinfo  {journal} {Phys.
				Rev. A}\ }\textbf {\bibinfo {volume} {88}},\ \bibinfo {pages} {042120}
		(\bibinfo {year} {2013})}\BibitemShut {NoStop}%
	\bibitem [{\citenamefont {{Ismail}}\ \emph {et~al.}(2016)\citenamefont
		{{Ismail}}, \citenamefont {{Kores}}, \citenamefont {{Geskus}},\ and\
		\citenamefont {{Pollnau}}}]{2016OExpr..2416366I}%
	\BibitemOpen
	\bibfield  {author} {\bibinfo {author} {\bibfnamefont {N.}~\bibnamefont
			{{Ismail}}}, \bibinfo {author} {\bibfnamefont {C.~C.}\ \bibnamefont
			{{Kores}}}, \bibinfo {author} {\bibfnamefont {D.}~\bibnamefont {{Geskus}}}, \
		and\ \bibinfo {author} {\bibfnamefont {M.}~\bibnamefont {{Pollnau}}},\ }\href
	{\doibase 10.1364/OE.24.016366} {\bibfield  {journal} {\bibinfo  {journal}
			{Opt. Express}\ }\textbf {\bibinfo {volume} {24}},\ \bibinfo {pages} {16366}
		(\bibinfo {year} {2016})}\BibitemShut {NoStop}%
	\bibitem [{\citenamefont {Poirson}\ \emph {et~al.}(1997)\citenamefont
		{Poirson}, \citenamefont {Bretenaker}, \citenamefont {Vallet},\ and\
		\citenamefont {Floch}}]{Poirson:97}%
	\BibitemOpen
	\bibfield  {author} {\bibinfo {author} {\bibfnamefont {J.}~\bibnamefont
			{Poirson}}, \bibinfo {author} {\bibfnamefont {F.}~\bibnamefont {Bretenaker}},
		\bibinfo {author} {\bibfnamefont {M.}~\bibnamefont {Vallet}}, \ and\ \bibinfo
		{author} {\bibfnamefont {A.~L.}\ \bibnamefont {Floch}},\ }\href {\doibase
		10.1364/JOSAB.14.002811} {\bibfield  {journal} {\bibinfo  {journal} {J. Opt.
				Soc. Am. B}\ }\textbf {\bibinfo {volume} {14}},\ \bibinfo {pages} {2811}
		(\bibinfo {year} {1997})}\BibitemShut {NoStop}%
	\bibitem [{\citenamefont {Akkermans}\ and\ \citenamefont
		{Dunne}(2012)}]{Akkermans2012}%
	\BibitemOpen
	\bibfield  {author} {\bibinfo {author} {\bibfnamefont {E.}~\bibnamefont
			{Akkermans}}\ and\ \bibinfo {author} {\bibfnamefont {G.~V.}\ \bibnamefont
			{Dunne}},\ }\href {\doibase 10.1103/physrevlett.108.030401} {\bibfield
		{journal} {\bibinfo  {journal} {Phys. Rev. Lett.}\ }\textbf {\bibinfo
			{volume} {108}},\ \bibinfo {pages} {030401} (\bibinfo {year}
		{2012})}\BibitemShut {NoStop}%
	\bibitem [{\citenamefont {Pinel}\ \emph {et~al.}(2015)\citenamefont {Pinel},
		\citenamefont {Everett}, \citenamefont {Hosseini}, \citenamefont {Campbell},
		\citenamefont {Buchler},\ and\ \citenamefont {Lam}}]{RN5}%
	\BibitemOpen
	\bibfield  {author} {\bibinfo {author} {\bibfnamefont {O.}~\bibnamefont
			{Pinel}}, \bibinfo {author} {\bibfnamefont {J.~L.}\ \bibnamefont {Everett}},
		\bibinfo {author} {\bibfnamefont {M.}~\bibnamefont {Hosseini}}, \bibinfo
		{author} {\bibfnamefont {G.~T.}\ \bibnamefont {Campbell}}, \bibinfo {author}
		{\bibfnamefont {B.~C.}\ \bibnamefont {Buchler}}, \ and\ \bibinfo {author}
		{\bibfnamefont {P.~K.}\ \bibnamefont {Lam}},\ }\href {\doibase
		10.1038/srep17633} {\bibfield  {journal} {\bibinfo  {journal} {Sci. Rep.}\
		}\textbf {\bibinfo {volume} {5}},\ \bibinfo {pages} {17633} (\bibinfo {year}
		{2015})}\BibitemShut {NoStop}%
	\bibitem [{\citenamefont {Carr}\ and\ \citenamefont
		{Purcell}(1954)}]{PhysRev.94.630}%
	\BibitemOpen
	\bibfield  {author} {\bibinfo {author} {\bibfnamefont {H.~Y.}\ \bibnamefont
			{Carr}}\ and\ \bibinfo {author} {\bibfnamefont {E.~M.}\ \bibnamefont
			{Purcell}},\ }\href {\doibase 10.1103/PhysRev.94.630} {\bibfield  {journal}
		{\bibinfo  {journal} {Phys. Rev.}\ }\textbf {\bibinfo {volume} {94}},\
		\bibinfo {pages} {630} (\bibinfo {year} {1954})}\BibitemShut {NoStop}%
	\bibitem [{\citenamefont {Viola}\ and\ \citenamefont
		{Lloyd}(1998)}]{Viola_1998}%
	\BibitemOpen
	\bibfield  {author} {\bibinfo {author} {\bibfnamefont {L.}~\bibnamefont
			{Viola}}\ and\ \bibinfo {author} {\bibfnamefont {S.}~\bibnamefont {Lloyd}},\
	}\href {\doibase 10.1103/physreva.58.2733} {\bibfield  {journal} {\bibinfo
			{journal} {Phys. Rev. A}\ }\textbf {\bibinfo {volume} {58}},\ \bibinfo
		{pages} {2733} (\bibinfo {year} {1998})}\BibitemShut {NoStop}%
	\bibitem [{\citenamefont {Degen}\ \emph {et~al.}(2017)\citenamefont {Degen},
		\citenamefont {Reinhard},\ and\ \citenamefont
		{Cappellaro}}]{RevModPhys.89.035002}%
	\BibitemOpen
	\bibfield  {author} {\bibinfo {author} {\bibfnamefont {C.~L.}\ \bibnamefont
			{Degen}}, \bibinfo {author} {\bibfnamefont {F.}~\bibnamefont {Reinhard}}, \
		and\ \bibinfo {author} {\bibfnamefont {P.}~\bibnamefont {Cappellaro}},\
	}\href {\doibase 10.1103/RevModPhys.89.035002} {\bibfield  {journal}
		{\bibinfo  {journal} {Rev. Mod. Phys.}\ }\textbf {\bibinfo {volume} {89}},\
		\bibinfo {pages} {035002} (\bibinfo {year} {2017})}\BibitemShut {NoStop}%
	\bibitem [{\citenamefont {Shahriar}\ \emph {et~al.}(2014)\citenamefont
		{Shahriar}, \citenamefont {Wang}, \citenamefont {Krishnamurthy},
		\citenamefont {Tu}, \citenamefont {Pati},\ and\ \citenamefont
		{Tseng}}]{Shahriar2014}%
	\BibitemOpen
	\bibfield  {author} {\bibinfo {author} {\bibfnamefont {M.}~\bibnamefont
			{Shahriar}}, \bibinfo {author} {\bibfnamefont {Y.}~\bibnamefont {Wang}},
		\bibinfo {author} {\bibfnamefont {S.}~\bibnamefont {Krishnamurthy}}, \bibinfo
		{author} {\bibfnamefont {Y.}~\bibnamefont {Tu}}, \bibinfo {author}
		{\bibfnamefont {G.}~\bibnamefont {Pati}}, \ and\ \bibinfo {author}
		{\bibfnamefont {S.}~\bibnamefont {Tseng}},\ }\href {\doibase
		10.1080/09500340.2013.865806} {\bibfield  {journal} {\bibinfo  {journal} {J.
				Mod. Opt}\ }\textbf {\bibinfo {volume} {61}},\ \bibinfo {pages} {351}
		(\bibinfo {year} {2014})}\BibitemShut {NoStop}%
	\bibitem [{SM()}]{SM}%
	\BibitemOpen
	\href@noop {} {}\bibinfo {note} {In the Supplementary Material, we present
		more details about: (i) the physical system and its model, (ii) analytical
		results for the temporal spinwave Fabry-Perot interferometry, and (iii)
		comparison among numerical, analytical and experimental results.}\BibitemShut
	{Stop}%
	\bibitem [{\citenamefont {Gorshkov}\ \emph {et~al.}(2007)\citenamefont
		{Gorshkov}, \citenamefont {Andr\'e}, \citenamefont {Fleischhauer},
		\citenamefont {S\o{}rensen},\ and\ \citenamefont
		{Lukin}}]{PhysRevLett.98.123601}%
	\BibitemOpen
	\bibfield  {author} {\bibinfo {author} {\bibfnamefont {A.~V.}\ \bibnamefont
			{Gorshkov}}, \bibinfo {author} {\bibfnamefont {A.}~\bibnamefont {Andr\'e}},
		\bibinfo {author} {\bibfnamefont {M.}~\bibnamefont {Fleischhauer}}, \bibinfo
		{author} {\bibfnamefont {A.~S.}\ \bibnamefont {S\o{}rensen}}, \ and\ \bibinfo
		{author} {\bibfnamefont {M.~D.}\ \bibnamefont {Lukin}},\ }\href {\doibase
		10.1103/PhysRevLett.98.123601} {\bibfield  {journal} {\bibinfo  {journal}
			{Phys. Rev. Lett.}\ }\textbf {\bibinfo {volume} {98}},\ \bibinfo {pages}
		{123601} (\bibinfo {year} {2007})}\BibitemShut {NoStop}%
	\bibitem [{\citenamefont {Chen}\ \emph {et~al.}(2000)\citenamefont {Chen},
		\citenamefont {Lin},\ and\ \citenamefont {Yu}}]{Chen_PRA_2000}%
	\BibitemOpen
	\bibfield  {author} {\bibinfo {author} {\bibfnamefont {Y.-C.}\ \bibnamefont
			{Chen}}, \bibinfo {author} {\bibfnamefont {C.-W.}\ \bibnamefont {Lin}}, \
		and\ \bibinfo {author} {\bibfnamefont {I.~A.}\ \bibnamefont {Yu}},\ }\href
	{\doibase 10.1103/PhysRevA.61.053805} {\bibfield  {journal} {\bibinfo
			{journal} {Phys. Rev. A}\ }\textbf {\bibinfo {volume} {61}},\ \bibinfo
		{pages} {053805} (\bibinfo {year} {2000})}\BibitemShut {NoStop}%
	\bibitem [{\citenamefont {Hemmer}\ \emph {et~al.}(1989)\citenamefont {Hemmer},
		\citenamefont {Shahriar}, \citenamefont {Natoli},\ and\ \citenamefont
		{Ezekiel}}]{Hemmer_1989}%
	\BibitemOpen
	\bibfield  {author} {\bibinfo {author} {\bibfnamefont {P.~R.}\ \bibnamefont
			{Hemmer}}, \bibinfo {author} {\bibfnamefont {M.~S.}\ \bibnamefont
			{Shahriar}}, \bibinfo {author} {\bibfnamefont {V.~D.}\ \bibnamefont
			{Natoli}}, \ and\ \bibinfo {author} {\bibfnamefont {S.}~\bibnamefont
			{Ezekiel}},\ }\href {\doibase 10.1364/JOSAB.6.001519} {\bibfield  {journal}
		{\bibinfo  {journal} {J. Opt. Soc. Am. B}\ }\textbf {\bibinfo {volume} {6}},\
		\bibinfo {pages} {1519} (\bibinfo {year} {1989})}\BibitemShut {NoStop}%
	\bibitem [{\citenamefont {Baumgart}\ \emph {et~al.}(2016)\citenamefont
		{Baumgart}, \citenamefont {Cai}, \citenamefont {Retzker}, \citenamefont
		{Plenio},\ and\ \citenamefont {Wunderlich}}]{Baumgart2016}%
	\BibitemOpen
	\bibfield  {author} {\bibinfo {author} {\bibfnamefont {I.}~\bibnamefont
			{Baumgart}}, \bibinfo {author} {\bibfnamefont {J.-M.}\ \bibnamefont {Cai}},
		\bibinfo {author} {\bibfnamefont {A.}~\bibnamefont {Retzker}}, \bibinfo
		{author} {\bibfnamefont {M.}~\bibnamefont {Plenio}}, \ and\ \bibinfo {author}
		{\bibfnamefont {C.}~\bibnamefont {Wunderlich}},\ }\href {\doibase
		10.1103/physrevlett.116.240801} {\bibfield  {journal} {\bibinfo  {journal}
			{Phys. Rev. Let.}\ }\textbf {\bibinfo {volume} {116}},\ \bibinfo {pages}
		{240801} (\bibinfo {year} {2016})}\BibitemShut {NoStop}%
	\bibitem [{\citenamefont {Chuchelov}\ \emph {et~al.}(2019)\citenamefont
		{Chuchelov}, \citenamefont {Tsygankov}, \citenamefont {Zibrov}, \citenamefont
		{Vaskovskaya}, \citenamefont {Vassiliev}, \citenamefont {Zibrov},
		\citenamefont {Yudin}, \citenamefont {Taichenachev},\ and\ \citenamefont
		{Velichansky}}]{Chuchelov2019}%
	\BibitemOpen
	\bibfield  {author} {\bibinfo {author} {\bibfnamefont {D.~S.}\ \bibnamefont
			{Chuchelov}}, \bibinfo {author} {\bibfnamefont {E.~A.}\ \bibnamefont
			{Tsygankov}}, \bibinfo {author} {\bibfnamefont {S.~A.}\ \bibnamefont
			{Zibrov}}, \bibinfo {author} {\bibfnamefont {M.~I.}\ \bibnamefont
			{Vaskovskaya}}, \bibinfo {author} {\bibfnamefont {V.~V.}\ \bibnamefont
			{Vassiliev}}, \bibinfo {author} {\bibfnamefont {A.~S.}\ \bibnamefont
			{Zibrov}}, \bibinfo {author} {\bibfnamefont {V.~I.}\ \bibnamefont {Yudin}},
		\bibinfo {author} {\bibfnamefont {A.~V.}\ \bibnamefont {Taichenachev}}, \
		and\ \bibinfo {author} {\bibfnamefont {V.~L.}\ \bibnamefont {Velichansky}},\
	}\href {\doibase 10.1063/1.5111312} {\bibfield  {journal} {\bibinfo
			{journal} {J. Appl. Phys.}\ }\textbf {\bibinfo {volume} {126}},\ \bibinfo
		{pages} {054503} (\bibinfo {year} {2019})}\BibitemShut {NoStop}%
	\bibitem [{\citenamefont {Siegman}(1986)}]{siegman1986lasers}%
	\BibitemOpen
	\bibfield  {author} {\bibinfo {author} {\bibfnamefont {A.}~\bibnamefont
			{Siegman}},\ }\href {https://books.google.co.kr/books?id=1BZVwUZLTkAC} {\emph
		{\bibinfo {title} {Lasers}}}\ (\bibinfo  {publisher} {University Science
		Books},\ \bibinfo {year} {1986})\BibitemShut {NoStop}%
	\bibitem [{\citenamefont {Fleischhauer}\ \emph {et~al.}(2005)\citenamefont
		{Fleischhauer}, \citenamefont {Imamoglu},\ and\ \citenamefont
		{Marangos}}]{RevModPhys.77.633}%
	\BibitemOpen
	\bibfield  {author} {\bibinfo {author} {\bibfnamefont {M.}~\bibnamefont
			{Fleischhauer}}, \bibinfo {author} {\bibfnamefont {A.}~\bibnamefont
			{Imamoglu}}, \ and\ \bibinfo {author} {\bibfnamefont {J.~P.}\ \bibnamefont
			{Marangos}},\ }\href {\doibase 10.1103/RevModPhys.77.633} {\bibfield
		{journal} {\bibinfo  {journal} {Rev. Mod. Phys.}\ }\textbf {\bibinfo {volume}
			{77}},\ \bibinfo {pages} {633} (\bibinfo {year} {2005})}\BibitemShut
	{NoStop}%
	\bibitem [{\citenamefont {S\'anchez}\ \emph {et~al.}(2008)\citenamefont
		{S\'anchez}, \citenamefont {L\'opez-Mon\'{\i}s},\ and\ \citenamefont
		{Platero}}]{PhysRevB.77.165312}%
	\BibitemOpen
	\bibfield  {author} {\bibinfo {author} {\bibfnamefont {R.}~\bibnamefont
			{S\'anchez}}, \bibinfo {author} {\bibfnamefont {C.}~\bibnamefont
			{L\'opez-Mon\'{\i}s}}, \ and\ \bibinfo {author} {\bibfnamefont
			{G.}~\bibnamefont {Platero}},\ }\href {\doibase 10.1103/PhysRevB.77.165312}
	{\bibfield  {journal} {\bibinfo  {journal} {Phys. Rev. B}\ }\textbf {\bibinfo
			{volume} {77}},\ \bibinfo {pages} {165312} (\bibinfo {year}
		{2008})}\BibitemShut {NoStop}%
	\bibitem [{\citenamefont {Donarini}\ \emph {et~al.}(2019)\citenamefont
		{Donarini}, \citenamefont {Niklas}, \citenamefont {Schafberger},
		\citenamefont {Paradiso}, \citenamefont {Strunk},\ and\ \citenamefont
		{Grifoni}}]{donarini2019coherent}%
	\BibitemOpen
	\bibfield  {author} {\bibinfo {author} {\bibfnamefont {A.}~\bibnamefont
			{Donarini}}, \bibinfo {author} {\bibfnamefont {M.}~\bibnamefont {Niklas}},
		\bibinfo {author} {\bibfnamefont {M.}~\bibnamefont {Schafberger}}, \bibinfo
		{author} {\bibfnamefont {N.}~\bibnamefont {Paradiso}}, \bibinfo {author}
		{\bibfnamefont {C.}~\bibnamefont {Strunk}}, \ and\ \bibinfo {author}
		{\bibfnamefont {M.}~\bibnamefont {Grifoni}},\ }\href {\doibase
		https://doi.org/10.1038/s41467-018-08112-x} {\bibfield  {journal} {\bibinfo
			{journal} {Nat. Commun.}\ }\textbf {\bibinfo {volume} {10}},\ \bibinfo
		{pages} {381} (\bibinfo {year} {2019})}\BibitemShut {NoStop}%
	\bibitem [{\citenamefont {Kelly}\ \emph {et~al.}(2010)\citenamefont {Kelly},
		\citenamefont {Dutton}, \citenamefont {Schlafer}, \citenamefont {Mookerji},
		\citenamefont {Ohki}, \citenamefont {Kline},\ and\ \citenamefont
		{Pappas}}]{kelly2010direct}%
	\BibitemOpen
	\bibfield  {author} {\bibinfo {author} {\bibfnamefont {W.~R.}\ \bibnamefont
			{Kelly}}, \bibinfo {author} {\bibfnamefont {Z.}~\bibnamefont {Dutton}},
		\bibinfo {author} {\bibfnamefont {J.}~\bibnamefont {Schlafer}}, \bibinfo
		{author} {\bibfnamefont {B.}~\bibnamefont {Mookerji}}, \bibinfo {author}
		{\bibfnamefont {T.~A.}\ \bibnamefont {Ohki}}, \bibinfo {author}
		{\bibfnamefont {J.~S.}\ \bibnamefont {Kline}}, \ and\ \bibinfo {author}
		{\bibfnamefont {D.~P.}\ \bibnamefont {Pappas}},\ }\href {\doibase
		10.1103/PhysRevLett.104.163601} {\bibfield  {journal} {\bibinfo  {journal}
			{Phys. Rev. Lett.}\ }\textbf {\bibinfo {volume} {104}},\ \bibinfo {pages}
		{163601} (\bibinfo {year} {2010})}\BibitemShut {NoStop}%
\end{thebibliography}

\begin{thebibliography}{7}%
	\makeatletter
	\providecommand \@ifxundefined [1]{%
		\@ifx{#1\undefined}
	}%
	\providecommand \@ifnum [1]{%
		\ifnum #1\expandafter \@firstoftwo
		\else \expandafter \@secondoftwo
		\fi
	}%
	\providecommand \@ifx [1]{%
		\ifx #1\expandafter \@firstoftwo
		\else \expandafter \@secondoftwo
		\fi
	}%
	\providecommand \natexlab [1]{#1}%
	\providecommand \enquote  [1]{``#1''}%
	\providecommand \bibnamefont  [1]{#1}%
	\providecommand \bibfnamefont [1]{#1}%
	\providecommand \citenamefont [1]{#1}%
	\providecommand \href@noop [0]{\@secondoftwo}%
	\providecommand \href [0]{\begingroup \@sanitize@url \@href}%
	\providecommand \@href[1]{\@@startlink{#1}\@@href}%
	\providecommand \@@href[1]{\endgroup#1\@@endlink}%
	\providecommand \@sanitize@url [0]{\catcode `\\12\catcode `\$12\catcode
		`\&12\catcode `\#12\catcode `\^12\catcode `\_12\catcode `\%12\relax}%
	\providecommand \@@startlink[1]{}%
	\providecommand \@@endlink[0]{}%
	\providecommand \url  [0]{\begingroup\@sanitize@url \@url }%
	\providecommand \@url [1]{\endgroup\@href {#1}{\urlprefix }}%
	\providecommand \urlprefix  [0]{URL }%
	\providecommand \Eprint [0]{\href }%
	\providecommand \doibase [0]{http://dx.doi.org/}%
	\providecommand \selectlanguage [0]{\@gobble}%
	\providecommand \bibinfo  [0]{\@secondoftwo}%
	\providecommand \bibfield  [0]{\@secondoftwo}%
	\providecommand \translation [1]{[#1]}%
	\providecommand \BibitemOpen [0]{}%
	\providecommand \bibitemStop [0]{}%
	\providecommand \bibitemNoStop [0]{.\EOS\space}%
	\providecommand \EOS [0]{\spacefactor3000\relax}%
	\providecommand \BibitemShut  [1]{\csname bibitem#1\endcsname}%
	\let\auto@bib@innerbib\@empty
	\bibitem [{\citenamefont {Shahriar}\ \emph {et~al.}(2014)\citenamefont
		{Shahriar}, \citenamefont {Wang}, \citenamefont {Krishnamurthy},
		\citenamefont {Tu}, \citenamefont {Pati},\ and\ \citenamefont
		{Tseng}}]{Shahriar2014}%
	\BibitemOpen
	\bibfield  {author} {\bibinfo {author} {\bibfnamefont {M.S.}\ \bibnamefont
			{Shahriar}}, \bibinfo {author} {\bibfnamefont {Ye}~\bibnamefont {Wang}},
		\bibinfo {author} {\bibfnamefont {Subramanian}\ \bibnamefont
			{Krishnamurthy}}, \bibinfo {author} {\bibfnamefont {Y.}~\bibnamefont {Tu}},
		\bibinfo {author} {\bibfnamefont {G.S.}\ \bibnamefont {Pati}}, \ and\
		\bibinfo {author} {\bibfnamefont {S.}~\bibnamefont {Tseng}},\ }\bibfield
	{title} {\enquote {\bibinfo {title} {Evolution of an {N}-level system via
				automated vectorization of the liouville equations and application to
				optically controlled polarization rotation},}\ }\href {\doibase
		10.1080/09500340.2013.865806} {\bibfield  {journal} {\bibinfo  {journal} {J.
				Mod. Opt}\ }\textbf {\bibinfo {volume} {61}},\ \bibinfo {pages} {351--367}
		(\bibinfo {year} {2014})}\BibitemShut {NoStop}%
	\bibitem [{\citenamefont {Baumgart}\ \emph {et~al.}(2016)\citenamefont
		{Baumgart}, \citenamefont {Cai}, \citenamefont {Retzker}, \citenamefont
		{Plenio},\ and\ \citenamefont {Wunderlich}}]{Baumgart2016}%
	\BibitemOpen
	\bibfield  {author} {\bibinfo {author} {\bibfnamefont {I.}~\bibnamefont
			{Baumgart}}, \bibinfo {author} {\bibfnamefont {J.-M.}\ \bibnamefont {Cai}},
		\bibinfo {author} {\bibfnamefont {A.}~\bibnamefont {Retzker}}, \bibinfo
		{author} {\bibfnamefont {M.{\hspace{0.167em}}B.}\ \bibnamefont {Plenio}}, \
		and\ \bibinfo {author} {\bibfnamefont {Ch.}\ \bibnamefont {Wunderlich}},\
	}\bibfield  {title} {\enquote {\bibinfo {title} {Ultrasensitive magnetometer
				using a single atom},}\ }\href {\doibase 10.1103/physrevlett.116.240801}
	{\bibfield  {journal} {\bibinfo  {journal} {Phys. Rev. Let.}\ }\textbf
		{\bibinfo {volume} {116}},\ \bibinfo {pages} {240801} (\bibinfo {year}
		{2016})}\BibitemShut {NoStop}%
	\bibitem [{\citenamefont {Steck}(2001)}]{steck2001rubidium}%
	\BibitemOpen
	\bibfield  {author} {\bibinfo {author} {\bibfnamefont {Daniel~A}\
			\bibnamefont {Steck}},\ }\href {http://steck.us/alkalidata} {\enquote
		{\bibinfo {title} {Rubidium 87 d line data},}\ } (\bibinfo {year}
	{2001})\BibitemShut {NoStop}%
	\bibitem [{\citenamefont {Chen}\ \emph {et~al.}(2000)\citenamefont {Chen},
		\citenamefont {Lin},\ and\ \citenamefont {Yu}}]{Chen_PRA_2000}%
	\BibitemOpen
	\bibfield  {author} {\bibinfo {author} {\bibfnamefont {Ying-Cheng}\
			\bibnamefont {Chen}}, \bibinfo {author} {\bibfnamefont {Chung-Wei}\
			\bibnamefont {Lin}}, \ and\ \bibinfo {author} {\bibfnamefont {Ite~A.}\
			\bibnamefont {Yu}},\ }\bibfield  {title} {\enquote {\bibinfo {title} {Roles
				of degenerate zeeman levels in electromagnetically induced transparency},}\
	}\href {\doibase 10.1103/PhysRevA.61.053805} {\bibfield  {journal} {\bibinfo
			{journal} {Phys. Rev. A}\ }\textbf {\bibinfo {volume} {61}},\ \bibinfo
		{pages} {053805} (\bibinfo {year} {2000})}\BibitemShut {NoStop}%
	\bibitem [{\citenamefont {Hemmer}\ \emph {et~al.}(1989)\citenamefont {Hemmer},
		\citenamefont {Shahriar}, \citenamefont {Natoli},\ and\ \citenamefont
		{Ezekiel}}]{Hemmer_1989}%
	\BibitemOpen
	\bibfield  {author} {\bibinfo {author} {\bibfnamefont {P.~R.}\ \bibnamefont
			{Hemmer}}, \bibinfo {author} {\bibfnamefont {M.~S.}\ \bibnamefont
			{Shahriar}}, \bibinfo {author} {\bibfnamefont {V.~D.}\ \bibnamefont
			{Natoli}}, \ and\ \bibinfo {author} {\bibfnamefont {S.}~\bibnamefont
			{Ezekiel}},\ }\bibfield  {title} {\enquote {\bibinfo {title} {Ac stark shifts
				in a two-zone {Raman} interaction},}\ }\href {\doibase
		10.1364/JOSAB.6.001519} {\bibfield  {journal} {\bibinfo  {journal} {J. Opt.
				Soc. Am. B}\ }\textbf {\bibinfo {volume} {6}},\ \bibinfo {pages} {1519--1528}
		(\bibinfo {year} {1989})}\BibitemShut {NoStop}%
	\bibitem [{\citenamefont {Chuchelov}\ \emph {et~al.}(2019)\citenamefont
		{Chuchelov}, \citenamefont {Tsygankov}, \citenamefont {Zibrov}, \citenamefont
		{Vaskovskaya}, \citenamefont {Vassiliev}, \citenamefont {Zibrov},
		\citenamefont {Yudin}, \citenamefont {Taichenachev},\ and\ \citenamefont
		{Velichansky}}]{Chuchelov2019}%
	\BibitemOpen
	\bibfield  {author} {\bibinfo {author} {\bibfnamefont {D.~S.}\ \bibnamefont
			{Chuchelov}}, \bibinfo {author} {\bibfnamefont {E.~A.}\ \bibnamefont
			{Tsygankov}}, \bibinfo {author} {\bibfnamefont {S.~A.}\ \bibnamefont
			{Zibrov}}, \bibinfo {author} {\bibfnamefont {M.~I.}\ \bibnamefont
			{Vaskovskaya}}, \bibinfo {author} {\bibfnamefont {V.~V.}\ \bibnamefont
			{Vassiliev}}, \bibinfo {author} {\bibfnamefont {A.~S.}\ \bibnamefont
			{Zibrov}}, \bibinfo {author} {\bibfnamefont {V.~I.}\ \bibnamefont {Yudin}},
		\bibinfo {author} {\bibfnamefont {A.~V.}\ \bibnamefont {Taichenachev}}, \
		and\ \bibinfo {author} {\bibfnamefont {V.~L.}\ \bibnamefont {Velichansky}},\
	}\bibfield  {title} {\enquote {\bibinfo {title} {Central ramsey fringe
				identification by means of an auxiliary optical field},}\ }\href {\doibase
		10.1063/1.5111312} {\bibfield  {journal} {\bibinfo  {journal} {J. Appl.
				Phys.}\ }\textbf {\bibinfo {volume} {126}},\ \bibinfo {pages} {054503}
		(\bibinfo {year} {2019})}\BibitemShut {NoStop}%
	\bibitem [{\citenamefont {Breschi}\ \emph {et~al.}(2009)\citenamefont
		{Breschi}, \citenamefont {Kazakov}, \citenamefont {Lammegger}, \citenamefont
		{Mileti}, \citenamefont {Matisov},\ and\ \citenamefont
		{Windholz}}]{PhysRevA.79.063837}%
	\BibitemOpen
	\bibfield  {author} {\bibinfo {author} {\bibfnamefont {E.}~\bibnamefont
			{Breschi}}, \bibinfo {author} {\bibfnamefont {G.}~\bibnamefont {Kazakov}},
		\bibinfo {author} {\bibfnamefont {R.}~\bibnamefont {Lammegger}}, \bibinfo
		{author} {\bibfnamefont {G.}~\bibnamefont {Mileti}}, \bibinfo {author}
		{\bibfnamefont {B.}~\bibnamefont {Matisov}}, \ and\ \bibinfo {author}
		{\bibfnamefont {L.}~\bibnamefont {Windholz}},\ }\bibfield  {title} {\enquote
		{\bibinfo {title} {Quantitative study of the destructive quantum-interference
				effect on coherent population trapping},}\ }\href {\doibase
		10.1103/PhysRevA.79.063837} {\bibfield  {journal} {\bibinfo  {journal} {Phys.
				Rev. A}\ }\textbf {\bibinfo {volume} {79}},\ \bibinfo {pages} {063837}
		(\bibinfo {year} {2009})}\BibitemShut {NoStop}%
	\bibitem [{\citenamefont {Taichenachev}\ \emph {et~al.}(2005)\citenamefont
		{Taichenachev}, \citenamefont {Yudin}, \citenamefont {Velichansky},\ and\
		\citenamefont {Zibrov}}]{Taichenachev2005a}%
	\BibitemOpen
	\bibfield  {author} {\bibinfo {author} {\bibfnamefont {A.~V.}\ \bibnamefont
			{Taichenachev}}, \bibinfo {author} {\bibfnamefont {V.~I.}\ \bibnamefont
			{Yudin}}, \bibinfo {author} {\bibfnamefont {V.~L.}\ \bibnamefont
			{Velichansky}}, \ and\ \bibinfo {author} {\bibfnamefont {S.~A.}\ \bibnamefont
			{Zibrov}},\ }\href {\doibase 10.1134/1.2142864} {\bibfield  {journal}
		{\bibinfo  {journal} {Jetp Lett.}\ }\textbf {\bibinfo {volume} {82}},\
		\bibinfo {pages} {398} (\bibinfo {year} {2005})}\BibitemShut {NoStop}%
\end{thebibliography}
%

\end{document}